\documentclass[12pt]{article}
\usepackage{amsfonts,mathtools}

\usepackage{dsfont}

\usepackage{caption}
\usepackage{hyperref}

\renewcommand{\baselinestretch}{1.2}

\usepackage[numbers]{natbib}
\newcommand{\tr}{\textrm{Tr}}

\begin{document}

\numberwithin{equation}{section}

\begin{titlepage}

\begin{center}
\hfill 
\\

\vspace{2cm}

{\Large\bf Fortuity on the conifold}

\vspace{2cm}

{\large Jaehyeok Choi$^{1}$, Sunjin Choi$^2$ and Seok Kim$^3$}

\vspace{0.7cm}

\textit{$^1$School of Physics, Korea Institute for Advanced Study,\\
85 Hoegi-ro, Dongdaemun-gu, Seoul 02455, Republic of Korea}\\

\vspace{0.2cm}

\textit{$^2$Kavli Institute for the Physics and Mathematics of the Universe (WPI),\\
The University of Tokyo Institutes for Advanced Study, The University of Tokyo,\\
Kashiwa, Chiba 277-8583, Japan}\\

\vspace{0.2cm}

\textit{$^3$Department of Physics and Astronomy \& Center for
Theoretical Physics,\\
Seoul National University, Seoul 08826, Korea}\\

\vspace{0.7cm}

E-mails: {\tt jaehyeokchoi@kias.re.kr, sunjin.choi@ipmu.jp,
seokkimseok@gmail.com }

\end{center}

\vspace{1cm}

\begin{abstract}

To study the BPS black hole states in typical 
$\mathcal{N}=1$ AdS$_5$/CFT$_4$ models, we formulate a classical  
cohomology problem in the Klebanov-Witten theory, which is defined at a strongly coupled RG fixed point. We set up its weakly coupled cohomology problem in the UV theory, 
suggesting a refined notion of monotonous and fortuitous cohomologies so that 
simple baryonic operators belong to the former. 
We construct infinitely many fortuitous cohomologies in the $SU(2)\times SU(2)$ theory 
and interpret some of them as $N=2$ versions of hairy black hole states dressed by 
baryonic condensates.  We also comment that the recently constructed fortuitous cohomologies in ABJM exhibit generalized monotonicity.

\end{abstract}

\end{titlepage}

\renewcommand{\baselinestretch}{1}

\tableofcontents

\renewcommand{\baselinestretch}{1.2}

\section{Introduction}

Studies of heavy BPS operators in holographic SCFTs have been initiated 
in recent years, to better understand the AdS black hole physics. 
Much of this work has been done in 4d $\mathcal{N}=4$ super-Yang-Mills (SYM) theory \cite{Chang:2022mjp,Choi:2022caq,Choi:2023znd,Budzik:2023vtr,Chang:2023ywj,Choi:2023vdm,
Chang:2024zqi,deMelloKoch:2024pcs,Gadde:2025yoa,Chang:2025mqp,Choi:2025bhi,Budzik:2025zvu,
Kim:2026rnx}. 
The basic framework in $\mathcal{N}=4$ SYM is to first explore classical
cohomologies \cite{Grant:2008sk,Chang:2013fba} for the weak-coupling BPS
operators, 
and then to distinguish the black hole states from gravitons  
at finite $N$ by the so-called fortuity criterion \cite{Chang:2024zqi}.
Representatives of fortuitous cohomologies are annihilated 
by a supercharge $Q$ by the trace relations of matrices at fixed $N$. 
On the other hand, the monotonous cohomologies admit universal representatives 
at arbitrary $N$. In $\mathcal{N}=4$ SYM, the supergraviton/black hole states 
belong to the monotonous/fortuitous classes, respectively.

In this paper, we suggest and study a similar program for the Klebanov-Witten (KW) 
theory \cite{Klebanov:1998hh}, expecting the ideas of this program to apply to 
other typical $\mathcal{N}=1$ AdS$_5$/CFT$_4$ models. 
This theory is an $SU(N)\times SU(N)$ quiver gauge theory, 
dual to the type IIB string theory on AdS$_5\times T^{1,1}$. We shall formulate 
the problem of weakly-coupled BPS states in terms of classical cohomologies, 
and then suggest a generalized fortuity criterion for the black hole 
states. Both steps demand more elaboration than in $\mathcal{N}=4$ SYM, 
as we shall highlight below.
We will also construct infinitely many black-hole-type cohomologies at $N=2$.

Unlike $\mathcal{N}=4$ SYM, many holographic SCFTs do not admit weakly-coupled limits. 
They are indirectly defined as RG fixed points, starting from relevant
deformations of UV SCFTs. We formulate the classical cohomology problem in the UV theory.
Starting from the conceptually straightforward definition in terms of the UV variables, 
we find a field redefinition to the more convenient `IR variables.' 
(A similar UV definition of classical cohomologies 
was explored for the Leigh-Strassler theory \cite{Choi:2025pqr}.) Whether 
these classical cohomologies are protected at the quantum level 
is a nontrivial issue \cite{Chang:2025mqp,Choi:2025bhi,Kim:2026rnx}, 
but we do not explore it in this paper.

In typical holographic  SCFTs, defining a good 
fortuity program for the black hole states is subtle due to the baryonic operators.
Baryons are determinant-type operators of matrix fields, whose 
$SU(N)$ gauge indices are contracted with the $\epsilon$ tensors. 
They explicitly depend on $N$ and do not fit into the naive definition 
of monotones. However, the simple baryons are 
dual to the D3-branes wrapping internal 3-cycles \cite{Gubser:1998fp,Berenstein:2002ke}, 
having nothing to do with the black hole states to the best of our knowledge. 
In this paper, we suggest a refined fortuity criterion which naturally puts these 
baryons into the generalized monotonous class, despite their apparent $N$ dependence. 
To illustrate the idea, consider the following operators with unit baryon charge, 
\begin{eqnarray}
  {\textstyle \frac{1}{N!}}\epsilon\epsilon(a^N)&=&(\det a)~{\bf 1}\\
  {\textstyle \frac{1}{(N-1)!}}\epsilon\epsilon(a^{N-1}\psi)&=&
  (\det a)~{\rm Tr}(a^{-1}\psi)\nonumber\\
  {\textstyle \frac{1}{(N-2)!}}\epsilon\epsilon(a^{N-2}\psi\chi)&=&
  (\det a)\left[{\rm Tr}(a^{-1}\psi){\rm Tr}(a^{-1}\chi)
  -{\rm Tr}(a^{-1}\psi a^{-1}\chi)\right]\ ,\nonumber
\end{eqnarray}
where $a$ is a scalar field and $\psi,\chi$ denote other 
fields. They can be written as products 
of the baryonic ground state operator $\det a$ 
which explicitly depends on $N$ and multi-trace operators which can be promoted to
universal $N$-independent forms. Although these multi-trace operators are not
legitimate local operators before multiplying $\det a$, 
they can be defined universally (as classical objects) for arbitrary $N$ at the 
price of introducing a formal letter $a^{-1}$. They can be used to construct generalized monotonous cohomologies in the baryonic sector, becoming $Q$-closed at any $N$. 
This idea is extended to multi-baryon charges in Section 3. Examples of 
baryonic monotonous cohomologies are presented in Section 4.  
The remaining cohomologies 
define our \textit{strongly fortuitous} cohomologies, 
which we propose as the black hole states.

The baryonic monotones are easier to study at $N=2$, due to an enhanced flavor
symmetry $SU(4) \supset SU(2)\times SU(2)\times U(1)_B$ which mixes mesons 
and baryons \cite{Klebanov:1998hh,Forcella:2008bb,Maruyoshi:2009uk}. Exploiting these structures, 
we shall present a characterization of generalized monotones at $N=2$ in Section 5.

We then construct the strongly fortuitous cohomologies in the KW theory at $N=2$. 
We first detect their existence by computing the index which subtracts the 
contributions from the graviton/baryon states, and then construct them 
using the ansatz of \cite{Choi:2023vdm}. While counting multi-graviton cohomologies 
(to be subtracted), we notice a curious partial failure of the counting by the 
`moduli space method' \cite{Choi:2023znd,Choi:2023vdm} and discuss its possible implications.
We also interpret the simplest fortuitous cohomology as a hairy black hole state, 
which suggests novel BPS hairy black holes in the AdS$_5\times S^5$ throat 
surrounded by baryonic condensate hair \cite{Klebanov:2007us}.

We also comment on the recently constructed fortuitous cohomologies 
\cite{Belin:2025hsg,Behan:2025hbx} in the ABJM theory \cite{Aharony:2008ug}, 
explaining that our refined notion of fortuity may reclassify them 
into monotones. Although the $U(N)$ type ABJM theory has no 
baryonic symmetries, multi-trace operators can factorize into baryon-antibaryon 
pairs at specific $N$ and exhibit generalized monotonicity. 
This may allude to an ambiguity in the fortuity program, to be 
fixed by (black hole) physics.

The rest of this paper is organized as follows. 
In Section 2, we set up the classical cohomology problem of the KW theory. 
Section 3 introduces the baryonic covering space and defines the refined notion 
of monotonicity and fortuity. Section 4 explains the mesonic and baryonic 
cohomologies in the monotonous class, and comments on a related 
cohomology problem in the 3d ABJM/BLG theories. In Section 5, we present 
new fortuitous/black hole cohomologies of the KW theory at $N=2$ and discuss
their possible interpretations.
Section 6 concludes with a discussion of future directions.

\section{Cohomologies of the Klebanov-Witten theory}

The Klebanov--Witten (KW) theory is an $\mathcal{N}=1$ superconformal field theory with gauge group $SU(N)\times SU(N)$ that arises at an interacting IR fixed point \cite{Klebanov:1998hh}. It is intrinsically strongly coupled, rather than a marginal deformation of a free gauge theory. The theory describes the low-energy dynamics of $N$ coincident D3-branes at the tip of the conifold. Its gravity dual is type $\rm IIB$ string theory on AdS$_5\times T^{1,1}$.

There are two $\mathcal{N}=1$ vector multiplets, $W_1$ and $W_2$, one for each gauge-group factor. The chiral multiplets $A_1,A_2$ transform in $(N,\bar N)$, while $B_1,B_2$ transform in $(\bar N,N)$. The superpotential is given by
\begin{equation}\label{superpotential}
    W=\frac{\lambda}{2}\epsilon^{ij}\epsilon^{kl}\tr A_iB_kA_jB_l
    =\lambda \tr(A_1B_1A_2B_2-A_1B_2A_2B_1)
    \;\;.
\end{equation}
Although this superpotential is nonrenormalizable in the UV, it is exactly marginal at the IR fixed point. One may therefore define the conifold theory by first setting $\lambda=0$ and flowing to the IR fixed point, and then turning on the superpotential deformation above \cite{Klebanov:1998hh}.

The continuous global symmetry is
\[
SU(2)\times SU(2)\times U(1)_R\times U(1)_B.
\]
The two $SU(2)$ factors rotate $A_i$ and $B_k$, respectively. In the convention $[r,Q_\alpha]=-Q_\alpha$, where $r$ is the superconformal $U(1)_R$ charge, 
all four chiral multiplets have $r=1/2$. Under the baryonic symmetry $U(1)_B$, $A_i$ have charge $+1$ and $B_k$ have charge $-1$.

The theory is also defined 
as the strongly-coupled IR fixed point of an RG flow triggered by a mass deformation 
of the weakly-coupled UV theory. As the UV theory, 
consider the $\mathcal{N}=2$ $U(N)\times U(N)$ theory with two hypermultiplets in $(N,\bar N)\oplus(\bar N,N)$. It describes $N$ coincident D3-branes at the orbifold fixed point of $\mathbb{C}^2/\mathbb{Z}_2\times\mathbb{C}$. The diagonal $U(1)$ of $U(N)\times U(N)$ is free and completely decoupled because the bifundamental matters 
are neutral under it. The relative $U(1)$ is IR-free and has a Landau pole in the UV. We therefore restrict attention to the interacting $SU(N)\times SU(N)$ sector.
In $\mathcal{N}=1$ language, the two hypermultiplets contain chirals $A_1,A_2$ in 
$(N,\bar N)$ and $B^1,B^2$ in $(\bar N,N)$. 
Decomposing the two $\mathcal{N}=2$ vector multiplets into the $\mathcal{N}=1$ multiplets, 
one obtains two $\mathcal{N}=1$ vector multiplets and 
adjoint chiral multiplets, respectively denoted $\Phi^{(1)}$ or $\Phi^{(2)}$. 
$\mathcal{N}=2$ supersymmetry demands the following superpotential, 
\begin{align}
    W_{\rm orb}=g_{YM}\tr\Phi^{(1)}A_mB^m+g_{YM}\tr \Phi^{(2)}B^mA_m\ .
\end{align}
Further introducing the following mass deformation to the superpotential 
\begin{align}
    \frac{M}{2}(\tr (\Phi^{(1)})^2-\tr (\Phi^{(2)})^2) ,
\end{align}
$\Phi^{(1)}$ and $\Phi^{(2)}$ become massive. Integrating them out replaces  
the adjoint chiral multiplets by the traceless part of $A_mB^m$ or $B^mA_m$, respectively.
After integrating them out, the superpotential in terms of the `light' bifundamental 
fields is given by
\begin{align}
    \frac{g_{YM}^2}{2M}(-\tr (A_mB^m)^2+\tr (B^mA_m)^2)
    =-\frac{g_{YM}^2}{M}\tr (A_1B^1A_2B^2-B^1A_1B^2A_2),
\end{align}
After converting to the conifold convention with $B^m=\epsilon^{m\hat i}B_{\hat i}$, 
one obtains
\begin{align}\label{conifold-superpotential}
    W=-\frac{g_{YM}^2}{M}\tr (A_1B_1A_2B_2-A_1B_2A_2B_1).
\end{align}
This is precisely the KW superpotential (\ref{superpotential}), 
with $\lambda=-g_{YM}^2/M$, so the mass-deformed orbifold theory flows 
to the KW fixed point.

\subsection{Cohomology problem}

In this paper, we are interested in the local gauge-invariant operators preserving 
2 supercharges, out of the $4$ Poincare supercharges 
$Q_\alpha$, $\overline{Q}_{\dot\alpha}$ and the $4$ conformal supercharges 
$S_\alpha$, $\overline{S}_{\dot\alpha}$. Here, $\alpha=\pm$ and $\dot\alpha=\pm$ 
are respectively the spinor indices of the Lorentz symmetry, or $\mathrm{Spin}(4)=SU(2)\times SU(2)$ of Euclidean QFT. In particular, we demand our BPS operators to be annihilated by 
$Q\equiv Q_-$ and $Q^\dag \equiv S^-$: note that they are mutually Hermitian conjugate 
in the radially quantized CFT. For those BPS operators, the scaling dimension $D$ of the 
operator (or the energy of the corresponding state) saturates the BPS bound, 
\begin{align}
    \{Q,Q^\dag\}=D-2j_1+\frac{3}{2}r \geq 0\ .
\end{align}
Here and below, $j_1$ and $j_2$ are respectively the Cartans of the two $SU(2)$'s 
on $\mathbb{R}^4$, and $r$ 
is the superconformal $U(1)_R$ charge normalized such that $Q_\alpha$ has charge $-1$.

Note that $Q$ is nilpotent, $Q^2=0$, from the algebra. Therefore, 
operators annihilated by $\{Q,Q^\dag\}=0$ are in 1-to-1 map to the 
cohomology classes of $Q$. Representatives of these cohomology classes
are given by $Q$-closed operators, identifying two $Q$-closed operators which 
differ by a $Q$-exact term. We shall be discussing the classification 
and constructions of these representatives of the $Q$-cohomologies.

Because the KW theory is defined at a strongly coupled IR fixed point, 
we have no explicit and practical way of defining its $Q$-cohomologies directly in IR. 
In this paper we shall discuss the \textit{classical} 
cohomology problem in the weakly-coupled UV SCFT with mass deformation, 
using the classical supercharge $Q$ of this theory. 
Under the RG flow, it may happen that the interactions lift some of the 
cohomologies constructed in the UV theory.
While we expect that such lifts may well happen for some operators, understanding 
the classical cohomologies would be an important first step towards better understanding 
the more challenging question of quantum cohomologies.

We shall first outline how we formulate the classical cohomologies from the UV 
description, and then present the technical details.

The starting point of our construction is the undeformed UV orbifold theory at $g_{YM}=M=0$. 
The free $Q$-cohomology contains letters (in the adjoint chiral multiplets)
which would become heavy in the IR after the mass deformation, 
and is therefore larger than the cohomology relevant to the KW theory. 
Even though we shall be working in the UV description, with `small' nonzero $M$,
we call these massive fields the heavy fields and the remaining fields 
the light fields.

We then turn on the gauge interactions and the mass deformation together, with $g_{YM}\neq0$ and $M\neq0$, and study the resulting classical $Q$-action. The interactions make this $Q$ nonlinear, while the mass deformation makes the adjoint chiral multiplets massive. After a field redefinition, we will show that every $Q$-closed operator containing a heavy letter is $Q$-exact. The cohomology consequently reduces to the light conifold letters with the $Q$-action in \eqref{eq:classical Q}.

At least two distinct quantum effects can modify this classical cohomology. First, even if the $Q$-transformation of each elementary field is kept fixed, a composite operator must be defined by renormalizing products of fields at coincident points. Interaction vertices generate short-distance contact terms in this renormalization, already in the undeformed theory with $M=0$. As a result, the action of $Q$ on a renormalized composite operator can contain multi-letter terms that are absent 
in the classical $Q$ which acts letter-by-letter and obeys the Leibniz rule.
Such anomaly-induced corrections can pair and lift classical cohomology classes, as recently demonstrated in $\mathcal{N}=4$ SYM \cite{Choi:2025bhi,Budzik:2025zvu,Kim:2026rnx}. Second, the component $Q$-action used here is an on-shell differential derived from the tree-level mass-deformed UV action. For chiral multiplets with a general K\"ahler potential, eliminating the auxiliary fields yields $Q\psi^i= F^i=-g^{i\bar j}\partial_{\bar j}\bar W$, where $g_{i\bar j}=\partial_i\partial_{\bar j}K$. Quantum corrections to the K\"ahler potential therefore modify the on-shell $Q$-action, or equivalently the expression for the Noether supercharge in terms of the renormalized component fields. This does not mean that the off-shell supersymmetry transformations or the supersymmetry algebra are renormalized; only their on-shell realization in the variables used for the cohomology problem changes. We leave both effects unaddressed in this paper.

The `light' UV BPS letters for the cohomologies of $Q=Q_-$ are given by
\[
\bar A^i,\quad \psi_{Ai+},\quad \bar B^{\hat i},\quad \psi_{B\hat i+},\quad
D_{+\dot\alpha},\quad \bar\lambda_{\dot\alpha}^{(k)},\quad F_{++}^{(k)}.
\]
In this and the next paragraphs, 
we shall first write down the $Q$ transformations that we shall use in this paper, 
which we shall argue in a moment below. The $Q$ transformation for the 
classical interacting field theory is (recall that $\lambda=-\frac{g_{\rm YM}^2}{M}$ 
from (\ref{conifold-superpotential}))
\begin{equation}
\begin{aligned}\label{eq:classical Q}
    [Q,\bar{A}^i]=&0=[Q,\bar{B}^{\hat i}]\;,\\
    \{Q,\psi_{Ai+}\}&=-\lambda \bar{B}^{\hat k}\bar{A}_i\bar{B}_{\hat k}\;,\\
    \{Q,\psi_{B{\hat k}+}\}&=\lambda \bar{A}^{i}\bar{B}_{\hat k}\bar{A}_{i}\;,\\
    \{Q,\bar{\lambda}_{\dot \beta}^{(i)}\}&=0\;,\\
    [Q,F_{++}^{(1)}]=g_{YM}&(\psi_{Ai+}\bar{A}^i-\bar{B}^{\hat i}\psi_{B\hat{i}+}-(trace))\;,\\
    [Q,F^{(2)}_{++}]=g_{YM}&
    (-\bar{A}^i\psi_{Ai+}+\psi_{B{\hat i}+}\bar{B}^{\hat i}-(trace))\;,\\
    [Q,D_{+\dot{\alpha}}O^{(i)}_{\mathrm{adj}}\}
    -&D_{+\dot{\alpha}}[Q,O^{(i)}_{\mathrm{adj}}\}
    =g_{YM}[\bar{\lambda}^{(i)}_{\dot{\alpha}},O^{(i)}_{\mathrm{adj}}\}
    \;,\;\\
    [Q,D_{+\dot{\alpha}}O_{\mathrm{bif}}\}
    -&D_{+\dot{\alpha}}[Q,O_{\mathrm{bif}}\}
    =g_{YM}\bar{\lambda}_{\dot{\alpha}}\cdot O_{\mathrm{bif}}
    \;.\;
\end{aligned}
\end{equation}
For a bifundamental operator $O_{\mathrm{bif}}$, fundamental in the $i$th 
gauge group and antifundamental in the $j$th, we use the notation
\[
\bar\lambda_{\dot\alpha}\cdot O_{\mathrm{bif}}
=\bar\lambda_{\dot\alpha}^{(i)}O_{\mathrm{bif}}
-(-1)^{|O_{\mathrm{bif}}|}O_{\mathrm{bif}}\bar\lambda_{\dot\alpha}^{(j)},
\]
where $|O_{\mathrm{bif}}|$ is its fermion number. In the transformations of $F_{++}^{(k)}$, the term $-(trace)$ removes the traceful component proportional to the $N\times N$ identity matrix. Note that because of this trace-part subtraction, the $Q$-action can increase the number of traces. Flavor indices are raised and lowered from the left with $\epsilon^{ij},\epsilon^{\hat i\hat j},\epsilon_{ij},\epsilon_{\hat i\hat j}$, with $\epsilon^{12}=1$. We reserve hatted indices for the second flavor $SU(2)$, except when the relevant group is already clear from the context.

It will be convenient to absorb the couplings into the following redefined fields:
\begin{equation}
\begin{aligned}\label{eq:redefined fields}
    a^i\equiv \bar{A}^i\;,\;\;
    b^{\hat i}\equiv \bar{B}^{\hat i}\;,\;\;
    \psi_{ai}\equiv \frac{\psi_{Ai+}}{\lambda}\;,\;\;
    \psi_{b\hat{i}}=\frac{\psi_{B\hat{i}+}}{\lambda}\;,\\
    \lambda_{1\dot \alpha}\equiv g_{YM}\bar{\lambda}_{\dot \alpha}^{(1)}\;,\;\;
    \lambda_{2\dot \alpha}\equiv g_{YM}\bar{\lambda}_{\dot \alpha}^{(2)}\;,\;\;
    f_1\equiv \frac{F^{(1)}_{++}}{g_{YM}\lambda}\;,\;\;
    f_2\equiv \frac{F^{(2)}_{++}}{g_{YM}\lambda}\;,
\end{aligned}
\end{equation}
In these variables, the classical transformations take the simpler form
\begin{equation}
\begin{aligned}\label{eq:Q-action}
    &Qa^i=Qb^{\hat i}=0\;,\\
    &Q\psi_{ai}=-b^{\hat k}a_ib_{\hat k}\;,\\
    &Q\psi_{b{\hat k}}=a^ib_{\hat k}a_i\;,\\
    &Q\lambda_{k\dot\alpha}=0\;,\\ 
    &[Q,D_{+\dot{\alpha}}]\bullet=\lambda_{\dot\alpha}\cdot\bullet\\
    &Qf_1=\psi_{ai}a^i-b^{\hat i}\psi_{b{\hat i}}-(trace)\;,\\
    &Qf_2=\psi_{b{\hat i}}b^{\hat i}-a^i\psi_{ai}-(trace)\;.
\end{aligned}
\end{equation}
Here and below, $QO$ abbreviates the graded commutator of $Q$ with $O$. 
The classical supercharge $Q$ obeys the graded Leibniz rule.

Now we explain the precise setup in which \eqref{eq:classical Q} should be 
understood. Very naively, (\ref{eq:classical Q}) may be derived from  
a `UV $\mathcal{N}=1$ theory' with four bifundamental chirals 
$A_i,B_{\hat j}$,
\[
K=\tr \bar A^i e^{2g_{YM}V\cdot}A_i+\tr \bar B^{\hat i}e^{2g_{YM}V\cdot}B_{\hat i},
\qquad
W=\frac{\lambda}{2}\tr A_iB_{\hat k}A^iB^{\hat k}.
\]
However, this is not a UV-renormalizable QFT 
because $\tr ABAB$ has dimension four. Therefore, 
we shall instead present a UV-complete derivation which starts from the conformal and
renormalizable $\mathcal{N}=2$ theory on $\mathbb{C}^2/\mathbb{Z}_2\times\mathbb{C}$, followed by a mass deformation \cite{Klebanov:1998hh}. 
The $Q$ transformation of \eqref{eq:classical Q}
will be derived from the classical $Q$ transformation of this UV theory.

In the UV orbifold theory, BPS letters of the free theory (at $g_{\rm YM}=0$, $M=0$) 
are given by
\begin{align}
    \bar{A}^m,\psi_{Am+},\bar{B}^m,\psi_{Bm+},D_{+\dot{\alpha}},\bar{\lambda}_{\dot \alpha}^{(k)},F^{(k)}_{++},\;\; \text{and}\;\; 
    \bar{\Phi}^{(k)},\tilde{\lambda}^{(k)}_{+},
\end{align}
where $\tilde\lambda_+^{(k)}$ denote the $\alpha=+$ components of the 
fermions in the adjoint chiral multiplets (which are gauginos 
in the $\mathcal{N}=2$ vector multiplets).
Relative to the KW theory, the additional heavy letters are $\bar\Phi^{(k)}$ and $\tilde\lambda_+^{(k)}$, together with their $D_{+\dot\alpha}$ derivatives. Our conventions are
\[
\bar B^m=\epsilon^{mn}\bar B_n,\qquad
\psi_{Bm+}=\epsilon_{mn}\psi_{B+}^n=\epsilon_{mn}[Q_+,B^n].
\]
Now with $g_{\rm YM}\neq 0$, $M\neq 0$, 
the classical $Q$ transforms these fields by
\begin{equation*}
\begin{aligned}
    \{Q,\psi_{Am+}\}&=g_{YM}(\bar{\Phi}^{(1)}\bar{B}_m+\bar{B}_m\bar{\Phi}^{(2)})\;,\\
    \{Q,\psi_{Bm+}\}&=g_{YM}(\bar{A}_m\bar{\Phi}^{(1)}+\bar{\Phi}^{(2)}\bar{A}_m)\;,\\
    \{Q,\tilde{\lambda}^{(1)}_+\}&=g_{YM}\bar{B}_m\bar{A}^m+M\bar{\Phi}^{(1)}\;,\\
    \{Q,\tilde{\lambda}^{(2)}_+\}&=g_{YM}\bar{A}^m\bar{B}_m-M\bar{\Phi}^{(2)}\;,\\
    [Q,F_{++}^{(1)}]=g_{YM}(\psi_{Am+}&\bar{A}^m-\bar{B}_{m}\psi_{B+}^m-(trace))+g_{YM}[\tilde{\lambda}_{+}^{(1)},\bar{\Phi}^{(1)}]\;,\\
    [Q,F^{(2)}_{++}]=g_{YM}
    (-\bar{A}^m&\psi_{Am+}+\psi_{B+}^m\bar{B}_m-(trace))+g_{YM}[\tilde{\lambda}_{+}^{(2)},\bar{\Phi}^{(2)}]\;.
\end{aligned}
\end{equation*}
The fields $\bar A^m,\bar B^m,\bar\Phi^{(k)}$, and $\bar\lambda_{\dot\alpha}^{(k)}$ are $Q$-closed, and the action on $D_{+\dot\alpha}$ is unchanged from \eqref{eq:classical Q}.
Note that in this paragraph, the indices $m=1,2$ 
are for the doublet of the $SU(2)_R$ symmetry of the $\mathcal{N}=2$ theory. 
Also note that at this point, the orbifold and conifold conventions for $B$ differ. In the orbifold setup of this paragraph, $(B^m)^\dagger=\bar B_m$, and hence $(B_m)^\dagger=-\bar B^m$; this choice makes $SU(2)_R$ manifest. In the conifold discussion we instead use $(B_m)^\dagger=\bar B^m$. Thus $(\bar B^m)_{\mathrm{orbifold}}$ corresponds to $-(\bar B^m)_{\mathrm{conifold}}$. Replacing $\bar B^m\to-\bar B^m$ by changing the convention, 
one obtains 
\begin{equation}
\begin{aligned}
    \{Q,\psi_{Am+}\}&=-g_{YM}(\bar{\Phi}^{(1)}\bar{B}_m+\bar{B}_m\bar{\Phi}^{(2)})\;,\\
    \{Q,\psi_{Bm+}\}&=g_{YM}(\bar{A}_m\bar{\Phi}^{(1)}+\bar{\Phi}^{(2)}\bar{A}_m)\;,\\
    \{Q,\tilde{\lambda}^{(1)}_+\}&=-g_{YM}\bar{B}_m\bar{A}^m+M\bar{\Phi}^{(1)}\;,\\
    \{Q,\tilde{\lambda}^{(2)}_+\}&=-g_{YM}\bar{A}^m\bar{B}_m-M\bar{\Phi}^{(2)}\;,\\
    [Q,F_{++}^{(1)}]=g_{YM}(\psi_{Am+}&\bar{A}^m-\bar{B}^{m}\psi_{Bm+}-(trace))+g_{YM}[\tilde{\lambda}_{+}^{(1)},\bar{\Phi}^{(1)}]\;,\\
    [Q,F^{(2)}_{++}]=g_{YM}
    (-\bar{A}^m&\psi_{Am+}+\psi_{Bm+}\bar{B}^m-(trace))+g_{YM}[\tilde{\lambda}_{+}^{(2)},\bar{\Phi}^{(2)}]\;,
\end{aligned}
\end{equation}
which we shall be using in the rest of this paper for discussing 
the $\mathcal{N}=1$ cohomologies. 

Recall that we only want to use the light fields to construct 
the cohomologies, because we want to study the BPS spectrum of the IR theory after 
the mass deformation. Morally, we would like to avoid using the heavy letters in the
adjoint chiral multiplets. 
However, since such naively defined heavy fields appear on the right-hand 
sides of the $Q$ transformations of the light fields, 
one would like to make suitable field redefinitions to consistently 
ignore certain heavy fields. The basic consistency requirement is that, 
if one divides the letters into `heavy' ones $H$ (not to be used to construct 
cohomologies) and the `light' ones $L$ (to be used), one should make sure that 
$f(L)\neq QH$ for any expressions made out of the light letters $L$. This is because, 
if such a relation $f(L)=QH$ holds for some $f(L)$, then one would incorrectly 
forget that $f(L)$ can be made $Q$-exact by ignoring the heavy letter $H$.

To start with, note that 
the $Q$ transformations of $\tilde\lambda_+^{(k)}$ can be solved for the adjoint scalars:
\begin{align}
    \bar{\Phi}^{(1)}=\frac{g_{YM}}{M}\bar{B}_m\bar{A}^m+\frac{1}{M}Q\tilde{\lambda}^{(1)}_+\;,\;\;
    \bar{\Phi}^{(2)}=-\frac{g_{YM}}{M}\bar{A}^m\bar{B}_m-\frac{1}{M}Q\tilde{\lambda}^{(2)}_+\;.
\end{align}
Substitution into the chiralino transformations yields
\begin{align*}
    Q\psi_{Am+}=&-\frac{g_{YM}^2}{M}(\bar{B}_n\bar{A}^n\bar{B}_m-\bar{B}_m\bar{A}^n\bar{B}_n)
    -\frac{g_{YM}}{M}(Q\tilde{\lambda}^{(1)}_+)\bar{B}_m+\frac{g_{YM}}{M}\bar{B}_m(Q\tilde{\lambda}^{(2)}_+),
    \nonumber\\
    Q\psi_{Bm+}=&\frac{g_{YM}^2}{M}(\bar{A}_m\bar{B}_n\bar{A}^n-\bar{A}^n\bar{B}_n\bar{A}_m)+\frac{g_{YM}}{M}\bar{A}_m(Q\tilde{\lambda}^{(1)}_+)
    -\frac{g_{YM}}{M}(Q\tilde{\lambda}^{(2)}_+)\bar{A}_m,\nonumber
\end{align*}
Equivalently, the $Q$-exact pieces can be absorbed into redefined chiralini:
\begin{equation}
\begin{aligned}
    Q\Big(\psi_{Am+}+\frac{g_{YM}}{M}(\tilde{\lambda}^{(1)}_+\bar{B}_m-\bar{B}_m\tilde{\lambda}^{(2)}_+)\Big)
    &=\frac{g^2_{YM}}{M}\bar{B}^n \bar{A}_m\bar{B}_n,\\
    Q\Big(\psi_{Bm+}+\frac{g_{YM}}{M}(\tilde{\lambda}^{(2)}_+\bar{A}_m-\bar{A}_m\tilde{\lambda}^{(1)}_+)\Big)
    &=-\frac{g_{YM}^2}{M}\bar{A}^n\bar{B}_m\bar{A}_n,
\end{aligned}
\end{equation}
We would like to regard the expressions inside $Q(\cdots)$ as the redefined 
chiralino fields. Similarly, the $Q$ transformations of the field strengths 
can be written as
\begin{equation}
\begin{aligned}
    &Q\big(F_{++}^{(1)}+\frac{g_{YM}}{M}(\tilde{\lambda}_+^{(1)})^2\big) \\
    ={}&g_{YM}\Big(\psi_{Am+}+\frac{g_{YM}}{M}\tilde{\lambda}_+\cdot \bar{B}_m\Big)\bar{A}^m
    -g_{YM}\bar{B}^m\Big(\psi_{Bm+}+\frac{g_{YM}}{M}\tilde{\lambda}_+\cdot \bar{A}_m\Big)-(trace),\\
    &Q\big(F_{++}^{(2)}-\frac{g_{YM}}{M}(\tilde{\lambda}_+^{(2)})^2\big)\\
    ={}&-g_{YM}\bar{A}^m\Big(\psi_{Am+}+\frac{g_{YM}}{M}\tilde{\lambda}_+\cdot \bar{B}_m\Big)
    +g_{YM}\Big(\psi_{Bm+}+\frac{g_{YM}}{M}\tilde{\lambda}_+\cdot \bar{A}_m\Big)\bar{B}^m-(trace),
\end{aligned}
\end{equation}
such that the right-hand sides depend on the redefined chiralino fields only. 
Here $\tilde\lambda\cdot\bar B=\tilde\lambda^{(1)}\bar B-\bar B\tilde\lambda^{(2)}$ and $\tilde\lambda\cdot\bar A=\tilde\lambda^{(2)}\bar A-\bar A\tilde\lambda^{(1)}$.
These results motivate the following combinations to be identified as the IR conifold letters:
\begin{equation}\label{light-letters}
\begin{aligned}
    \bar{A}_{IR,m}=\Big(\frac{-\lambda M}{g_{YM}^2}&\Big)^{-1/4}\bar{A}_m,
    \qquad
    \bar{B}_{IR,m}=\Big(\frac{-\lambda M}{g_{YM}^2}\Big)^{-1/4}\bar{B}_m,\\
    \psi_{IR,Am+}=\Big(&\frac{-\lambda M}{g_{YM}^2}\Big)^{1/4}
    \Big(\psi_{Am+}+\frac{g_{YM}}{M}\tilde{\lambda}_+\cdot \bar{B}_m\Big),
    \\
    \psi_{IR,Bm+}=\Big(&\frac{-\lambda M}{g_{YM}^2}\Big)^{1/4}
    \Big(\psi_{Bm+}+\frac{g_{YM}}{M}\tilde{\lambda}_+\cdot \bar{A}_m\Big),\\
    F^{(1)}_{IR,++}=F^{(1)}_{++}+\frac{g_{YM}}{M}&(\tilde{\lambda}_+^{(1)})^2,\qquad
    F^{(2)}_{IR,++}=F_{++}^{(2)}-\frac{g_{YM}}{M}(\tilde{\lambda}_+^{(2)})^2,
\end{aligned}
\end{equation}
These IR combinations obey \eqref{eq:classical Q}. They will be our `light letters' 
in the UV setup for the cohomology.

To complete the setup, the heavy letters can be chosen as
\begin{equation}\label{heavy-letters}
\begin{aligned}
    \zeta^{(1)}=M^{1/2}\Big(\bar{\Phi}^{(1)}-\frac{g_{YM}}{M}\bar{B}_m\bar{A}^m\Big),\qquad \chi^{(1)}=M^{-1/2}\tilde{\lambda}^{(1)}_+,\\
    \zeta^{(2)}=M^{1/2}\Big(\bar{\Phi}^{(2)}+\frac{g_{YM}}{M}\bar{A}^m\bar{B}_m\Big),\qquad \chi^{(2)}=-M^{-1/2}\tilde{\lambda}^{(2)}_+,
\end{aligned}
\end{equation}
The action of $Q$ on these heavy letters is given by 
\begin{align}
    Q\zeta^{(k)}=0\ \ ,\ \ Q\chi^{(k)}=\zeta^{(k)}.
\end{align}
This implies that 
keeping the `light' letters (\ref{light-letters}) and not using the 
`heavy' letters (\ref{heavy-letters}) is consistent.
Also note that, apart from the overall factors of $M$, $g_{\rm YM}$ in the 
definitions of the redefined letters, the formal $\frac{g_{\rm YM}}{M}\rightarrow 0$ 
limit yields the naively defined light/heavy letters as discussed above. 
This intuitively justifies the consistently chosen $L$'s and $H$'s above from the 
perspective of the UV weakly-coupled theory.

So far, we argued that the truncated cohomology problem using the redefined letters is consistent. 
One can make a stronger statement by discussing the cohomologies which contain 
the heavy fields $H=\{\zeta^{(k)},\chi^{(k)}\}$. That is, one can argue 
that cohomologies containing $H$ are $Q$-exact, i.e. trivial \cite{Choi:2025pqr}. 
We start by defining an auxiliary fermionic operation $S_H$ by
\begin{align}
    S_H\zeta^{(k)}=\chi^{(k)},\qquad
    S_H\chi^{(k)}=0,\qquad S_H(\textrm{all other letters})=0\ .
\end{align}
We extend \(S_H\) to derivative descendants by requiring it to commute with \(D_{+\dot\alpha}\). Since \([Q,D_{+\dot\alpha}]\) acts through the light gaugino \(\lambda_{\dot\alpha}\), which is annihilated by \(S_H\), the relation \(\{Q,S_H\}=N_H\) continues to hold on derivative descendants.
Its action on composite operators is fixed by the graded Leibniz rule. It then follows that $\{Q,S_H\}$ acts as the identity on 
the heavy letters and annihilates the light letters. On a composite operator, it therefore counts the total number $N_H$ of heavy letters:
\begin{align}
    N_H=\{Q,S_H\}.
\end{align}
Since $[Q,N_H]=0$, a $Q$-closed operator $O$ may be chosen to have a definite value $N_H=k$. If $k>0$, then
\begin{align}
    O=\frac{\{Q,S_H\}}{k}O=Q(S_HO/k).
\end{align}
So $Q$-closed operators containing heavy letters are $Q$-exact.

Equation \eqref{eq:Q-action} shows that the classical $Q$-action preserves the number $\dot n$ of dotted Lorentz spinor indices (such as $\dot\alpha$ in \eqref{eq:Q-action}):
letters without dotted indices map to expressions without dotted indices, while letters carrying dotted indices map within the dotted sector. Thus, $\dot n$ is conserved by the classical cohomology problem.\footnote{
This conservation law is accidental and is broken by the one-loop correction of order $g_{YM}^2$ to the supercharge \cite{Choi:2025bhi,Budzik:2025zvu}. The same type of correction is present in $\mathcal{N}=1$ gauge theories, including the KW theory.}
We may therefore consistently restrict the classical cohomology problem 
to any fixed-$\dot n$ sector. In Section \ref{Section:fortuitous}, when we construct new fortuitous 
cohomologies we shall focus on the so-called BMN sector defined by $\dot n=0$
\cite{Choi:2023znd,Kim:2003rza,Berenstein:2002jq}. 
The BMN letters are $a^i,b^{\hat i},\psi_{ai},\psi_{b\hat i},f_1,f_2$. 
Their $Q$-transformations are given by 
\begin{equation}
\begin{aligned}
    &Qa^i=Qb^{\hat i}=0\;,\\
    &Q\psi_{ai}=-b^{\hat k}a_ib_{\hat k}\;,\\
    &Q\psi_{b{\hat k}}=a^ib_{\hat k}a_i\;,\\
    &Qf_1=\psi_{ai}a^i-b^{\hat i}\psi_{b{\hat i}}-(trace)\;,\\
    &Qf_2=\psi_{b{\hat i}}b^{\hat i}-a^i\psi_{ai}-(trace)\;.
    \label{eq:BMN-sector-Q}
\end{aligned}
\end{equation}

We also discuss the superconformal index, which counts these cohomologies 
with $-1$ sign for fermionic ones. 
This index is a partition function with $(-1)^F$ inserted, 
which counts the local operators annihilated by $Q=Q_-$. It is defined by 
\begin{align}\label{eq:index-definition}
    \mathcal{I}(t,y,u,v,b)=\tr (-1)^F t^{3(2j_1-r)}y^{2j_2}
    u^{2m_1}v^{2m_2}b^{B}\;.
\end{align}
The quantum numbers $j_1$ and $j_2$ are the half-integrally normalized Cartan generators of $SU(2)\times SU(2)=\mathrm{Spin}(4)$. The first Lorentz $SU(2)$ acts on undotted indices $\alpha$, and the second acts on dotted indices $\dot\alpha$. The charge $r$ is the superconformal R-charge at the IR fixed point. 
The charges $m_1,m_2$ generate the Cartan subgroup of the flavor symmetry $SU(2)\times SU(2)$, and $B$ denotes baryon number.
Only the local operators on $\mathbb{R}^4$ annihilated by $Q_-$ and $S^-$ 
contribute to the index, since the charges appearing in the trace of 
(\ref{eq:index-definition}) commute with these supercharges.

Since the BPS states and the cohomologies are in 1-to-1 map, 
one may regard the index as counting the cohomologies. 
Furthermore, if one takes $Q$ to be the classical supercharge as discussed in 
this section, the index may be regarded as counting the classical cohomologies. 
(Those which may pair and get lifted by quantum effects are canceled.)

As the number of dotted indices $\dot n$ is a conserved quantity in the classical cohomology problem, one may define a refined index which counts the classical cohomologies with a fixed $\dot n$. (At quantum level $\dot{n}$ is no longer a conserved quantity, so the $\dot{n}$-refined index is not protected at the quantum level.)
Focusing on the $\dot{n}=0$ sector, one obtains the BMN index, which we use in Section \ref{Section:fortuitous}. Note that 
$\dot{n}=0$ implies $j_2=0$. The BMN index is defined by
\begin{align}\label{eq:BMN-index-definition}
    \mathcal{I}_{\rm BMN}(t,u,v,b)=\tr_{\rm BMN} (-1)^F t^{3(2j_1-r)}u^{2m_1}v^{2m_2}b^{B}\;,
\end{align}
where the trace is restricted to the operators with $\dot n=0$. 

Appendix \ref{appendix:finite-N-index} presents the matrix integral formulae 
for these indices.

\subsection{Truncation to $\mathcal{N}=4$ SYM cohomologies}\label{subsec:KW-to-SYM}

In this subsection, we consider a Higgs-branch (more specifically, baryonic) vacuum of the KW theory that flows in the IR to $SU(N)$ $\mathcal{N}=4$ SYM \cite{Klebanov:2007us}.
This RG flow will motivate us to develop a consistent truncation of the cohomologies 
of the KW theory, down to those of the $SU(N)$ SYM. Such a truncation will be 
useful in Section 5 for arguing non-$Q$-exactness of certain cohomologies in the 
KW theory, and will also provide their physical interpretation.

Recall that KW theory describes the dynamics of $N$ D3-branes on the conifold. 
When the branes are located at the singular point of the conifold, one obtains the 
KW SCFT at low energy. On the other hand, if the $N$ D-branes 
are located at a smooth point of the conifold, the low-energy interacting theory is $SU(N)$ $\mathcal{N}=4$ SYM because the transverse geometry is locally $\mathbb{C}^3$. 
Therefore, turning on such a nonzero vacuum expectation value in the KW theory, 
one expects an RG flow along this Higgs/baryonic branch to $SU(N)$ $\mathcal N=4$ SYM.

To implement this explicitly in the KW theory, recall that the $SU(N)\times SU(N)$ KW theory together with the decoupled free $U(1)$ vector multiplet describes the full low-energy D3-brane system, including its center-of-mass motion.\footnote{We will not distinguish carefully between $U(N)=(SU(N)\times U(1))/\mathbb{Z}_N$ and $SU(N)\times U(1)$. This global distinction does not affect the local operators relevant to the cohomology problem.} At a vacuum where all $A_i$ and $B_{\hat j}$ are proportional to 
the $N\times N$ identity matrix, the relative gauge field $A_\mu^{(1)}-A_\mu^{(2)}$ becomes massive by combining with one traceless chiral fluctuation. The remaining three traceless chiral superfields transform in the adjoint representation of the unbroken diagonal $SU(N)$ and furnish the three adjoint chiral multiplets of $\mathcal{N}=4$ SYM.

The trace components of the four original chiral multiplets remain massless. They decouple in the IR because their only interactions inherited from the UV superpotential are quartic and hence irrelevant. Three of them combine with the free $U(1)$ vector multiplet into the abelian $\mathcal{N}=4$ multiplet describing the center-of-mass motion in $\mathbb{C}^3$. The fourth is associated with the complexified K\"ahler modulus of the conifold: it parametrizes the complexified resolution, whose real part can be identified with $\mathcal U$ in \eqref{eq:D-term}. This additional chiral is present because the gauge group is $SU(N)\times SU(N)$, allowing the resolution or baryonic modulus.

Motivated by the above RG flow on the moduli space, we would like to establish 
a consistent truncation of the cohomology problem of the KW theory. 
The basic idea is explained in Section 2.1. If one can split the letters 
of the cohomology problem into `heavy' ones $H$ and the `light' ones $L$, 
in the sense that no combinations of $L$'s can be written as $f(L)=QH$, then 
not using $H$'s in the cohomology problem is consistent.

For definiteness, we pick the special point on the moduli space to be 
\begin{align}
    A_1=0,\quad A_2=0,\quad B_1=0,\quad B_2=v \mathds{1}_N. 
\end{align}
Using the \(U(1)_B\) symmetry, we take \(v>0\) without loss of generality.
This configuration obeys both the F-term and D-term conditions \eqref{eq:D-term}, 
with $\mathcal U=-|v|^2\neq0$. We parametrize the fluctuations of 
$\bar A^1,\bar A^2,\bar B^1,\bar B^2$ around this point by
\begin{equation}
\begin{aligned}\label{eq:scalar decomposition}
    \bar{A}^1&=v^{-1/3}(X+x\mathds{1}_N), \\
    \bar{A}^2&=v^{-1/3}(Y+y\mathds{1}_N),\\ 
    \bar{B}^1&=v^{-1/3}(Z+z\mathds{1}_N),\\
    \bar{B}^2=v&\mathds{1}_N+v^{-1/3}(w\mathds{1}_N+W),
\end{aligned}
\end{equation}
where the matrices $X,Y,Z,W$ are traceless and $x,y,z,w$ are trace modes. The factor $v^{-1/3}$ ensures that the fluctuations $X,Y,Z,W$ and $x,y,z,w$ 
all have mass dimension one.

In this setup, we take the `light' letters to contain $L\supset \{X,Y,Z\}$ 
and the `heavy' letters to contain $H\supset \{x,y,z,w,W\}$. 
Here, heavy letters merely denote those not to be used in the truncated cohomology problem. $W$ is indeed a heavy field in IR in the RG sense, 
but $x,y,z,w$ are light and IR-free, 
which we put into $H$ because we shall not use them. 
(`Light' or `heavy' fields here follow the similar notions 
in the consistent truncation of supergravity.)

We now present the light-heavy decomposition of all the remaining letters. 
We first decompose $3$ chiralinos $\psi_{a1}$, $\psi_{a2}$, $\psi_{b1}$ 
to the traceless parts $\psi_1$, $\psi_2$, $\psi_3$ and the trace parts as 
\begin{equation}\label{eq:fermion decomposition}
    \psi_{a1}=v^{1/3}\psi_1+\frac{1}{N}\tr\psi_{a1}{\bf 1}_N\ ,\ 
    \psi_{a2}=v^{1/3}\psi_2+\frac{1}{N}\tr\psi_{a2}{\bf 1}_N\ ,\ 
    \psi_{b1}=v^{1/3}\psi_3+\frac{1}{N}\tr\psi_{b1}{\bf 1}_N\ ,\\
\end{equation}
and define the fermion $\chi$ from the last chiralino $\psi_{b2}$ by 
\begin{equation}\label{eq:fermion decomposition2}
  \psi_{b2}=\frac{1}{2v} \big(\{X,\psi_1\}+\{Y,\psi_2\}-\{Z,\psi_3\}\big)+\chi\ .
\end{equation}
Other adjoint fields are decomposed as 
\begin{equation}
\begin{aligned}    
    f&\equiv f_1+f_2,&f_H&\equiv f_1-f_2,\\
    \lambda_{\dot\alpha}
    &\equiv\frac{1}{2}(\lambda_{1\dot\alpha}+\lambda_{2\dot\alpha}),&
    \lambda_{H\dot\alpha}
    &\equiv\lambda_{1\dot\alpha}-\lambda_{2\dot\alpha}.
\end{aligned}
\end{equation}
Together with the scalar decomposition \eqref{eq:scalar decomposition}, our
complete assignments of $L$ and $H$ are 
\begin{equation}\label{eq:LH assignment}
\begin{aligned}
L={}&\{X,Y,Z,\psi_1,\psi_2,\psi_3,f,\lambda_{\dot\alpha}\},\\
H={}&\{x,y,z,w,W,\tr\psi_{a1},\tr\psi_{a2},\tr\psi_{b1},
\chi,f_H,\lambda_{H\dot\alpha}\}.
\end{aligned}
\end{equation}
The nonlinear term on the right hand side of (\ref{eq:fermion decomposition2}) is 
introduced for the consistency of turning off $H$, which will be explained below.

Substituting \eqref{eq:scalar decomposition} to the matter fermions' $Q$ 
transformation yields
\begin{equation}
\begin{aligned}
Q\psi_{a1}
&=b^2a^2b^1-b^1a^2b^2
=v^{1/3}[Y,Z]+(\text{heavy}),\\
Q\psi_{a2}
&=b^1a^1b^2-b^2a^1b^1
=v^{1/3}[Z,X]+(\text{heavy}),\\
Q\psi_{b1}
&=a^1b^2a^2-a^2b^2a^1
=v^{1/3}[X,Y]+(\text{heavy}),\\
Q\psi_{b2}
&=a^2b^1a^1-a^1b^1a^2
=v^{-1}(YZX-XZY)+(\text{heavy}),
\end{aligned}
\end{equation}
where the terms denoted by `(heavy)' contain one or more letters from $H$.
The first three lines and \eqref{eq:fermion decomposition} imply
\begin{equation}
\begin{aligned}
Q\psi_1&=[Y,Z]+(\text{heavy}),&
Q\psi_2&=[Z,X]+(\text{heavy}),&
Q\psi_3&=[X,Y]+(\text{heavy}).
\end{aligned}
\end{equation}
That is, the first terms on the right-hand sides should be attributed to the
transformations of $\psi_{1,2,3}$ since they are defined to be the traceless 
parts of the matter fermions. So if it is the case that the heavy letters 
$H$ can consistently be turned off (as we shall show shortly), 
the induced transformations after the truncation are
\begin{equation}\label{eq:light chiralino Q}
\left.Q\psi_1\right|_{H=0}=[Y,Z],\qquad
\left.Q\psi_2\right|_{H=0}=[Z,X],\qquad
\left.Q\psi_3\right|_{H=0}=[X,Y].
\end{equation}
Here and below, $\left.O\right|_{H=0}$ denotes the truncated SUSY transformation 
in which all the letters in $H$ are set to zero.
As for the light field strength $f$, one obtains
\begin{equation}
\begin{aligned}
Qf
&=Q(f_1+f_2)\\
&=[\psi_{a1},a^1]+[\psi_{a2},a^2]
+[\psi_{b1},b^1]+[\psi_{b2},b^2]\\
&=[\psi_1,X]+[\psi_2,Y]+[\psi_3,Z]+(\text{heavy}),
\end{aligned}
\end{equation}
and hence
\begin{equation}\label{eq:light field strength Q}
\left.Qf\right|_{H=0}
=[\psi_1,X]+[\psi_2,Y]+[\psi_3,Z].
\end{equation}
The light scalars and gaugino satisfy
\begin{equation}
QX=QY=QZ=Q\lambda_{\dot\alpha}=0.
\end{equation}
For letters containing derivatives, for instance,
\begin{equation}
\begin{aligned}
Q\big(v^{1/3}D_{+\dot\alpha}a^1\big)
&=v^{1/3}\big(\lambda_{2\dot\alpha}a^1
-a^1\lambda_{1\dot\alpha}\big)\\
&=[\lambda_{\dot\alpha},X]
-\frac{1}{2}\{\lambda_{H\dot\alpha},X\}
+(\text{heavy}),
\end{aligned}
\end{equation}
so that
\begin{equation}
\left.Q\big(v^{1/3}D_{+\dot\alpha}a^1\big)\right|_{H=0}
=[\lambda_{\dot\alpha},X],
\end{equation}
and similar calculations yield $[\lambda_{\dot\alpha},Y]$ and
$[\lambda_{\dot\alpha},Z]$ for the other two light scalars, and so on.
The projected transformations \eqref{eq:light chiralino Q} and
\eqref{eq:light field strength Q}, together with the scalar, gaugino, and
derivative transformations above, are precisely the classical $Q$-action of
$\mathcal N=4$ SYM written as an $\mathcal N=1$ vector multiplet and three
adjoint chiral multiplets. In the normalization used here, the three chiralino
transformations are the F-term transformations associated with the conjugate
cubic superpotential $\bar W_{\mathcal N=4}=\tr X[Y,Z]$. 

We next verify that turning off the heavy letters $H$ is consistent. 
One should check that no expressions of $L$'s can be written as $QH$.
First, the following heavy letters are $Q$-closed:
\begin{equation}
Qx=Qy=Qz=Qw=QW=0,\qquad Q\lambda_{H\dot\alpha}=0.
\end{equation}
For the trace components of the three chiralini, one obtains
\begin{equation}
\begin{aligned}
Q\tr\psi_{a1}&=\frac{1}{v}\tr W[Y,Z],&
Q\tr\psi_{a2}&=\frac{1}{v}\tr W[Z,X],&
Q\tr\psi_{b1}&=\frac{1}{v}\tr W[Y,X].
\end{aligned}
\end{equation}
Every expression on the right contains the heavy scalar $W$, so cannot generate 
an expression of $L$'s only. The remaining
heavy fermion satisfies
\begin{equation}
\begin{aligned}
Q\chi
={}&Q\psi_{b2}-\frac{1}{2v}
\big(\{X,Q\psi_1\}+\{Y,Q\psi_2\}-\{Z,Q\psi_3\}\big),\\
\left.Q\chi\right|_{H=0}
={}&\frac{1}{v}(YZX-XZY)\\
&-\frac{1}{2v}\big(
\{X,[Y,Z]\}+\{Y,[Z,X]\}-\{Z,[X,Y]\}\big)=0.
\end{aligned}
\end{equation}
So there are no terms in $Q\chi$ which contain light fields only. 
Finally, the relative field strength satisfies
\begin{equation}
\begin{aligned}
Qf_H
={}&\{\psi_{a1},a^1\}+\{\psi_{a2},a^2\}
-\{\psi_{b1},b^1\}-\{\psi_{b2},b^2\}-(\text{trace}),\\
\left.Qf_H\right|_{H=0}
={}&\{\psi_1,X\}+\{\psi_2,Y\}-\{\psi_3,Z\}\\
&-\big(\{X,\psi_1\}+\{Y,\psi_2\}-\{Z,\psi_3\}\big)=0.
\end{aligned}
\end{equation}
We have thus checked $\left.Qh\right|_{H=0}=0$ for every
heavy letter $h$. Setting all the letters in $H$ to zero is therefore
consistent. 
If $T$ denotes this truncation, the calculations above imply
\begin{equation}
TQ_{\mathrm{KW}}=Q_{\mathcal N=4}T,
\end{equation}
meaning that $T$ maps $Q_{\rm KW}$-closed operators to $Q_{\mathcal N=4}$-closed operators, and $Q_{\rm KW}$-exact ones to $Q_{\mathcal N=4}$-exact.
We will use this map in Section \ref{Section:fortuitous} to establish that a
particular family of fortuitous classes is not exact.

\section{Generalized monotonicity and strong fortuity}

\subsection{From ordinary to baryonic covering spaces}

We begin by recalling the notions of fortuity and monotonicity introduced in \cite{Chang:2024zqi}. We shall then extend them to theories with baryonic operators.

In $\mathcal{N}=4$ SYM theory with gauge group $SU(N)$, 
all the fields are in the adjoint representation. Therefore, 
if the rank $N$ epsilon tensor is used to contract row indices of 
matrices in a gauge-invariant operator, there should always be another epsilon 
which can contract the column indices. 
Such a pair of epsilon tensors can be rewritten as
$\epsilon_{i_1\cdots i_N}\epsilon^{j_1\cdots j_N}
=N!\delta^{[j_1}_{i_1}\cdots\delta^{j_N]}_{i_N}$, 
implying that the gauge-invariant operator can be written as a linear combination 
of multi-trace operators. Thus, every gauge-invariant operator
admits a multi-trace expression. On the gravity side, this is consistent with the absence of a non-trivial three-cycle in $S^5$ on which a D3-brane could wrap.

The corresponding \emph{covering space} $\widetilde{\mathcal H}$ is the space of multi-trace operators before imposing finite-$N$ trace relations \cite{Chang:2024zqi}. If $I_N$ denotes the space of such relations, the finite-rank Hilbert space is $\mathcal H_N\simeq\widetilde{\mathcal H}/I_N$. At fixed finite energy there are no trace relations at infinite rank, so one may identify $\widetilde{\mathcal H}$ with the large-$N$ operator space $\mathcal H_\infty$. We denote the projection to rank $N$ by
$\pi_N:\widetilde{\mathcal H}\to\mathcal H_N$.

The projection induces a map from the covering-space cohomology to the finite-$N$ cohomology.\footnote{
The induced map is well-defined because $\pi_N\widetilde Q=Q_N\pi_N$: the projection is a cochain map and therefore sends $\widetilde Q$-closed operators to $Q_N$-closed operators and $\widetilde Q$-exact operators to $Q_N$-exact operators. Here $\widetilde Q$ and $Q_N$ denote the supercharges acting on $\widetilde{\mathcal H}$ and $\mathcal H_N$, respectively.
}
A \emph{monotonous cohomology} class is, by definition, a class in the image of
\[
\pi_N^*:H^*(\widetilde{\mathcal H})\longrightarrow H^*(\mathcal H_N).
\]
Single-trace monotone classes correspond to the BPS components of single-particle Kaluza--Klein states of type IIB supergravity on AdS$_5\times S^5$; products of these classes are therefore often called multi-gravitons. A \emph{fortuitous cohomology} class is a non-trivial finite-$N$ class linearly independent of the monotone classes. In the convention of \cite{Chang:2024zqi}, two fortuitous classes are identified if they differ by a monotone class, while the corresponding fortuitous BPS states are taken to be orthogonal to the monotone BPS states. The existence of fortuitous classes means that, although $\pi_N$ is surjective on operator spaces, the induced map
$\pi_N^*:H^n(\widetilde{\mathcal H})\to H^n(\mathcal H_N)$ need not be surjective.

For a monotone representative, the $Q$-variation already vanishes in the covering space, without using finite-$N$ trace relations. Its projection is consequently $Q$-closed at every rank.\footnote{
If $\widetilde O\in\widetilde{\mathcal H}$ represents a $\widetilde Q$-cohomology class, then $\pi_N(\widetilde O)$ is $Q_N$-closed for every $N$.
If $\pi_N(\widetilde O)$ is not $Q_N$-exact, i.e. if 
$(\widetilde O+I_N)\cap\widetilde Q\widetilde{\mathcal H}$ is empty, 
then $\pi_{N+1}(\widetilde O)$ is not $Q_{N+1}$-exact either because $I_{N+1}\subset I_N$. Conversely, the relation space grows as $N$ decreases, so a monotone class may become exact at sufficiently small $N$ when $\widetilde O+I_N$ begins to intersect $\widetilde Q\widetilde{\mathcal H}$.}
By contrast, let $O\in\mathcal H_N$ represent a fortuitous class and choose a lift $\widetilde O\in\widetilde{\mathcal H}$. Its $Q$-variation lies in the relation space, $\widetilde Q\widetilde O\in I_N$, because $O$ is closed only after the finite-$N$ quotient. The lift is ambiguous up to $\widetilde O\to\widetilde O+r$ with $r\in I_N$, but this does not change the conclusion because $\widetilde Qr\in I_N$.
Moreover, because the finite-$N$ class represented by $O$ does not lie in
$\operatorname{Im}\pi_N^{*}$, the element
$\widetilde Q\widetilde O+\widetilde Qr$ represents a non-trivial class in
$H^{*}(I_N)$.
These classes form the kernel of the map on cohomology induced by the inclusion $I_N\hookrightarrow\widetilde{\mathcal H}$ \cite{Chang:2024zqi}.

This construction has to be modified when bifundamental matter produces genuine baryons.
In the KW theory, for example,
\begin{equation}\label{minimal-baryon-present}
  \epsilon_{p_1\cdots p_N}\epsilon^{\hat p_1\cdots\hat p_N}
  (A_{i_1})^{p_1}{}_{\hat p_1}\cdots
  (A_{i_N})^{p_N}{}_{\hat p_N}
\end{equation}
uses epsilon tensors of two different gauge groups and cannot be rewritten as a multi-trace. 
So the multi-trace space alone cannot serve as a covering space whose quotient yields the full finite-$N$ Hilbert space.

To illustrate how a notion of rank-independent covering space can nevertheless arise, 
consider the following examples of minimal and excited baryon cohomologies, 
\begin{equation}
\begin{aligned}
\epsilon\epsilon\,a^N
&=N!\det a,\\
\epsilon\epsilon\,a^{N-1}\psi_b^{\hat i}
&=(N-1)!\det a\,\tr(a^{-1}\psi_b^{\hat i}),\\
\epsilon\epsilon\,a^{N-2}a_2\psi_b^{\hat i}
&=(N-2)!\det a\Big[
\tr(a^{-1}a_2)\tr(a^{-1}\psi_b^{\hat i})
-\tr(a^{-1}a_2a^{-1}\psi_b^{\hat i})\Big], 
\end{aligned}
\end{equation}
where $a \equiv a_1=\bar A_1$ and $a_2=\bar{A}_2$ are defined in \eqref{eq:redefined fields}, 
and $\epsilon\epsilon$ schematically denotes the two epsilon-tensor contractions 
as in (\ref{minimal-baryon-present}). 
The forms of these operators depend on the specific value of $N$, 
so they do not define a single rank-independent element of the ordinary covering space 
labeled by $N=1,2,\cdots$. So it is unclear how one would form a covering space 
which contains these operators. However, apart from trivial $N$-dependent normalization 
factors like $N!$, note that the manifestly $N$-dependent parts are all factored out 
into the `ground state operator' $\det a$. 
The remaining part of the operators can be recast in the multi-trace forms with 
$N$-independent coefficients,
\begin{equation}\label{excited-baryon-covering-present}
  {\bf 1}\ \ ,\ \ {\rm Tr}(a^{-1}\psi_b^{\hat{i}})\ \ ,\ \ 
  {\rm Tr}(a^{-1}a_2){\rm Tr}(a^{-1}\psi_b^{\hat{i}})-{\rm Tr}(a^{-1}a_2a^{-1}\psi_b^{\hat{i}})\ ,
\end{equation}
at the cost of introducing an extra formal letter $a^{-1}$. 
The last operators can be promoted to an infinite sequence of operators labeled 
by $N$. The baryonic covering space that we 
will suggest in this section is defined by the multi-trace operators that include 
the inverse of certain scalar letters (in the examples above, $a^{-1}$) that describe the
baryonic ground states. 

The inverse letter is given by $a^{-1}\equiv \frac{{\rm Cof}(a)}{\det a}$, where 
the cofactor matrix is given by
\begin{equation}
  {{\rm Cof}(a)^{p}}_{\hat{p}}\equiv \frac{1}{(N-1)!}
  \epsilon_{\hat{p}\hat{p}_1\cdots \hat{p}_{N-1}}
  \epsilon^{pp_1\cdots p_{N-1}}
  {a^{\hat{p}_1}}_{{p}_1}\cdots{a^{\hat{p}_{N-1}}}_{{p}_{N-1}}\ .
\end{equation}
Since $a^{-1}$ has $\det a$ in the denominator, operators containing $a^{-1}$ 
are not legitimate local operators. 
They can be promoted to legitimate operators only after multiplying sufficiently 
many ground-state factors $\det a$. We are suggesting that the `excited state factor'
like (\ref{excited-baryon-covering-present}) admits $N$-independent definitions.
As we shall explain in Section 3.2, using the fermionic integral formulation, 
all baryonic excitations admit such decompositions. 

The ground state factor $\det a$ is trivially $Q$-closed. Therefore, to construct 
$Q$-closed baryonic operators, all one has to ensure is that 
the operators (\ref{excited-baryon-covering-present}) are $Q$-closed. 
In fact, one can easily check that they are all $Q$-closed without using 
the specific matrix sizes, qualifying them to be called \textit{generalized monotone} 
cohomologies. For instance, note that 
$Q{\rm Tr}(a^{-1}\psi_b^{\hat{i}})\sim {\rm Tr}(a^{-1}(ab^{\hat{i}}a_2-a_2b^{\hat{i}}a))=0$ for any $N$.

The ordinary large-$N$ space $\mathcal H_\infty$ of local operators 
remains meaningful only in the mesonic sector. Its finite-$N$ quotient therefore recovers only the mesonic sector,
$\mathcal H_\infty/I_N=\mathcal H_{N,0}$, rather than the full Hilbert space
\[
\mathcal H_N=\bigoplus_B\mathcal H_{N,NB},
\]
where $\mathcal H_{N,NB}$ denotes the sector of baryon number $NB$. 

To describe the nontrivial baryonic sector with $B\neq 0$, 
we shall define a separate \emph{baryonic covering space} by
\begin{equation}
\widetilde{\mathcal{H}}_{B}\;\equiv\;\lim_{N\to \infty}\mathcal{H}_{N,NB}=\mathcal{H}_{\infty,\infty\cdot B},
\end{equation}
where the limit keeps finite-energy excitations above the rank-dependent baryonic ground state. 
This can be understood as the multi-trace operators including new formal 
letters like $a^{-1}$ above, and at this stage should not itself be interpreted as a space of physical local operators. See Section 3.2 for a more comprehensive definition 
and explanations. At rank $N$, the physical sector is recovered as
\begin{equation}
\mathcal{H}_{N,NB}\;\simeq\;\widetilde{\mathcal{H}}_{B}\big/I_{N,NB},
\end{equation}
where $I_{N,NB}$ is the rank-$N$ relation space, defined more explicitly below. Combining all baryon numbers, one obtains 
$\widetilde{\mathcal H}=\bigoplus_B\widetilde{\mathcal H}_B$ and
$I_N=\bigoplus_B I_{N,NB}$, so that
$\mathcal H_N=\bigoplus_B\widetilde{\mathcal H}_B/I_{N,NB}
=\widetilde{\mathcal H}/I_N$.

\subsection{Baryonic covering space from fermionic integrals}

In this subsection, we introduce a fermionic integral representation
\cite{Gaiotto:2021xce} of the baryonic operators, which will naturally provide 
the notion of baryonic covering space.
Although we present the construction for the KW theory, the same idea 
should apply to more general theories with topological baryons.

\subsubsection{Baryons as fermionic integrals}
In the sector with baryon number $B$, we introduce the following fermionic variables 
\begin{equation}
\psi^{\alpha}_{\hat p},\qquad \chi_{\alpha}^{p},\qquad \alpha=1,\dots,|B|,\qquad p,\hat p = 1,2,...,\infty\ .
\end{equation}
The $\infty$ in the ranges of $p,\hat{p}$ comes from $N=\infty$, which realizes the
baryonic covering space that we would like to define. 
If one wishes, one may regard $\infty$ as a sufficiently large $N$. 
For simplicity, we will assume $B<0$. (The case with $B>0$ can be discussed 
similarly, with $a$ below replaced by $b\equiv b_{\hat{1}}$.)
Let us start by representing the baryonic ground state $(\det a)^{|B|}$ as
\begin{equation}
(\det a)^{|B|}=
\int [d\psi d\chi]\;
\exp\!\left(\psi^{\alpha}_{\hat p}\,a^{\hat p}{}_{p}\,\chi^{ p}_{\alpha}\right),\qquad 
[d\psi d\chi]=\prod_{\alpha=1}^{|B|}\Bigl(\prod_{\hat{p}\ge 1} d\psi^{\alpha}_{\hat p}\Bigr)\Bigl(\prod_{ p\ge 1} d\chi^{ p}_{\alpha}\Bigr)\;.
\end{equation}
Since $a$ has baryon number $-1$, we assign baryon number $1/2$ to each of $\psi^\alpha_{\hat p}$ and $\chi^{p}_\alpha$. 
Each factor $d\psi_{\hat p}^\alpha d\chi^{p}_\alpha$ in the measure has baryon number $-1$. 
This operator may look rather formal at $N=\infty$, in which case one can imagine 
sufficiently large $N$. In fact there are many more baryonic ground states, 
which replaces some $a\equiv a_1$ by $a_2$ within each determinant. 
In our formalism below, replacing some $a$ by $a_2$ will be formally treated 
as excited baryons. In other words, our fermionic integral formalism 
lacks manifest $SU(2)$ covariance.\footnote{One may 
restore the covariance by introducing $a_\alpha(u)\equiv a_i u^i_\alpha$ 
for each $\alpha=1,\cdots,|B|$, and then replacing the Gaussian integrand by 
$e^{\psi^\alpha a \chi_\alpha}\rightarrow e^{\sum_\alpha\psi^\alpha a_\alpha(u)\chi_\alpha}$. After constructing the $u$-dependent 
operators with this replacement, the coefficients of the $u$ expansion 
will yield the baryonic operators. }

Now we discuss the excited baryons. First consider the
following excited baryon state
\begin{equation}\label{excited-baryon-B1}
  \epsilon^{p_1\cdots p_N}\epsilon_{\hat{p}_1\cdots \hat{p}_N}
  {a^{\hat{p}_1}}_{{p}_1}\cdots{a^{\hat{p}_{N-1}}}_{{p}_{N-1}}{X^{\hat{p}_N}}_{{p}_N}\ ,
\end{equation}
where $X^{\hat{p}}{}_{p}$ is a bifundamental (possibly composite) operator.
This operator can be represented by the following fermionic integral 
\begin{equation}
  \int [d\psi d\chi](\psi^\alpha X\chi_\beta)
  e^{\psi^\gamma a\chi_\gamma}\ ,
\end{equation}
where $(\psi^\alpha X\chi_\beta)
\;\equiv\;\psi^{\alpha}_{\hat p}\,X^{\hat p}{}_{p}\,\chi^{p}_{\beta}$. 
Because the Gaussian weight and the integration measure are $U(|B|)$-invariant, 
the integral with a single bilinear insertion is proportional to $\delta^\alpha_\beta$. 
This yields an excited baryon of the form (\ref{excited-baryon-B1}) multiplied by 
the ground state operator $(\det a)^{|B|-1}$.
More generally, one can consider the integral given by
\begin{equation}
  \int [d\psi d\chi] e^{\psi^\alpha a \chi_\alpha}
  (\psi^{\alpha_1} X_1\chi_{\beta_1})\cdots (\psi^{\alpha_m} X_m\chi_{\beta_m})\ .
\end{equation}
This operator replaces $m$ occurrences 
of $a$ in $(\det a)^{|B|}$ by $X$'s. The $\alpha$ and $\beta$ indices of this insertion should
form a $U(|B|)$-invariant tensor \cite{Gaiotto:2021xce}.
The diagonal case $\alpha=\beta$ describes an excitation within a single 
determinant, whereas the insertions with $\alpha\neq \beta$ 
would merge two determinant factors. Morally, this would describe an open string 
represented by $X$ connecting the $\alpha$'th and the $\beta$'th D3-branes.

The operators described in the previous paragraph would contain 
a ground state factor $(\det a)^{|B|}$, which depends on `$N=\infty$,' 
times the excitation parts which belong to the baryonic covering space
$\widetilde{\mathcal{H}}_B$ (to be more precisely defined below). 
At this stage, it is easy to understand the connection to 
the intuitive  definition that we presented in Section 3.1. Integrating over the fermions
using Feynman rules, $a^{-1}$ will play the role of the propagator, and the operator 
one obtains takes a multi-trace form which contains 
$X$'s and $a^{-1}$. Formally regarding $a^{-1}$ as 
an extra letter, the excitation part of the operator has a smooth $N=\infty$ limit.

The full definition of $\widetilde{\mathcal{H}}_B$ 
at fixed charge $B$ should also include the multiplications of mesons 
which carry $B=0$. Also, as we shall explain in Section 3.2.2, 
there are redundancies in the definition of these operators in terms of 
fermionic integrals, in that different integrands may yield the same operator. 
We shall explain all these aspects in Section 3.2.2 to complete the definition 
of the baryonic covering space.

\subsubsection{Ward identities and the covering space}

In this section, we discuss the redundancies in the fermionic integral definition 
of baryonic operators. Let $\mathcal U_B$ be the vector space spanned by 
\begin{equation}\label{eq: U_B mono}
(\psi^{\alpha_1} X_1\chi_{\beta_1})\cdots (\psi^{\alpha_m} X_m\chi_{\beta_m})
\,\tr (Y_1)\cdots\tr (Y_l),
\qquad m,l\ge 0,
\end{equation}
where $\alpha_a,\beta_a\in\{1,\dots,|B|\}$. Each $(X_a)^{\hat p}{}_{p}$ is a word of baryon number $-1$, while each $Y_i$ is a word of baryon number zero with either index structure $(Y_i)^p{}_q$ or $(Y_i)^{\hat p}{}_{\hat q}$. 
The fermionic integral defines a linear map
\begin{equation}
\mathfrak{I}_{B}:\ \mathcal{U}_{B}\longrightarrow \mathcal{H}_{\infty,\infty\cdot B},
\qquad
\mathfrak{I}_{B}(\mathcal{O})
\;\equiv\;
\int [d\psi d\chi]\;
e^{(\psi^\alpha a \chi_\alpha)}\;\mathcal{O}
\end{equation}
which defines the baryonic covering space $\widetilde{\mathcal{H}}_B$. 
This map is not injective: distinct elements 
of $\mathcal{U}_B$ can have the same image because the fermionic integral obeys 
Ward identities, as we shall now explain.

For any polynomial, or more generally formal power series, $\mathcal F$ of the auxiliary fermions and physical letters, integration of a total derivative vanishes:
\begin{equation}
0=
\int [d\psi d\chi]\;
\frac{\partial}{\partial \psi^{\alpha}_{\hat p}}
\Bigl(
e^{(\psi^\beta a \chi_\beta)}\,\mathcal{F}
\Bigr),
\qquad
0=
\int [d\psi d\chi]\;
\frac{\partial}{\partial \chi^{p}_{\alpha}}
\Bigl(
e^{(\psi^\beta a \chi_\beta)}\,\mathcal{F}
\Bigr).
\end{equation}
Expanding these derivatives yields
\begin{equation}
\int [d\psi d\chi]\;
e^{(\psi^\beta a \chi_\beta)}\,
\Bigl(
(a\chi_{\alpha})^{\hat p}\,\mathcal{F}
+\frac{\partial \mathcal{F}}{\partial \psi^{\alpha}_{\hat p}}
\Bigr)=0,
\qquad
\int [d\psi d\chi]\;
e^{(\psi^\beta a \chi_\beta)}\,
\Bigl(
(\psi^{\alpha}a)_{p}\,\mathcal{F}
-\frac{\partial \mathcal{F}}{\partial \chi^{p}_{\alpha}}
\Bigr)=0,
\end{equation}
where $(a\chi_{\alpha})^{\hat p}\equiv a^{\hat p}{}_{q}\chi^{q}_{\alpha}$ and $(\psi^{\alpha}a)_{ p}\equiv \psi^{\alpha}_{\hat q}a^{\hat q}{}_{ p}$.
These Ward identities, equivalently the Schwinger--Dyson equations of the auxiliary system, allow a factor $a\chi_\alpha$ or $\psi^\alpha a$ to be replaced by a derivative acting on the remaining insertions. In the large-$N$ limit with finitely many insertions, they generate all linear dependencies among the bilinears. More precisely, let $W_B\subset\mathcal U_B$ be the span of
\begin{equation}
e^{-(\psi^\beta a \chi_\beta)}\frac{\partial}{\partial \psi^{\alpha}_{\hat p}}\Bigl(
e^{(\psi^\beta a \chi_\beta)}\,\mathcal{F}
\Bigr),
\qquad
e^{-(\psi^\beta a \chi_\beta)}\frac{\partial}{\partial \chi^{p}{}_{\alpha}}\Bigl(
e^{(\psi^\beta a \chi_\beta)}\,\mathcal{F}
\Bigr),
\end{equation}
Then $\ker\mathfrak I_B=W_B$. To see this, first regulate the auxiliary index range to a finite cutoff $N$. Denote all Grassmann variables $\psi^\alpha_{\hat p},\chi^{p}{}_\alpha$ collectively by $\Theta_i$, and set
\begin{equation}
S=\psi^\alpha_{\hat p} a^{\hat p}{}_{ p}\chi^{ p}_\alpha.
\end{equation}
Suppose that
\begin{equation}
\int [d\psi d\chi]_N\,e^{S}\,\mathcal O=0.
\end{equation}
Then $F:=e^S\mathcal O$ has no top Grassmann component, since the integral extracts precisely that coefficient. Every term in $F$ therefore omits at least one variable $\Theta_i$. 
It follows that
\begin{equation}
F=\sum_i \frac{\partial G_i}{\partial \Theta_i}
\end{equation}
for suitable $G_i$. Defining $H_i=e^{-S}G_i$, this can be rewritten as
\begin{equation}
\mathcal O=\sum_i e^{-S}\frac{\partial}{\partial\Theta_i}\!\left(e^{S}H_i\right).
\end{equation}
Thus, any vanishing auxiliary-fermion integral is a Ward identity, unless the integrand itself vanishes (e.g. by finite-$N$ relations).
Even the vanishing of the insertion $(\psi^\alpha X\chi_{\beta})$ 
with $\alpha\neq \beta$ can be recast as a Ward identity.

The Ward identities also relate explicit mesonic traces to fermion bilinears. Let $Y^{ p}{}_{ q}$ be a bosonic word carrying indices of the first $SU(N)$ gauge group, and take 
\begin{equation}
\mathcal{F}=(Y\chi_\gamma)^{ p}\equiv Y^{ p}{}_{ q}\,\chi^{ q}{}_{\gamma}.
\end{equation}
The identity generated by $\partial/\partial\chi^{ p}_\alpha$ then yields
\begin{equation}
\int [d\psi d\chi]\;e^{(\psi^\beta a\chi_\beta)}\,
(\psi^\alpha aY\chi_\gamma)
\;=\;
\tr(Y)\,\delta^\alpha_\gamma
\int [d\psi d\chi]\;e^{(\psi^\beta a\chi_\beta)}.
\end{equation}
Thus, within the integral, the multiplication of a meson $\tr Y$ may be 
exchanged for a bilinear with $X=aY$.

Bilinears alone do not, however, generate arbitrary multi-trace insertions. In the $B=1$ sector, for example,
\[
\int e^{(\psi a\chi)}(\psi aY_1\chi)(\psi aY_2\chi)
\sim
(\tr Y_1\tr Y_2-\tr Y_1Y_2)\int e^{(\psi a\chi)}.
\]
No polynomial in bilinears produces the first term $\tr Y_1\tr Y_2\int e^{(\psi a\chi)}$ by itself.
More generally, multiplications of at most $|B|$ single-trace meson factors can arise from integrations with bilinears, whereas expressions with more single-trace meson factors occur only in particular combinations with terms containing fewer traces.

A minor subtlety arises for $Y=\mathbf{1}$. In this case, the Ward identity contains
$\delta^{ p}{}_{ p}=\tr\mathbf{1}$, which is not defined if the gauge-index range is taken to be infinite from the outset. With a finite large-rank regulator, however, $\tr\mathbf{1}=N$, and the normalized insertion
$N^{-1}(\psi^\alpha a\chi_\gamma)$ has the well-defined large-$N$ limit $\delta^\alpha_\gamma$. When other insertions, collectively denoted by $\mathcal{F}$, are present, there is in addition an order-$1/N$ correction,
$-\frac{1}{N}\psi^{\alpha}_{\hat p}\frac{\partial}{\partial \psi^\gamma_{\hat p}}\mathcal{F}$.
This is reminiscent of the phenomenon discussed in \cite{Witten:2021unn}: the Hamiltonian divided by $N$ is central in the strict large-$N$ limit, while perturbative $1/N$ corrections remove this centrality and lead to a type $\mathrm{II}_\infty$ operator algebra.

With these Ward identities understood, we therefore 
obtain the following isomorphism 
\begin{equation}
\widetilde{\mathcal{H}}_B=\mathcal{H}_{\infty,\infty\cdot B}\;\simeq\;\mathcal{U}_{B}\big/ W_{B},
\end{equation}
through the induced map
\begin{equation}
\widetilde{\mathfrak{I}}_{B}:\ \mathcal{U}_B/W_B\longrightarrow \mathcal{H}_{\infty,\infty\cdot B},
\qquad
\widetilde{\mathfrak{I}}_{B}\bigl([\mathcal{O}]\bigr)
\;=\;
\int [d\psi d\chi]\;
e^{(\psi^\alpha a \chi_\alpha)}\;\mathcal{O},
\end{equation}
which is well-defined and invertible because $W_B=\ker\mathfrak I_B$. Equivalently, one may take $\mathcal U_B/W_B$ itself as the definition of the baryonic covering space, without referring to the formal infinite-energy sector $\mathcal H_{\infty,\infty\cdot B}$. There is no need to impose $U(|B|)$ invariance separately on $\mathcal U_B$, since the quotient by $W_B$ already enforces it. No finite-$N$ relations have yet been imposed, 
which will enter only through the projection discussed in the next section.

\subsubsection{Finite $N$ projection and `trace relations'}
At each rank $N$, define
\begin{equation}
\pi_{N,NB}:\ \widetilde{\mathcal{H}}_{B}\longrightarrow \mathcal{H}_{N,NB}
\end{equation}
by inserting the $N\times N$ matrices to the operators in the covering space 
given in terms of $a^{-1}$, $X_a$'s and $Y_l$'s in (\ref{eq: U_B mono}).
Alternatively, the operators can be defined by the fermion integrals, 
now restricting $p,\hat p$ to $1,\ldots,N$. In the latter definition, 
the ground state factor $(\det a)^{|B|}$ is multiplied to those 
defined by the first definition. 
The rank-$N$ relation space is the kernel
\begin{equation}
I_{N,NB}\;\equiv\;\ker(\pi_{N,NB}).
\end{equation}
Hence,
\begin{equation}
\mathcal{H}_{N,NB}\;\simeq\;\widetilde{\mathcal{H}}_{B}/I_{N,NB}.
\end{equation}

As mentioned earlier, the fermion integral can be evaluated by 
the Feynman rules, using the Wick contractions with the propagator 
$\langle\psi^\alpha_{\hat p}\chi^{ p}_\beta\rangle
=\delta^\alpha_\beta(a^{-1})^{ p}{}_{\hat p}$.
The resulting operator is $(\det a)^{|B|}$ times a multi-trace operator built 
from $a^{-1}$ and the other inserted words. For example,
\begin{equation}\label{eq: Wick example}
\int [d\psi d\chi]\;e^{(\psi^\gamma a\chi_\gamma)}\,(\psi^\alpha X\chi_\beta)
\;\propto\;
\delta^\alpha_\beta\,(\det a)^{|B|}\,\tr(a^{-1}X),
\end{equation}
while products of bilinears would yield sums of multi-traces containing factors such as
\begin{equation}
(\det a)^{|B|}\,\tr(a^{-1}X_1a^{-1}X_2\cdots)\, .
\end{equation}
These Wick-contraction identities hold at every rank and therefore also in the covering space. The additional identities that arise only after projection are the usual finite-$N$ trace relations among the resulting multi-traces.\footnote{
Because the definition of $\mathcal U_B$ in \eqref{eq: U_B mono} includes independent mesonic multi-trace insertions, $I_{N,NB}$ contains the mesonic trace-relation space $I_{N,0}$. It also contains relations involving multi-traces with the additional formal letter $a^{-1}$. As explained above, however, bilinears do not generate arbitrary multi-traces once the number of trace factors exceeds $|B|$. Consequently, not every trace relation involving $a^{-1}$ belongs to $I_{N,NB}$. For example, multiplying \eqref{eq:Procesi} by a multi-trace containing $a^{-1}$ need not produce an element of $I_{N,NB}$.}

As an example, consider the $B=-1$ sector. There are $2N$ auxiliary fermions, so a product of $N+1$ bilinears vanishes identically. To express this as a trace relation, take $X_1,\ldots,X_r$ to be bosonic and set $Y_i=a^{-1}X_i$. Standard Wick contraction gives
\begin{equation}\label{eq: Wick B=1}
\begin{aligned}
    \langle (\psi X_1\chi)...(\psi X_r\chi)\rangle 
    &=r!(X_1)^{{p}_1}{}_{\hat{p}_1}\cdots (X_r)^{p_r}{}_{\hat{p}_r} (a^{-1})^{[\hat{p}_1}{}_{{p}_1}\cdots (a^{-1})^{\hat{p}_r]}{}_{{p}_r}\\
    &=\sum_{\sigma\in S_r}\text{sgn}(\sigma) \prod_{c\in \mathrm{Cycles(\sigma)}}
    \tr \Big(\prod_{i\in c} Y_i\Big)\;,
\end{aligned}
\end{equation}
where
$\langle\cdots\rangle=(\det a)^{-1}\int[d\psi d\chi]_N
e^{(\psi a\chi)}(\cdots)$.
For general $|B|$, each term on the right-hand side is accompanied by
$\delta^{\alpha_1}_{\beta_{\sigma(1)}}\cdots
\delta^{\alpha_r}_{\beta_{\sigma(r)}}$.
At $r=N+1$, one obtains
\begin{equation}\label{eq:Procesi}
    0=\sum_{\sigma\in S_{N+1}}\text{sgn}(\sigma) \prod_{c\in \mathrm{Cycles(\sigma)}}
    \tr \Big(\prod_{i\in c} Y_i\Big)\;.
\end{equation}
This is the polarized Cayley--Hamilton identity. It follows by applying the ordinary Cayley--Hamilton identity to
$Y_1,Y_1+Y_2,\ldots,Y_1+\cdots+Y_N$
and extracting the fully symmetrized product in the $Y_i$ \cite{Choi:2023znd}. Allowing the $Y_i$ to range over all words, relations of the form \eqref{eq:Procesi} generate all trace relations \cite{Procesi1976TheIT}.

\subsubsection{Baryonic fortuity and monotonicity}
Because $Q$ preserves baryon number, it acts separately on each $\widetilde{\mathcal H}_B$ and on each finite-$N$ sector $\mathcal H_{N,NB}$. Denote these actions by $\widetilde Q_B$ and $Q_{N,NB}$. They are compatible with the projection,
\[
\pi_{N,NB}\widetilde Q_B=Q_{N,NB}\pi_{N,NB},
\]
so $\pi_{N,NB}$ is a cochain map. We suppress the subscripts on the supercharges whenever the relevant space is clear.

A \emph{baryonic monotonous cohomology} class is defined, in direct analogy with the mesonic case, as an element of
\[
\operatorname{Im}\!\left(
\pi_{N,NB}^*:H^*(\widetilde{\mathcal H}_B)
\longrightarrow H^*(\mathcal H_{N,NB})
\right).
\]
Its $Q$-closedness in the covering space may use the auxiliary-fermion Ward identities, which are already part of the quotient $\mathcal U_B/W_B$, but it does not use any finite-$N$ relation. Equivalently, after Wick-contracting the bilinears as in \eqref{eq: Wick B=1}, the representative takes the form $(\det a)^{|B|}$ times a multi-trace and is $Q$-closed without invoking trace relations.

The \(B=-1\) monotone operators need not generate all operator structures with \(B<-1\).
Roughly speaking, the multi-baryon excitations in which the open strings $X$'s are
suspended between different determinants cannot be realized by multiplying $B=-1$
operators. This differs from the mesonic setting, where the single-trace monotones generate the full monotone sector. 

For $B\leq -3$, for example, one obtains
\begin{align}\label{eq: Wick-to-Single}
\langle (\psi^3W_1\chi_1)(\psi^1 W_2 \chi_2)(\psi^2 W_3\chi_3)\rangle
=(-1)^{|W_1|+|W_2|+|W_3|}\tr (a^{-1}W_1a^{-1}W_2a^{-1}W_3) 
\end{align}
where $|W_i|$ denotes the fermion number of the word $W_i$.
This operator cannot be generated by $B=-1$ operators, because 
\eqref{eq: Wick B=1} shows that a $B=-1$ integral always produces 
an expression in which all $W_i$'s are symmetrized inside the trace, 
unlike (\ref{eq: Wick-to-Single}).
For $B=-2$, the following example cannot be generated by $B=-1$ operators.
\begin{equation}
\begin{aligned}
\langle (\psi^2W_1\chi_1)(\psi^1 W_2 \chi_2)(\psi^2 W_3\chi_2)\rangle
=(-1)^{|W_1|+|W_2|+|W_3|}&\big[\tr (a^{-1}W_1a^{-1}W_2a^{-1}W_3)\\
&-\tr (a^{-1}W_1 a^{-1}W_2) \tr (a^{-1}W_3) \big],
\end{aligned}
\end{equation}
Although the examples above show that multi-baryonic operator structures need not be generated by \(B=-1\) monotone operators, we have not found an example of multi-baryonic monotonous cohomology that cannot be generated by single-baryonic monotones. We leave further investigation of multi-baryonic monotones to future work.

A \emph{baryonic fortuitous cohomology} class is a non-trivial class in $H^*(\mathcal H_{N,NB})$ that is linearly independent of the baryonic monotone classes, or equivalently an element of
\[
H^*(\mathcal H_{N,NB})/\operatorname{Im}\pi^*_{N,NB}.
\]
If $O=\pi_{N,NB}(\widetilde O)$ represents such a class, then
$\widetilde Q\widetilde O\in I_{N,NB}$. Changing the lift by an element of $I_{N,NB}$ preserves this property because
$\widetilde QI_{N,NB}\subset I_{N,NB}$.
Thus, the class becomes closed only after finite-$N$ relations are imposed,
and $\widetilde Q\widetilde O$ defines a non-trivial class in
$H^{*}(I_{N,NB})$, since otherwise the finite-$N$ class would lie in
$\operatorname{Im}\pi_{N,NB}^{*}$.

The dependence of $Q$-closure/exactness on rank parallels the mesonic case, because the space of relations $I_{N,NB}$ becomes strictly smaller for higher $N$.
Suppose a monotone class
\[
\pi_{N,NB}^*([\widetilde O_m])
=[\pi_{N,NB}(\widetilde O_m)]
\]
is non-exact at $N=N_1$. It remains non-exact for every $N_2>N_1$, because
$I_{N_2,N_2B}\subset I_{N_1,N_1B}$. 
This inclusion is defined in the common covering space, even though the projected operators carry different baryon numbers $N_1B$ and $N_2B$. At lower rank the relation space grows, so additional monotone classes may become exact.

The same argument shows that a fortuitous class that is non-exact at $N_1$ cannot become exact at larger rank. Its closure, however, is eventually lost. If $\widetilde O_f$ is a lift with $\pi_{N,NB}(\widetilde O_f)=O_f$, then
$\widetilde Q\widetilde O_f\in I_{N,NB}$. For sufficiently large $N'>N$, this element no longer belongs to $I_{N',N'B}$, so $\pi_{N',N'B}(\widetilde O_f)$ is not $Q$-closed. Such an $N'$ exists because the finite-rank relation spaces disappear in the large-$N$ limit.

We use the term \emph{generalized monotone} for a cohomology family admitting a
$Q$-closed construction that follows a uniform pattern as $N$ varies. This class
includes the ordinary mesonic monotones, the baryonic monotones defined above,
and nontrivial products of baryonic monotones. In particular, when the two
factors have opposite baryon numbers, each factor is represented in its
respective baryonic covering space, and their factorized product inherits a
uniform $N$-dependence. At the classical level, its $Q$-closure follows directly
from that of the two factors and the Leibniz rule of the $Q$ action. 
More general modifications of
$\det(ab)$ may also exhibit such generalized-monotone behavior, but no systematic
classification of them is needed for the examples below, and we will not pursue
it here.\footnote{
Note that modifications of $\det ab$ are more general than the product of a modification of $\det a$ and a modification $\det b$.
Any factorized product (with $B=0$) of a baryon and anti-baryon can be expressed as a modification of $(\det ab)^k$ for some suitable power $k$.
But for example, $\det ab \;\tr (ab)^{-1}\sim \epsilon\epsilon (ab)^{N-1}\mathbf{1}_N$ cannot be expressed as a product of baryon and anti-baryon.
However, this example is an ordinary mesonic monotone, and we have not found an example of such non-factorizable generalized monotonous cohomology with $B=0$ that is independent of ordinary mesonic monotones. We adopt the working assumption that there is no non-factorizable generalized monotone at \(N=2\), which is consistent with the monotone counting in Section \ref{Section:fortuitous}.
} We call a finite-$N$ cohomology \emph{strongly fortuitous} if it does not
admit an extension to any generalized-monotone family.

\section{Monotonous cohomologies}\label{Section: monotone}
We now construct explicit representatives of generalized monotonous cohomologies in Klebanov--Witten theory.
The mesonic monotonous cohomologies correspond to BPS Kaluza--Klein (KK) particle states in AdS$_5\times T^{1,1}$, which we loosely refer to as gravitons. The complete KK particle spectrum was found in \cite{Ceresole:1999zs,Ceresole:1999ht}.
The KK particles in AdS$_5\times T^{1,1}$ belong to several types of 
short multiplets of $SU(2,2|1)$.
For each short multiplet relevant to the index, there is, up to the action of the commuting global and spacetime symmetries, a distinguished conformal-primary state satisfying $\{Q,Q^{\dag}\}=D-2j_1+\frac{3}{2}r=0$. Here, $j_1$ is the eigenvalue of the Cartan generator for the left $SU(2)$ spin.
Table \ref{gravity-multiplets} summarizes the KK particles that form short multiplets and helps us 
to identify the explicit forms of the possible graviton cohomologies.

By contrast, the spectrum of baryonic monotonous cohomologies is not 
completely known.
Only the (anti)chiral-ring sector of baryonic BPS states has been studied \cite{Gubser:1998fp,Berenstein:2002ke,Beasley:2002xv,Forcella:2007wk}.
To find explicit representatives of these cohomologies, 
we therefore perform a direct search.
In practice, we search for single-trace $Q$-cohomology classes while allowing the additional ``letters'' $(a_1)^{-1}$ and $(b_1)^{-1}$. We can then recover ordinary baryonic representatives by multiplying by suitable powers of $\det a_1$ and $\det b_1$. For any such single-trace expression, one can find a suitable fermionic integral that produces it after Wick contractions by extending \eqref{eq: Wick-to-Single} to an arbitrary number of insertions:
\begin{align}\label{eq: Wick-to-Single_r}
    \langle (\psi^rW_1\chi_1)(\psi^1W_2\chi_2)\cdots(\psi^{r-1} W_r\chi_r)\rangle
    =(-1)^{r-1+\sum_i |W_i|}\tr (a^{-1}W_1a^{-1}W_2\cdots a^{-1}W_r),
\end{align}

\subsection{Mesonic monotonous cohomologies}\label{sec:mesonic-monotones}

Linearized fluctuations of type IIB supergravity fields on AdS$_5\times T^{1,1}$ form representations of $SU(2,2|1)\times SU(2)_1\times SU(2)_2$, where $SU(2)_1\times SU(2)_2\times U(1)_R$ is the isometry group of $T^{1,1}=(SU(2)\times SU(2))/U(1)$. Table \ref{gravity-multiplets} lists the single-particle supergravity states in short multiplets that contain states contributing to the index. (For example, short multiplets $L\bar{A}_i$ are not listed.) We use the convention of \cite{Cordova:2016emh}: the shortening conditions $A_1,A_2,B_1,L$ are associated with $Q_\alpha$, whereas $\bar A_1,\bar A_2,\bar B_1,\bar L$ are associated with $\bar Q_{\dot\alpha}$. The labels $[j;\bar j]$ denote the usual $SU(2)\times SU(2)$ spins of the superconformal primary in the half-integer convention. This differs from \cite{Cordova:2016emh}, where the spins are integer-valued. The quantities $j$ and $\bar j$ are not Cartan eigenvalues; we denote those eigenvalues by $j_1$ and $j_2$.

Using this Kaluza--Klein analysis \cite{Ceresole:1999zs,Ceresole:1999ht}, the $N=\infty$ superconformal index was first computed in \cite{Nakayama:2006ur}. The result agrees with the field-theory computation in \cite{Gadde:2010en}.
Appendix \ref{infinite N index} reviews the computation of the $N=\infty$ index from the KK spectrum.

For a $B_1$ multiplet, the superconformal primary $O$ has vanishing left spin and satisfies $Q_\alpha O=0$. It is therefore an anti-chiral operator with $\Delta=-\frac32 r$. For an $A_1$ multiplet, the superconformal primary $O_{(\alpha_1\cdots\alpha_{2j})}$ satisfies $\epsilon^{\alpha\beta}Q_\alpha O_{(\beta\alpha_2\cdots\alpha_{2j})}=0$ and has dimension $\Delta_{\mathrm{h.w.}}=2+2j-\frac32 r_{\mathrm{h.w.}}$. Its descendant $Q_+O_{+\cdots +}$ satisfies the BPS condition $\{Q_-,S^-\}=D-2j_1+\frac32 r=0$, while no other conformal primary in the same multiplet does.
The $A_1$ shortening applies for $j\geq \frac12$. When the left spin vanishes, $j=0$, the shortening condition is instead denoted by $A_2$: the superconformal primary $O$ satisfies $Q_\alpha Q^\alpha O=0$ and has scaling dimension $\Delta_{\mathrm{h.w.}}=2-\frac32 r$. The descendant $Q_+O$ again satisfies $\{Q_-,S^-\}=0$ and is therefore the state that contributes to the index. Finally, $L$ denotes a long multiplet, for which unitarity requires $\Delta>2+2j-\frac32 r$. The barred shortening conditions $\bar A_1,\bar A_2,\bar B_1,\bar L$ are obtained by suitably replacing $Q_\alpha$, $j$, $j_1$, and $-r$ by 
$\bar Q_{\dot\alpha}$, $\bar j$, $j_2$, and $r$, respectively.

\begin{table}[!h]
\begin{centering}
\begin{tabular}{|l|l|l|l|l||l|}
\hline
\multicolumn{1}{|c|}{AdS$_5$} 
& CFT superfield (schematic)
& $s_1$
& $s_2$
& Multiplet 
& \multicolumn{1}{l|}{Comment}
\tabularnewline
\hline
\hline
Graviton 
& $\tr \bar{W}_{2\dot{\alpha}}e^{V_2}W_{2\alpha}e^{-V_2}(\bar{A}\bar{B})^k$
& $\frac{k}{2}$
& $\frac{k}{2}$
& $A_1\overline{L}[\frac{1}{2};\frac{1}{2}]^{(r=-k)}_{\Delta=3-\frac{3}{2}r}$
& $k\geq 1$
\tabularnewline
\cline{2-6}
& $\tr \bar{W}_{2\dot{\alpha}}e^{V_2}W_{2\alpha}e^{-V_2}$
& $0$
& $0$
& $A_1\overline{A}_1[\frac{1}{2};\frac{1}{2}]^{(r=0)}_{\Delta=3}$
& stress tensor
\tabularnewline
\hline
\hline
Gravitino$_{{\rm I}}$  
& $\tr \bar{W}_{2\dot{\alpha}}(\bar{A}\bar{B})^k$
& $\frac{k}{2}$
& $\frac{k}{2}$
& $B_1\overline{L}[0;\frac{1}{2}]^{(r=-k-1)}_{\Delta=-\frac{3}{2}r}$
& ($\tr \bar{W}=0$)
\tabularnewline
\cline{2-6}
& $\tr \bar{W}_{2\dot{\alpha}}e^{V_2}Be^{-V_1}\bar{B}(\bar{A}\bar{B})^k$
& $\frac{k}{2}$ 
& $\frac{k+2}{2}$ 
& $A_2\overline{L}[0;\frac{1}{2}]^{(r=-k-1)}_{\Delta=2-\frac{3}{2}r}$ 
& 
\tabularnewline
\cline{2-6}
& $\tr \bar{W}_{2\dot{\alpha}}\bar{A}e^{V_1}Ae^{-V_2}(\bar{A}\bar{B})^k$
& $\frac{k+2}{2}$ 
& $\frac{k}{2}$ 
& $A_2\overline{L}[0;\frac{1}{2}]^{(r=-k-1)}_{\Delta=2-\frac{3}{2}r}$ 
& 
\tabularnewline
\hline
\hline
Gravitino$_{{\rm III}}$ 
& $\tr e^{V_2}W_{2\alpha}e^{-V_2}(\bar{A}\bar{B})^k$ 
& $\frac{k}{2}$ 
& $\frac{k}{2}$ 
& $A_1\overline{L}[\frac{1}{2};0]^{(r=-k+1)}_{\Delta=3-\frac{3}{2}r}$ 
& ($\tr W=0$)
\tabularnewline
\hline
\hline
Gravitino$_{{\rm IV}}$ 
& $\tr e^{V_2}W_{2\alpha}e^{-V_2}
\bar{W}_{2\dot\alpha}\bar{W}_2^{\dot{\alpha}}(\bar{A}\bar{B})^k$ 
& $\frac{k}{2}$ 
& $\frac{k}{2}$ 
& $A_1\overline{L}[\frac{1}{2};0]^{(r=-k-1)}_{\Delta=3-\frac{3}{2}r}$ 
& 
\tabularnewline
\hline
\hline
Vector$_{{\rm I}}$ 
& $\tr (\bar{A}\bar{B})^k$ 
& $\frac{k}{2}$ 
& $\frac{k}{2}$ 
& $B_1\overline{L}[0;0]^{(r=-k)}_{\Delta=-\frac{3}{2}r}$ 
& $k\geq 1$
\tabularnewline
\cline{2-6}
& $\tr e^{V_2}Be^{-V_1}\bar{B}(\bar{A}\bar{B})^k$
& $\frac{k}{2}$ 
& $\frac{k+2}{2}$ 
& $A_2\overline{L}[0;0]^{(r=-k)}_{\Delta=2-\frac{3}{2}r}$ 
& $k\geq 1$, 
\tabularnewline
\cline{2-6}
& $\tr e^{V_2}Be^{-V_1}\bar{B}$
& $0$
& $1$
& $A_2\overline{A}_2[0;0]^{(r=0)}_{\Delta=2}$
& $SU(2)_2$ current
\tabularnewline
\cline{2-6}
& $\tr \bar{A}e^{V_1}Ae^{-V_2}(\bar{A}\bar{B})^k$
& $\frac{k+2}{2}$ 
& $\frac{k}{2}$ 
& $A_2\overline{L}[0;0]^{(r=-k)}_{\Delta=2-\frac{3}{2}r}$ 
& $k\geq 1$, 
\tabularnewline
\cline{2-6}
& $\tr \bar{A}e^{V_1}Ae^{-V_2}$
& $1$
& $0$
& $A_2\overline{A}_2[0;0]^{(r=0)}_{\Delta=2}$
& $SU(2)_1$ current
\tabularnewline
\hline
\hline
Vector$_{{\rm III}}$ 
& $\tr \bar{W}_{2\dot\alpha}\bar{W}_2^{\dot{\alpha}}(\bar{A}\bar{B})^k$ 
& $\frac{k}{2}$ 
& $\frac{k}{2}$ 
& $B_1\overline{L}[0;0]^{(r=-k-2)}_{\Delta=-\frac{3}{2}r}$ 
& 
\tabularnewline
\cline{2-6}
& $\tr e^{V_2}Be^{-V_1}\bar{B}\bar{W}_{2\dot\alpha}\bar{W}_2^{\dot{\alpha}}(\bar{A}\bar{B})^k$ 
& $\frac{k}{2}$ 
& $\frac{k+2}{2}$ 
& $A_2\overline{L}[0;0]^{(r=-k-2)}_{\Delta=2-\frac{3}{2}r}$ 
& 
\tabularnewline
\cline{2-6}
& $\tr \bar{A}e^{V_1}Ae^{-V_2}\bar{W}_{2\dot\alpha}\bar{W}_2^{\dot{\alpha}}(\bar{A}\bar{B})^k$ 
& $\frac{k+2}{2}$ 
& $\frac{k}{2}$ 
& $A_2\overline{L}[0;0]^{(r=-k-2)}_{\Delta=2-\frac{3}{2}r}$
& 
\tabularnewline
\hline
\end{tabular}
\par\end{centering}
\caption{\label{gravity-multiplets}Short multiplets appearing in the Kaluza--Klein reduction of type IIB supergravity on AdS$_5\times T^{1,1}$. The leftmost column lists the names of the families of multiplets in AdS$_5$ \cite{Ceresole:1999zs,Ceresole:1999ht}. The fields $\bar W_{1\dot\alpha}$ and $\bar W_{2\dot\alpha}$ are the anti-chiral field-strength superfields of the vector multiplets for the two gauge groups, while $V_1$ and $V_2$ denote the corresponding real vector superfields. Finally, $s_1$ and $s_2$ are the quantum numbers under the global symmetry $SU(2)\times SU(2)$. Unless stated in the Comment column, $k\geq 0$.}
\end{table}

In addition to the multiplets listed in Table \ref{gravity-multiplets}, there are two more short multiplets at $N=\infty$.
Their contributions cancel because one multiplet corresponds to an exactly marginal deformation and the other to a conserved current.

One of the two short multiplets omitted from Table \ref{gravity-multiplets} is a linear combination of gaugino bilinears. The ``Vector$_{\rm III}$'' family includes a tower of the schematic form $\tr \bar{W}_{2\dot{\alpha}}\bar{W}^{\dot\alpha}_2 (\bar{A}\bar{B})^k$ for $k\geq 0$. If we instead use the other gaugino, the operator $\tr \bar{W}_{1\dot{\alpha}}\bar{W}^{\dot\alpha}_1 (\bar{B}\bar{A})^k$ is $Q$-cohomologous to the one containing $\bar{W}_{2\dot{\alpha}}$. This follows from relations such as $Q(D_{\dot\alpha}\bar{A})=\bar{W}_{2\dot{\alpha}}\bar{A}-\bar{A}\bar{W}_{1\dot{\alpha}}$.
However, at $k=0$, the two gaugino bilinears, $\tr \bar{W}_{2\dot{\alpha}}\bar{W}_2^{\dot{\alpha}}$ and $\tr \bar{W}_{1\dot{\alpha}}\bar{W}_1^{\dot{\alpha}}$, are independent.
This reflects the existence of two exactly marginal operators that preserve the global $SU(2)\times SU(2)$ symmetry \cite{Klebanov:1998hh}. One is the difference of the gauge kinetic terms, and the other is a linear combination of the superpotential and the sum of the gauge kinetic terms. The remaining linear combination of the superpotential and the sum of the gauge kinetic terms is not exactly marginal because of the Konishi anomaly \cite{Konishi:1983hf,Konishi:1985tu}.

The second multiplet omitted from Table \ref{gravity-multiplets} is $A_2\bar{A}_2[0;0]_{\Delta=2}^{(r=0)}$, the $U(1)_B$ current multiplet:
\begin{align}
    \tr A_i e^{-V_2}\bar A^i e^{V_1}-\tr e^{V_2}B_k e^{-V_1}\bar B^k \; .
\end{align}
A cohomology representative of the BPS operator in this multiplet is
\begin{align}
    \tr (\psi_{ai}a^i- b^{\hat i}\psi_{b\hat{i}})\;.
\end{align}
This operator is readily seen to be \(Q\)-closed. It represents a nontrivial cohomology class because it belongs to the conserved \(U(1)_B\) current multiplet. In \eqref{eq:Q-action}, it is precisely the trace part subtracted from \(Qf_1\).
On the gravity side, this is the Betti multiplet associated with the non-exact closed two-form on $T^{1,1}$ \cite{Ceresole:1999zs}.
To see its origin, consider the fluctuation of the RR four-form field $a_{\mu\alpha\beta\gamma}$, where $\mu$ is an AdS$_5$ index and $\alpha,\beta,\gamma$ are indices on $T^{1,1}$. The Lorenz-type gauge condition $D^\alpha a_{\mu\alpha\beta\gamma}=0$ can be dualized on $T^{1,1}$ to $d_5 (*_5a_\mu)=0$.
Hence, the internal Hodge dual $*_5 a_\mu$ may be expanded in closed two-forms. 
Since $T^{1,1}$ has non-vanishing Betti numbers $b_2=b_3=1$, there exists a closed two-form that is not exact.
Moreover, the continuous isometry group acts trivially on de Rham cohomology. Indeed, for a closed form $\omega$ and a Killing vector $k$, the Lie derivative
\[
\mathcal{L}_k\omega=d(\iota_k\omega)+\iota_k d\omega=d(\iota_k\omega),
\]
is exact. Therefore, the Betti multiplet is a singlet under the global symmetry. In particular, unlike the multiplets in Table \ref{gravity-multiplets}, the Betti multiplet does not come with a tower structure.

Table \ref{gravity-multiplets} helps us identify explicit representatives of the $Q$-cohomology classes. Its second column provides the schematic superfield expressions for the corresponding \(\mathcal{N}=1\) multiplets. Because these expressions are schematic, they need not satisfy the relevant shortening conditions. For anti-chiral superfields, the condition is simple enough that the schematic expressions satisfy it automatically: they are annihilated by the superspace covariant derivative with an undotted index. By contrast, the schematic expressions for the $A_i\bar L$-type or $A_i\bar{A}_i$-type superfields in Table \ref{gravity-multiplets} need not satisfy the required shortening conditions. Thus, the would-be BPS operator $Q_+ O^{\mathrm{h.w.}}_{+\cdots +}$ is not necessarily $Q$-closed,
where $O^{\mathrm{h.w.}}_{\alpha_1\cdots \alpha_{2j}}$ is the naive schematic expression of the highest weight of $A_i\bar{L}$ (or $A_i\bar{A}_i$) multiplets.

As an illustration, consider the lowest multiplet in the ``Graviton'' family, \(A_1\bar A_1[\frac12;\frac12]^{(r=0)}_{\Delta=3}\), namely the stress-tensor multiplet of the \(\mathcal N=1\) superconformal algebra. To identify the corresponding \(Q\)-cohomology class, one must construct a \(Q\)-closed operator containing
\begin{align}
    \tr \lambda_{1\dot\alpha}f_1
    \qquad \text{and} \qquad
    \tr \lambda_{2\dot\alpha}f_2.
\end{align}
These terms can arise as a $Q_+$-descendant of the naive highest weight.
Other operators with the same quantum numbers include
\begin{align}
    \tr (D_{\dot\alpha}a^i)\psi_{ai}\;, \qquad
    \tr (D_{\dot\alpha}b^{\hat i})\psi_{b\hat{i}}\;, \qquad
    \tr a^i (D_{\dot\alpha}\psi_{ai})\;, \qquad
    \tr b^{\hat i}(D_{\dot\alpha}\psi_{b\hat{i}})\;.
\end{align}
One then finds a \(Q\)-closed linear combination containing the \(\tr \lambda f\) terms:
\begin{align}
    &\bigl(\tr \lambda_{1\dot\alpha}f_1+\tr\lambda_{2\dot\alpha}f_2\bigr)
    -\frac{3}{4}\bigl(\tr (D_{\dot\alpha}a^i)\psi_{ai}
    +\tr (D_{\dot\alpha}b^{\hat i})\psi_{b\hat{i}}\bigr) \notag\\
    &\hspace{4cm}
    +\frac{1}{4}\bigl(\tr a^i (D_{\dot\alpha}\psi_{ai})
    +\tr b^{\hat i}(D_{\dot\alpha}\psi_{b\hat{i}})\bigr).
\end{align}
This \(Q\)-closed operator represents the nontrivial cohomology class in the stress-tensor multiplet and is therefore not \(Q\)-exact.
This is not obtained simply by acting with \(Q_+\) on the schematic highest-weight expression inferred from Table \ref{gravity-multiplets}.

For the remainder of this discussion, we focus on the multiplets in Table \ref{gravity-multiplets} that lie in the BMN sector \eqref{eq:BMN-sector-Q}.
In Table \ref{gravity-multiplets}, the ``Vector$_{\mathrm{I}}$'' and ``Gravitino$_{\mathrm{III}}$'' families lie in the BMN sector. Explicit representatives of the mesonic monotonous cohomologies in the Vector$_{\mathrm{I}}$ family are
\begin{equation}
\begin{aligned}\label{eq: vector1 family}
    \tr a^{(i_1}b^{(\hat{j}_1} a^{i_2}b^{\hat{j}_2}\cdots a^{i_k)}b^{\hat{j_k})},\qquad (k\geq 1)\\
    \tr \psi_b^{(\hat{j}_1}b^{\hat{j}_2}a^{(i_1}b^{\hat{j}_3}\cdots a^{i_k)}b^{\hat{j}_{k+2})},\qquad (k\geq 0)\\
    \tr a^{(i_1}\psi_a^{i_2}a^{i_3}b^{(\hat{j}_1}\cdots a^{i_{k+2})}b^{\hat{j}_k)},\qquad (k\geq 0)
\end{aligned}
\end{equation}
where the two sets of global-symmetry indices are symmetrized separately: all unhatted indices are symmetrized with one another, as are all hatted indices.
The cohomology representatives in the Gravitino$_{\rm III}$ family are
\begin{equation}
\begin{aligned}\label{eq: gravitino3 family}
    \tr \Big[(a^{(i_1}f_1+f_2a^{(i_1})b^{(\hat{j}_1}a^{i_2}b^{\hat{j}_2}\cdots a^{i_k)}b^{\hat{j}_k)}
    -\sum_{l=0}^{k-1} a^{(i_1}b^{(\hat{j}_1}\cdots a^{i_l}b^{\hat{j_l}}\psi_b^{\hat{j}_{l+1}}b^{\hat{j}_{l+2}}a^{i_{l+1}}\cdots b^{\hat{j}_{k})}a^{i_{k-1}}\psi_a^{i_{k})}\Big]
\end{aligned}
\end{equation}
where $k\geq 1$, and the unhatted and hatted indices are each symmetrized among themselves.
Except for the cohomology corresponding to $U(1)_B$ current multiplet, these exhaust the BMN mesonic single-trace monotonous cohomologies.
We leave the construction of explicit representatives for non-BMN mesonic monotonous cohomologies to future work.

At finite $N$, due to the presence of trace relations, the single-trace cohomologies above \eqref{eq: vector1 family},\eqref{eq: gravitino3 family} for sufficiently large $k$ split into sums of multi-trace graviton cohomologies.
This is nontrivial, especially for \eqref{eq: gravitino3 family}, because in principle a single-trace cohomology may split into a sum of multi-traces that are not themselves cohomologies. 

It would be sufficient to consider the highest-weight components of 
$SU(2)\times SU(2)$ flavor symmetry, which we choose to be the ones with all 
lowered indices set to $1$. In terms of $a=a_1$, $b=b_1$, $\psi_a\equiv \psi_{a1}$, 
$\psi_b\equiv \psi_{b1}$, we write the highest-weight components of the BMN gravitons as
\begin{equation}
\begin{aligned}
    u_k\equiv \tr (ab)^k,\qquad
    v_{a;k+1}\equiv \tr a\psi_{a}(ab)^k,\qquad
    v_{b;k+1}\equiv \tr \psi_{b}b(ab)^k,\\
    w_k\equiv \tr f_1(ba)^k+\tr f_2(ab)^k-\sum_{m=0}^{k-1}\tr (ab)^m\psi_b(ab)^{k-m-1}\psi_a.
\end{aligned}
\end{equation}

First, one can show that $u_k$ with $k>N$ can be expressed as 
polynomials in $u_1,...,u_N$ for $k>N$. To see this, first note that 
any $N\times N$ matrix $M$ satisfies the Cayley-Hamilton identity
\begin{equation}
\begin{aligned}\label{eq: CayleyHamilton}
    M^N=\sum_{k=0}^{N-1}c_{k}M^k,\qquad (M^0= \mathbf{1}_N)
\end{aligned}
\end{equation}
where the coefficients $c_k$ are polynomials in $\tr M,...,\tr M^{N-k}$.
The coefficient $c_0=(-1)^{N-1}\det M$ can be expressed as a polynomial in $\tr M,...,\tr M^N$ by taking trace of this identity \eqref{eq: CayleyHamilton}.
For example, for $N=2$, we have $M^2=(\tr M)M-\frac12((\tr M)^2-\tr M^2)\mathbf{1}_2$.
More generally,
\begin{equation}\label{eq: CH_l}
    M^{N+l}=\sum_{k=0}^{N-1}c_{k}^{(l)}M^{k},\qquad l\geq 0.
\end{equation}
where the coefficients $c_{k}^{(l)}$ are polynomials in $\tr M,...,\tr M^{m}$, with $m=\min\{N-k+l, N\}$. These coefficients satisfy the following recurrence relation:
\begin{equation}
    c_{k}^{(l+1)}=c_{N-1}^{(l)}c_k^{(0)}+c_{k-1}^{(l)},\qquad (c_k^{(0)}=c_k,\;c_{-1}^{(l)}=0,\;\;0\leq k\leq N-1).
\end{equation}
Applying this to $M=ab$, we see that $u_{N+l}$ can be expressed as a polynomial in $u_1,...,u_N$ for any $l\geq 1$.
\begin{equation}\label{eq: u_k generating}
    u_{N+l}=\sum_{k=0}^{N-1}c_{k}^{(l)}u_k,\qquad (u_0=N)\ .
\end{equation}
Here, $c_{k}^{(l)}$ are polynomials in $u_1,...,u_{m}$, with $m=\min\{N-k+l, N\}$.
Instead of using $u_N$, we equivalently can use $\det ab$ to generate the $u_{N+l}$ for $l\geq 0$.

For the $v_{a;k}$ and $v_{b;k}$, we can use the identity obtained by acting $P^{i}{}_{j}\frac{\partial}{\partial M^{i}{}_{j}}$ on \eqref{eq: CH_l}:
\begin{equation}
\begin{aligned}
    &PM^{N+l-1}+\cdots M^{N+l-1}P
    =\sum_{k=1}^{N-1}c_{k}^{(l)}(PM^{k-1}+\cdots +M^{k-1}P)
    +\sum_{k=0}^{N-1}d_k^{(l)}M^{k},
\end{aligned}
\end{equation}
where $d_k^{(l)}=P^{i}{}_{j}\frac{\partial c_{k}^{(l)}}{\partial M^{i}{}_{j}}$ and $l\geq 0$.
Setting $M=ab$ and $P=a\psi_a$ (resp. $P=\psi_b b$), we see that $v_{a;N+l}$ (resp. $v_{b;N+l}$) can be expressed as polynomials in $u_1,...,u_N$ and $v_{a;1},...,v_{a;N}$ (resp. $v_{b;1},...,v_{b;N}$) for any $l\geq 1$.
\begin{equation}
\begin{aligned}\label{eq: v_k generating}
    (N+l)v_{a;N+l}=\sum_{k=1}^{N-1}c_{k}^{(l)}kv_{a;k}+\sum_{k=0}^{N-1}d_{a;k}^{(l)}u_k,\qquad
    (N+l)v_{b;N+l}=\sum_{k=1}^{N-1}c_{k}^{(l)}kv_{b;k}+\sum_{k=0}^{N-1}d_{b;k}^{(l)}u_k.
\end{aligned}
\end{equation}
where $d_{a;k}^{(l)}=d_{k}^{(l)}\big|_{M=ab,P=a\psi_a}$ and $d_{b;k}^{(l)}=d_{k}^{(l)}\big|_{M=ab,P=\psi_b b}$.
Here, $d_{a;k}^{(l)}$ (resp. $d_{b;k}^{(l)}$) are polynomials in $u_1,...,u_m$ and $v_{a;1},...,v_{a;m}$ (resp. $v_{b;1},...,v_{b;m}$) with $m=\min\{N-k+l,N\}$. Of course, $d_{a;k}^{(l)}$ (resp. $d_{b;k}^{(l)}$) are linear in $v_{a;1},...,v_{a;m}$ (resp. $v_{b;1},...,v_{b;m}$).
The identities above hold for \(l\geq0\). However, they provide new reduction relations for \(v_{a;N+l}\) and \(v_{b;N+l}\) only when \(l\geq1\), since the \(l=0\) relation already involves \(v_{a;N}\) (resp. \(v_{b;N}\)) through the coefficients \(d_{a;k}^{(0)}\) (resp. \(d_{b;k}^{(0)}\)).

For the $w_k$, consider
\begin{equation}
    (ab)^{N+l}=(ab)(ab)^{N+l-1}=\sum_{k=1}^{N}c_{k-1}^{(l-1)}(ab)^k,
\end{equation}
and act $(\psi_b)^{\hat p}{}_{p}\frac{\partial}{\partial a^{\hat p}{}_{p}}$ on both sides. We obtain
\begin{equation}
\begin{aligned}\label{eq: psps}
    (\psi_b)^{\hat p}{}_{p}\frac{\partial}{\partial a^{\hat p}{}_{p}}(ab)^{N+l}
    &=\psi_bb(ab)^{N+l-1}+\cdots +(ab)^{N+l-1}\psi_bb\\
    &=\sum_{k=1}^{N}c_{k-1}^{(l-1)}\big(\psi_bb(ab)^{k-1}+\cdots +(ab)^{k-1}\psi_bb\big)
    +\sum_{k=1}^{N}d_{b;k-1}^{(l-1)}(ab)^k,
\end{aligned}
\end{equation}
Then multiplying $b^{-1}\psi_a$ from the right and taking the trace, we get
\begin{equation}
\begin{aligned} \label{eq: trpsps}
    \sum_{m=0}^{N+l-1}\tr (ab)^m\psi_b(ba)^{N+l-m-1}\psi_a=
    \sum_{k=1}^{N}c_{k-1}^{(l-1)}\sum_{m=0}^{k-1}\tr (ab)^m\psi_b(ba)^{k-m-1}\psi_a
    +\sum_{k=1}^{N}d_{b;k-1}^{(l-1)}\tr (ab)^{k-1}a\psi_a.
\end{aligned}
\end{equation}
And the terms in $w_{N+l}$ that contain the field strengths $f_1$ and $f_2$ can be expressed as
\begin{equation}
\begin{aligned}
    \tr f_1(ba)^{N+l}+\tr f_2(ab)^{N+l}
    &=\sum_{k=1}^{N}c_{k-1}^{(l-1)}\big(\tr f_1(ba)^k+\tr f_2(ab)^k\big).
\end{aligned}
\end{equation}
Combining the two results above, we see that $w_{N+l}$ can be expressed as a polynomial in $u_1,...,u_N$, $v_{a;1},...,v_{a;N}$, $v_{b;1},...,v_{b;N}$, and $w_1,...,w_N$ for any $l\geq 1$.
\begin{equation}
\begin{aligned}
    w_{N+l}=\sum_{k=1}^{N}c_{k-1}^{(l-1)}w_k-\sum_{k=1}^{N}d_{b;k-1}^{(l-1)}v_{a;k}.
\end{aligned}
\end{equation}

Therefore, the finite $N$ generators of the BMN gravitons are the 
$SU(2)\times SU(2)$ multiplets obtained from the following highest weight states,  
\begin{equation}
    u_1,...,u_N,\;\;\; v_{a;1},...,v_{a;N},\;\;\; v_{b;1},...,v_{b;N},\;\;\;  w_1,...,w_N.
\end{equation}
In other words, BMN graviton cohomologies are polynomials in these generators.

An equivalent choice of generators consists of the flavor rotations of $u_1,...,u_{N-1}$, $v_{a;1},...,v_{a;N-1}$, $v_{b;1},...,v_{b;N-1}$, and $w_1,...,w_{N-1}$, together with the following generalized monotones:
\begin{equation}
\begin{aligned}
    \det ab,\qquad \epsilon\epsilon a^{N}\epsilon\epsilon b^{N-1}\psi_a,\qquad \epsilon\epsilon a^{N-1}\psi_b\epsilon\epsilon b^{N},\qquad
    \epsilon\epsilon a^{N-1}\psi_b\epsilon\epsilon b^{N-1}\psi_a.
\end{aligned}
\end{equation}
where $\epsilon\epsilon a^{N}\equiv \epsilon_{\hat{p}_1...\hat{p}_N}\epsilon^{p_1...p_N}a^{\hat p_1}{}_{p_1}\cdots a^{\hat p_N}{}_{p_N}$, and similarly for the other three generalized monotones.
$\epsilon\epsilon a^{N-1}\psi_b$ and $\epsilon\epsilon b^{N-1}\psi_a$ are the highest-weight components of the baryonic monotonous cohomologies defined in \eqref{eq:spinning-baryonic-classes} and \eqref{eq:spinning-baryonic-classes2}, respectively.
Using $\epsilon\epsilon a^N \epsilon\epsilon b^{N-1}\psi_a$ (resp. $\epsilon\epsilon a^{N-1}\psi_b \epsilon\epsilon b^N$) instead of $v_{a;N}$ (resp. $v_{b;N}$) is equivalent because of the $l=0$ specialization of the identity \eqref{eq: v_k generating}. Note that 
\begin{equation}
\begin{aligned}
    d_{a;0}^{(0)}\propto\epsilon\epsilon(ab)^{N-1}(a\psi_a)\ \propto\epsilon\epsilon a^N \epsilon\epsilon b^{N-1}\psi_a,
    \qquad
    d_{b;0}^{(0)}
    \propto \epsilon\epsilon(ab)^{N-1}(\psi_bb)\
    \propto \epsilon\epsilon a^{N-1}\psi_b \epsilon\epsilon b^N,
\end{aligned}
\end{equation}
Using $\epsilon\epsilon a^{N-1}\psi_b \epsilon\epsilon b^{N-1}\psi_a$ instead of $w_N$ is equivalent because of the following identity.
\begin{equation}
\begin{aligned}
    w_{N+l}=\sum_{k=1}^{N-1}(c_{k}^{(l)}w_k-d_{b;k}^{(l)}v_{a;k})-d_{b;0}^{(l)}\tr b^{-1}\psi_a,\qquad (l\geq 0)
\end{aligned}
\end{equation}
The last term with $b^{-1}$ is well-defined because it is multiplied by $d_{b;0}^{(l)}\propto \det b$.
This identity can be derived by repeating the procedure around \eqref{eq: psps} and \eqref{eq: trpsps}, but starting with $(ab)^{N+l}=\sum_{k=1}^{N-1}c_{k}^{(l)}(ab)^k$ instead of $(ab)^{N+l}=\sum_{k=1}^{N}c_{k-1}^{(l-1)}(ab)^k$.
At $l=0$, we have
\begin{equation}
\begin{aligned}\label{eq: w_N to gm}
    w_N&=\sum_{k=1}^{N-1}(c_{k}^{(0)}w_k-d_{b;k}^{(0)}v_{a;k})-d_{b;0}^{(0)}\tr b^{-1}\psi_a\\
    &=\sum_{k=1}^{N-1}(c_{k}^{(0)}w_k-d_{b;k}^{(0)}v_{a;k})+\frac{(-1)^N}{(N-1)!^2}\epsilon\epsilon a^{N-1}\psi_b \epsilon\epsilon b^{N-1}\psi_a.
\end{aligned}
\end{equation}

\subsection{Baryonic monotonous cohomologies}\label{sec:baryonic-monotones}

In this subsection, we first discuss the scalar (anti)chiral-ring sector, for which a general description is available.  
We then present several low-lying spinning baryonic classes at general $N$ whose specialization to $N=2$ will be used in the BMN cohomology calculation of Section~\ref{Section:fortuitous}.  A complete classification of the non-scalar baryonic cohomologies will not be attempted here.

As explained at the beginning of Section~\ref{Section: monotone}, 
any single trace operator containing finitely many formal letters $a^{-1}$ 
can be rewritten as a fermionic integral as \eqref{eq: Wick-to-Single_r}. 
A single-trace operator with $r$ inverse letters requires at most $r$ determinant factors 
to be multiplied, for it to be a legitimate local operator.  
Multiplying fewer factors of determinant may suffice, however, for a specific linear combination of multi-traces, which arises as Wick contraction of $r$ bilinear insertions of $|B|<r$ fermionic integral. 

This observation provides a practical way to search for baryonic monotonous cohomologies.  We may work directly with multi-trace expressions containing the additional letter $a^{-1}$ and convert them into determinant forms afterward. The conversion is not unique, but it is always possible.  Also, since any baryonic operator can be written as $(\det a)^{|B|}$ times a multi-trace expression that may contain $a^{-1}$, this description does not miss any baryonic operator.

\subsubsection{Scalar (anti)chiral-ring sector}

As reviewed in Appendix~\ref{appendix:scalar-baryons}, general scalar baryons with $|B|=1$ take the form
\begin{equation}
\begin{aligned}
\epsilon_{\hat{p}_1...\hat{p}_N}\epsilon^{p_1...p_N} (a_{I_1;J_1})^{\hat p_1}{}_{p_1}\cdots (a_{I_N;J_N})^{\hat p_N}{}_{p_N},
\end{aligned}
\end{equation}
where the flavor indices collected in $I$ and $J$ are symmetrized separately:
\begin{equation}
\begin{aligned}
a_{I;J}=a_{(i_1}b_{(\hat i_1}\cdots a_{i_m}b_{\hat i_m)}a_{i_{m+1})},
\qquad
I=(i_1,...,i_{m+1}),\;\;\; J=(\hat{i}_1,...,\hat{i}_m)\ .
\end{aligned}
\end{equation}
An explicit operator-level classification of the general scalar chiral ring with $|B|>1$ 
is considerably more complicated and will not be attempted here.  
Its spectrum, however, can be counted using generating-function and moduli-space methods \cite{Forcella:2007wk,Forcella:2008bb}.  As reviewed in Appendix~\ref{appendix:moduli-space}, this moduli space can be described in terms of the eigenvalues of the scalar matrices.  The same sector has also been studied from the gravity side \cite{Gubser:1998fp,Berenstein:2002ke,Beasley:2002xv}.  In particular, quantizing the moduli space of holomorphic surfaces \cite{Mikhailov:2000ya} describing supersymmetric wrapped D3-branes implies that every cohomology class made of 
scalars with $|B|>1$ is generated by products of $|B|$ unit-baryon classes \cite{Beasley:2002xv}.
This statement is nontrivial from the field-theory viewpoint.  As explained around equation~\eqref{eq: Wick-to-Single}, general baryonic operators with $|B|>1$ need not be generated by operators with $|B|=1$; the absence of non-product cohomology classes 
is a special property of the scalar chiral ring sector.

In a special case, to be specified below, we could prove from 
our setup of multi-trace operators with additional formal letter $a^{-1}$ 
that the baryonic chiral rings are generated by the $|B|=1$ cohomologies.
We shall consider the case with $B<0$ here: similar studies for $B>0$ can be 
done by exchanging $a$ and $b$.

We consider multi-trace $Q$-cohomologies built from $a,a_2,b_1,b_2$ and the additional letter $a^{-1}$, subject to $aa^{-1}=1$.  Single-trace operators 
can be constructed with the following adjoint combinations of 
the first gauge group:
\begin{equation}
\begin{aligned}
a^{-1}a_2,\qquad b_1a,\qquad b_2a,\qquad b_1a_2,\qquad b_2a_2.
\end{aligned}
\end{equation}
They commute with one another in the scalar (anti)chiral ring.  For example, the commutators involving $a^{-1}a_2$ are $Q$-exact because
\begin{equation}
\begin{aligned}
    Q(a^{-1}\psi_{b\hat{i}})&=a^{-1}(a_2b_{\hat{i}}a-ab_{\hat{i}}a_2)=[a^{-1}a_2,b_{\hat{i}}a],\\
    Q(a^{-1}\psi_{b\hat{i}}a^{-1}a_2)&=a^{-1}(a_2b_{\hat{i}}a-ab_{\hat{i}}a_2)a^{-1}a_2=[a^{-1}a_2,b_{\hat{i}}a_2]\ .
\end{aligned}
\end{equation}
$Q$-exactness of other commutators are also easily derived 
from the second and third lines of (\ref{eq:Q-action}).
Thus, within cohomology, the factors inside each trace may be freely reordered. 
Furthermore, if there is an $a^{-1}a_2$ factor in the trace, any factor $b_{\hat i}a$ can be moved to its left and replaced as $b_{\hat i}a\,a^{-1}a_2=b_{\hat i}a_2$. If there are at least as many $b_{\hat i}a$ factors as $a^{-1}a_2$ factors, all explicit inverse letters can be removed.  Otherwise, all $b_{\hat i}a$ factors can be replaced by $b_{\hat i}a_2$, leaving only the unpaired $a^{-1}a_2$ factors.
The most general single-trace representative containing explicit $a^{-1}$ can therefore be written as
\begin{equation}\label{eq:chiral ring a-1}
\begin{aligned}
\tr (a^{-1}a_2)^{n_0}(b_1a_2)^{n_1}(b_2a_2)^{n_2}
\end{aligned}
\end{equation}
with $n_0\geq 1$ and $n_1,n_2\geq 0$, 
where all the adjoint factors commute inside the trace. 
When there is no explicit $a^{-1}$, the most general single-trace representative is instead
\begin{equation}\label{eq:chiral ring meson}
\begin{aligned}
\tr (b_1a)^{n_1}(b_2a)^{n_2}(b_1a_2)^{n_3}(b_2a_2)^{n_4}
\end{aligned}
\end{equation}
with $n_1,n_2,n_3,n_4\geq 0$.  This is an ordinary mesonic single trace, and products of such traces can dress a baryon arbitrarily many times.

Therefore, any baryonic cohomology in this sector can be expressed as polynomials in the single-traces \eqref{eq:chiral ring a-1} and \eqref{eq:chiral ring meson}, multiplied by an appropriate number of determinants $(\det a)^{|B|}$. They are all baryonic monotone classes unless they become exact by finite-$N$ relations. 
At finite $N$, these obey trace relations, and the resulting finite-$N$ counting was carried out in \cite{Forcella:2007wk}.

As explained in Section 3, the minimal number $d\equiv |B|$ of determinants 
to be multiplied is no larger than the number $r$ of inverse letters 
$a^{-1}$ appearing in the multi-trace operator, $d\leq r$.
In general, there are multi-trace operators for which $d<r$: we have seen 
such examples in Section 3.
Within the chiral ring sector, we shall 
consider the multi-trace operators with $d=r$.
In this case, we will now argue within the field theory 
that the chiral ring is generated by unit-baryon cohomology classes.

Consider first a single-trace operator of the form
(\ref{eq:chiral ring a-1}) containing $r=n_0$ inverse letters, multiplied by
$(\det a)^r$. Since the scalar factors commute within
the trace, it can be written in the form
\[
\tr(Y_{(1}\cdots Y_{r)}),\qquad Y_i=a^{-1}X_i.
\]
The single-trace terms appearing in the Wick contraction \eqref{eq: Wick B=1} of $B=-1$ fermionic integral with $r$ insertions $X_i$ are of this form, up to the determinant factor. 

We can therefore use the $B=-1$ integral with these $r$
insertions, multiplied by $(\det a)^{r-1}$, to subtract the desired
single-trace term.  This expression is a product of one excited $B=-1$
cohomology and $r-1$ ground-state baryons.  Its Wick contraction also
produces terms containing two or more traces, so these terms remain after
the first subtraction.

Consider next one of the resulting double-trace terms.  The $r$ inverse
letters are divided between its two traces, and each trace contains fewer
than $r$ inverse letters.  As above, each trace appears as the single-trace
part of a $B=-1$ fermionic integral with the corresponding insertions.  The
product of these two integrals, multiplied by $(\det a)^{r-2}$, therefore
contains the desired double-trace term.  Subtracting a suitable multiple
cancels that term and produces only terms with three or more traces.

The same procedure can then be repeated in increasing order of the number
of traces.  A term with $k$ traces is cancelled using a product of $k$
unit-baryon integrals, multiplied by $(\det a)^{r-k}$.  This subtraction can
only produce terms with more than $k$ traces.  The procedure terminates at
$k=r$, where every trace contains one inverse letter and is directly
represented by an integral with one insertion.  Every expression used in
these subtractions is a product of $r$ unit-baryon cohomologies, including
the required ground-state factors.  It follows that the original
single-trace operator is a linear combination of such products.

For a general multi-trace operator with $d=r$, the same argument can be
applied to each trace containing inverse letters.  Traces without inverse
letters are ordinary mesonic dressings and may be included in any one of
the unit-baryon factors.  Thus, every scalar cohomology in the $d=r$ sector
is generated by products of $d$ unit-baryon cohomologies.

A general multi-trace operator with $r$ inverse letters may 
require $d<r$ determinant factors.  The generic right-hand side of 
equation~\eqref{eq: Wick B=1}, for which $d=1$, is an example.  We do not attempt a direct field-theory proof of unit-baryon generation in this more general case.  
However, based on the moduli-space counting and the gravity dual studies of D3-branes 
quoted at the beginning of this subsection, one should be able to 
make a general argument for $d<r$.

\subsubsection{Spinning baryonic monotones}

Having discussed in the previous subsection the chiral ring sector of the baryonic 
cohomologies, made of the scalar letters only, next we turn to the 
spinning cohomology classes. We shall not attempt to systematically classify them, 
but present a few low-lying excited states at general $N$. 
Our selection is guided by the $N=2$ BMN calculation
in Section~\ref{Section:fortuitous}.

Consider the following two families of baryonic excited states at $B=-1$, 
\begin{equation}\label{eq:spinning-baryonic-classes}
    \epsilon\epsilon a_{i_1}\cdots a_{i_{N-1}}\psi_b^{\hat{i}}\ \ ,\ \ 
    \epsilon\epsilon a_{i_1}\cdots a_{i_{N-2}} a^i (a_if_1+f_2a_i)
    +\frac{N}{2}\epsilon\epsilon a_{i_1}\cdots a_{i_{N-2}}\psi^{\hat{i}}_b\psi_{b\hat{i}}\ ,
\end{equation}
and also the two families at $B=1$ which can be obtained from the above by 
exchanging the two $SU(N)$ gauge groups and $a\leftrightarrow b$,
\begin{eqnarray}\label{eq:spinning-baryonic-classes2}
  \epsilon\epsilon b_{\hat{i}_1}\cdots b_{\hat{i}_{N-1}}\psi_a^i\ \ ,\ \ 
    \epsilon\epsilon b_{\hat{i}_1}\cdots b_{\hat{i}_{N-2}}b^{\hat{i}}(b_{\hat{i}}f_2+f_1b_{\hat{i}})
    -\frac{N}{2}\epsilon\epsilon b_{\hat{i}_1}\cdots b_{\hat{i}_{N-2}}\psi_a^i\psi_{ai}\ .
\end{eqnarray}
Note that the free flavor indices carried by the 
scalars are totally symmetrized due to the contractions of their gauge indices 
with the $\epsilon$ tensor. We will show that the operators 
(\ref{eq:spinning-baryonic-classes}) at $B=-1$ are $Q$-closed and not 
$Q$-exact. Similar calculations can be done for those at $B=1$.

For the $B=-1$ operators (\ref{eq:spinning-baryonic-classes}), 
it suffices to consider those in which all the scalars with uncontracted 
flavor indices are $a\equiv a_1$, which are the highest-weight states of 
the totally symmetric representation of the $SU(2)$ flavor symmetry.

We use the auxiliary fermionic integrals introduced in
Section~3.2. Recall that the ground state baryon is given by
\begin{equation}
\begin{aligned}
    \epsilon\epsilon a\cdots a
    =N!\det a
    =N! \int[d\psi d\chi]e^{\psi a\chi}\;.
\end{aligned}
\end{equation}
Excited baryons are represented by bilinear insertions in the
integral.  We use the notation
$\int_{\psi,\chi}\equiv \int[d\psi d\chi]e^{\psi a\chi}$ and write
$F_i\equiv a_i f_1+f_2a_i$.  The two spinning representatives 
(\ref{eq:spinning-baryonic-classes}) are given by
\begin{align}
    \epsilon\epsilon a\cdots a \psi_b^{\hat{i}}
    =&-(N-1)!\int_{\psi,\chi}(\psi\,\psi^{\hat{i}}_{b}\chi)\;,\\
    \epsilon\epsilon a\cdots a a^i F_i
    +\frac{N}{2}\epsilon\epsilon a\cdots a
    \psi_b^{\hat{i}}\psi_{b\hat{i}}
    =&\,(N-2)!\int_{\psi,\chi}\left[(\psi a^i\chi)(\psi F_i\chi)
    +\frac{N}{2}(\psi\,\psi_b^{\hat{i}}\chi)
    (\psi\,\psi_{b\hat{i}}\chi)\right].
    \label{eq: non-BH baryon as integral 2}
\end{align}
The minus sign on the right-hand side of the first line comes from moving the
physical fermion $\psi_b$ through an auxiliary fermion.

We now perform the Wick contractions.  For two insertions, with the order of
fermionic letters kept as displayed, the relevant identity is
\begin{align}\label{eq:two-insertion-Wick}
\int_{\psi,\chi}(\psi X\chi)(\psi Y\chi)
=\det a\left[
\tr(a^{-1}X)\tr(a^{-1}Y)
-\tr(a^{-1}Xa^{-1}Y)\right].
\end{align}
The one-fermion class becomes
\begin{align}\label{eq:one-fermion-baryon-Wick}
\epsilon\epsilon a\cdots a\psi_b^{\hat i}
=(N-1)!\det a\,\tr(a^{-1}\psi_b^{\hat i}).
\end{align}
For the last $B=-1$ class with two insertions, using the tracelessness of $f_1,f_2$ and
$a^ib_{\hat i}a_i=a_2b_{\hat i}a-ab_{\hat i}a_2$, the identity \eqref{eq:two-insertion-Wick} yields
\begin{align}
&\epsilon\epsilon a\cdots a a^iF_i
+\frac{N}{2}\epsilon\epsilon a\cdots a
\psi_b^{\hat i}\psi_{b\hat i}
\nonumber\\
&=N(N-2)!\det a\Big(
-\tr\!\left[a^{-1}(a_2f_1+f_2a_2)\right]
\nonumber\\&\hspace{4.1cm}
+\frac12\left[
\tr(a^{-1}\psi_b^{\hat i})\tr(a^{-1}\psi_{b\hat i})
-\tr(a^{-1}\psi_b^{\hat i}a^{-1}\psi_{b\hat i})
\right]\Big).
\label{eq:two-fermion-baryon-Wick}
\end{align}

These expressions are $Q$-closed without using finite-$N$ trace relations.
For \eqref{eq:one-fermion-baryon-Wick}, this follows immediately from
\begin{align}
Q\tr(a^{-1}\psi_{b\hat i})
=\tr\!\left(a^{-1}(a_2b_{\hat i}a-ab_{\hat i}a_2)\right)=0.\label{eq:one-fermion-closed}
\end{align}
For \eqref{eq:two-fermion-baryon-Wick}, single-trace and double-trace parts are $Q$-closed individually.
\begin{align}
Q\left[
\tr(a^{-1}\psi_b^{\hat i})\tr(a^{-1}\psi_{b\hat i})
\right]&=0,\\
Q\Big(
-\tr\!\left[a^{-1}(a_2f_1+f_2a_2)\right]
-\frac12\tr(a^{-1}\psi_b^{\hat i}a^{-1}\psi_{b\hat i})
\Big)&=0.\label{eq:single-trace-baryon-closed}
\end{align}
The first equality follows from \eqref{eq:one-fermion-closed} and the
Leibniz rule. The second follows by substituting \eqref{eq:Q-action}; the
trace-subtraction terms and terms containing $\psi_{ai}$ in $Qf_1$ and $Qf_2$ cancel. The terms that contain $\psi_{b\hat{i}}$ in $Qf_{1,2}$ then cancel with the second term in \eqref{eq:single-trace-baryon-closed}. Both calculations use
only $a^{-1}a=1$ and trace cyclicity, and no finite-$N$ trace relation. 
(Note also that this separate \(Q\)-closedness makes the monotone nature of these representatives manifest, since their \(Q\)-closure does not rely on any finite-\(N\) trace relation.)

The factor $N/2$ in \eqref{eq:spinning-baryonic-classes} has a simple
Wick-contraction origin.  One contraction of the $(\psi a^i\chi)(\psi F_i\chi)$ term contains
$\tr(a^{-1}a)=N$.  The coefficient $N/2$ of the two-fermion term matches
this factor and leaves the relative coefficient $1/2$ appearing in
\eqref{eq:two-fermion-baryon-Wick} and
\eqref{eq:single-trace-baryon-closed}.

The single-trace and double-trace expressions are
therefore separately $Q$-closed in the baryonic covering space.  The
double-trace pattern, however, does not determine a
unique physical baryonic operator.  Multiplying it by $\det a$ yields the
double-trace contribution to the $B=-1$ two-insertion integral, whereas
multiplying it by $(\det a)^2$ yields the $B=-2$ product of two one-insertion baryons. 
Thus, the same $Q$-closed trace pattern can occur with different
determinant dressings and hence in different baryon-number sectors.

Non-exactness follows from a simple consistent truncation that sets $b^{\hat i}=\psi_{ai}=0$, i.e. not using them regarded as `heavy' letters, together 
with also not using the non-BMN letters (gauginos and letters with derivatives). 
Every right-hand side in \eqref{eq:Q-action} then vanishes 
(in particular $Q$-action on $\psi_{ai}$ vanishes), 
implying that the truncation is consistent. 
Had an operator been $Q$-exact, it should have been $Q$-exact even after 
this consistent truncation. However, the operators 
(\ref{eq:spinning-baryonic-classes}) remain nonzero after
the truncation and therefore cannot be $Q$-exact (because all $Q$ transformation 
vanishes after this truncation). This implies that they cannot be $Q$-exact 
before the truncation either.

\subsection{A digression to ABJM/BLG}\label{section:abjm}

In this section, we discuss the cohomologies of 3d SCFTs which are 
closely related to those of the KW theory. We shall discuss the 
classical cohomologies of 3-dimensional ABJM theory \cite{Aharony:2008ug} or 
the BLG theory \cite{Bagger:2007jr,Gustavsson:2007vu}. The ABJM theory 
has $U(N)_k\times U(N)_{-k}$ or $SU(N)_{k}\times SU(N)_{-k}$ gauge groups 
and CS levels, while the BLG theory may be regarded as a specialization of the ABJM theory 
to $SU(2)_k\times SU(2)_{-k}$. 
Despite various differences between the ABJM/BLG theories and the KW theory, 
starting from different dimensionalities, 
they have certain formal similarities. They have the same 
quiver structures, thus the same bifundamental chiral multiplet
contents in the 4d $\mathcal{N}=1$ (3d $\mathcal{N}=2$) formalism. 
Their superpotentials also take the same functional form.
However, the nature of the vector multiplets and the gauge interactions 
(propagating 4d $\mathcal{N}=1$ vector multiplet vs. auxiliary CS vector multiplet) 
are different. 

The $Q$-cohomology problems of the 4d and 3d systems also have 
related similarities and differences. 
The 3d cohomologies are discussed mostly at large $k$, 
at weak-coupling \cite{Belin:2025hsg,Behan:2025hbx}. 
Among the BPS letters, the bifundamental part ($a_i$, $b_{\hat{i}}$, 
$\psi_{ai}$, $\psi_{b\hat{i}}$) is the same. On the other hand, while 
the field strengths $f_{1,2}$ and the gauginos $\lambda_{1,2\dot\alpha}$ are independent 
letters in 4d, they are composite (thus auxiliary) operators in 3d.
Finally, while the 4d theories admit two BPS derivatives, the 3d theories 
have only one BPS derivative. 
The $Q$-transformation rules of the 4d/3d theories are the same only if 
one can restrict to the sector which does not use the derivatives and the fields 
in the vector multiplets. This restriction cannot be consistently imposed 
in the sense of our Section 2. However, there are examples of 
3d cohomology representatives \cite{Belin:2025hsg,Behan:2025hbx} which 
can be discussed within this sector and can thus be related 
to the KW cohomologies. 
While the connection is more straightforward for the $SU(N)$ theory, 
certain cohomologies in the $U(N)$ theory can also be understood from the perspective 
of the KW theory.

One purpose of these comparisons is to emphasize that the notion of 
fortuity/monotonicity may be subtle in the sense of our generalized monotonicity. 
The subtlety will be found even in the $U(N)\times U(N)$ ABJM theory 
without $U(1)_B$ baryon symmetry, where certain multi-trace operators can 
be factorized into gauge non-invariant baryon-antibaryon pairs. 
One guiding idea of the fortuity program is to define the
fortuitous class as novel $N$-dependent cohomologies which would exhibit certain 
(black-hole-like) erratic features. So it might be desirable to set a stricter  
criterion for fortuity if one can observe extra regular patterns and use them 
to enlarge the monotonous class. 
Here our aim is not suggesting a definite/complete fortuity criterion for
the ABJM theories of the $U(N)$ type, but rather to point out that there may
be ambiguities in defining good fortuity rules, to be fixed by physical criteria.

A fortuitous cohomology of the $U(2)_k\times U(2)_{-k}$ ABJM theory was constructed in \cite{Belin:2025hsg}. It transforms in the $(\mathbf{2},\mathbf{2})$ representation of the $SU(2)\times SU(2)\subset SU(4)$ global symmetry. More recently, \cite{Behan:2025hbx} constructed two fortuitous cohomologies in the $U(3)_k\times U(3)_{-k}$ theory. One is a singlet under $SU(2)\times SU(2)$, and the other transforms in the $(\mathbf{3},\mathbf{3})$ representation.
Their operator counting also shows that the $U(4)_k\times U(4)_{-k}$ theory has two
fortuitous cohomologies with $E=\frac{9}{2}$ and $J=\frac{1}{2}$. They transform in the $(\mathbf{2},\mathbf{2})$ and $(\mathbf{4},\mathbf{4})$ representations, respectively.

We now construct a fortuitous cohomology of ABJM theory at general rank $N$. It has quantum numbers
\begin{equation}
E_{\mathrm{ABJM}}=N+\frac{1}{2},\qquad J=\frac{1}{2},
\end{equation}
and transforms in the $(\mathbf{N},\mathbf{N})$ representation of $SU(2)\times SU(2)$. The construction uses baryonic monotonous cohomologies of KW theory, so it can also be 
viewed as a generalized monotone in ABJM theory. For $N=2,3,4$, this family reproduces the fortuitous cohomologies in the $(\mathbf{2},\mathbf{2})$, $(\mathbf{3},\mathbf{3})$, and $(\mathbf{4},\mathbf{4})$ representations found in \cite{Belin:2025hsg,Behan:2025hbx}.

Consider the following four baryonic monotones 
of the $SU(N)\times SU(N)$ KW theory:
\begin{equation}
\begin{aligned}
    \mathbf{A}_{i_1\cdots i_N}=\epsilon\epsilon
    a_{i_1}\cdots a_{i_N},
    \qquad &
    \mathbf{B}_{\hat{i}_1\cdots \hat{i}_{N}}=\epsilon\epsilon b_{\hat{i}_1}\cdots b_{\hat{i}_{N}},
    \\ 
    \mathbf{A}_{\psi,\hat{i}i_1\cdots i_{N-1}}=\epsilon\epsilon  \psi_{b\hat{i}} a_{i_1}\cdots a_{i_{N-1}},
    \qquad &
    \mathbf{B}_{\psi ,i\hat{i}_1\cdots \hat{i}_{N-1}}=\epsilon\epsilon
    \psi_{ai}b_{\hat{i}_1}\cdots b_{\hat{i}_{N-1}}.
\end{aligned}
\end{equation}
They can also be regarded as gauge-invariant operators of the $SU(N)$ 
type ABJM theory, but not of the $U(N)$ type ABJM because the overall $U(1)$ 
charge is not zero.
From these, consider the following four products $\mathbf{AB}$, $\mathbf{AB}_{\psi}$, $\mathbf{A}_\psi\mathbf{B}$, and $\mathbf{A}_\psi\mathbf{B}_\psi$:
\begin{equation}\label{dibaryon}
\begin{aligned}
    \mathbf{AB}&:\; (\mathbf{N+1},\mathbf{N+1}),\\
    \mathbf{AB}_{\psi}&:\; (\mathbf{N+2},\mathbf{N})\oplus(\mathbf{N},\mathbf{N}),\\
    \mathbf{A}_\psi\mathbf{B}&:\; (\mathbf{N},\mathbf{N+2})\oplus (\mathbf{N},\mathbf{N}),\\
    \mathbf{A}_\psi\mathbf{B}_\psi &:\; (\mathbf{N+1},\mathbf{N+1})\oplus(\mathbf{N+1},\mathbf{N-1})\oplus(\mathbf{N-1},\mathbf{N+1})\oplus(\mathbf{N-1},\mathbf{N-1}).
\end{aligned}
\end{equation}
They can all be decomposed into mesons.  
All of these can be viewed as gauge-invariant, $Q$-closed operators of the $U(N)_k\times U(N)_{-k}$ ABJM theory, because the overall $U(1)$ gauge charges carried by the ${\bf A}$ and ${\bf B}$ type baryons cancel. In the gravity dual on AdS$_4\times\mathbb{CP}^3$, 
they correspond to D4-branes wrapping a $\mathbb{CP}^2$ cycle twice \cite{Murugan:2011zd}, one along 
the $a$ direction and another along the $b$ direction with opposite orientation. 
The different orientations of the two wrappings allow them to 
topologically unwind to mesons. They can all be regarded as generalized monotones 
even in the $U(N)$ theory.

The two $(\mathbf N,\mathbf N)$ components in
$\mathbf{AB}_{\psi}$ and $\mathbf A_{\psi}\mathbf B$ can mix.  We shall show
that their sum is fortuitous in the ordinary mesonic classification, whereas
their difference is an ordinary mesonic monotone.  Below, we refer to ordinary
mesonic monotones, including their multi-graviton products, simply as graviton
cohomologies.  Both combinations retain the factorized pattern of generalized
monotones.  The other seven combinations in (\ref{dibaryon}) are expected to be graviton
cohomologies, although their explicit representatives will not be needed below.\footnote{
This is nontrivial for $\mathbf{A}_\psi\mathbf{B}_\psi$, but we expect them to be gravitons because there is no known non-graviton cohomology found with the same quantum number as $\mathbf{A}_\psi\mathbf{B}_\psi$ at $N=2,3,4$ \cite{Belin:2025hsg,Behan:2025hbx}.
The $\mathbf{A}_\psi\mathbf{B}_\psi$ in representation $(\mathbf{N+1},\mathbf{N+1})$ is a graviton in KW theory because of the equation \eqref{eq: w_N to gm}.
}
The sum of the two $(\bf{N},\bf{N})$ (which will exhibit ordinary fortuity) is
\begin{equation}\label{eq:general-ABJM-fortuitous}
F_{i_1\cdots i_{N-1}
\hat{i}_1\cdots\hat{i}_{N-1}}
=
\mathbf{A}_{i_1\cdots i_{N-1}\,i}
\mathbf{B}^{i}_{\psi,\hat{i}_1\cdots\hat{i}_{N-1}}
+\mathbf{A}^{\hat i}_{\psi,i_1\cdots i_{N-1}}
\mathbf{B}_{\hat{i}\hat{i}_1\cdots\hat{i}_{N-1}}.
\end{equation}

At $N=2$, $F_{i\hat{i}}$ of (\ref{eq:general-ABJM-fortuitous}) can be written as 
a multi-trace operator as follows:
\begin{equation}
\begin{aligned}\label{eq:F-N2-expansion}
    F_{i\hat i}&=\epsilon\epsilon a_i a_j \epsilon\epsilon \psi_a^j b_{\hat i}
    +\epsilon\epsilon \psi_{b}^{\hat j}a_i \epsilon\epsilon b_{\hat j}b_{\hat i}\\
    &=\tr a_i\psi_{a}^j \tr a_j b_{\hat i} +\tr a_i b_{\hat i}\tr a_j\psi_a^j
    -\tr a_i\psi_a^ja_j b_{\hat i}-\tr a_ib_{\hat i}a_j\psi_a^j\\
    &\quad+\tr b_{\hat i}\psi_{b}^{\hat j}\tr b_{\hat j}a_i
    +\tr b_{\hat i}a_i \tr b_{\hat j}\psi_b^{\hat j}
    -\tr b_{\hat i}\psi_b^{\hat j}b_{\hat j}a_i
    -\tr b_{\hat i}a_i b_{\hat j}\psi_b^{\hat j}.
\end{aligned}
\end{equation}
On the other hand, the ABJM fortuitous cohomology of \cite{Belin:2025hsg} is 
given by
\begin{equation}
\begin{aligned}
    O_{f,i\hat{i}}=&\tr \psi_{aj}a^j b_{\hat{i}}a_i+\tr \psi_{b\hat{j}}b^{\hat{j}}a_i b_{\hat{i}}
    -\frac{3}{4}\tr b_{\hat{i}}a_i (\tr \psi_{aj}a^j+\tr \psi_{b\hat{j}}b^{\hat{j}}).
\end{aligned}
\end{equation}
This is proportional to $F_{i\hat{i}}$ up to gravitons and a $Q$-exact term, 
\begin{equation}
\begin{aligned}
    2O_{f,i\hat{i}}=F_{i\hat{i}}+\tr \psi_{a(i}a_{j)}\tr a^j b_{\hat{i}}+\tr \psi_{b(\hat{i}}b_{\hat{j})}\tr a_i b^{\hat{j}}-Q\tr \psi_{ai}\psi_{b\hat{i}}.
\end{aligned}
\end{equation}

As a side comment, we also present the full set of BPS operators in the KW theory 
and the ABJM theory with the same quantum numbers as $O_{f,i\hat{i}}$ 
(or $F_{i\hat{i}}$):
\begin{equation}
\begin{aligned}
    E_{\rm ABJM}=5/2,\qquad J=1/2,\qquad q_3=2,\qquad (\mathbf{2},\mathbf{2}),\qquad \text{for ABJM},\\
    E_{\rm KW}=7/2,\qquad j_1=1/2,\qquad j_2=0, \qquad r=-2, \qquad (\mathbf{2},\mathbf{2}),\qquad \text{for KW}.
\end{aligned}
\end{equation}
Here, $(\mathbf2,\mathbf2)$ in the two theories mean the following. 
In KW, it is the representation of the flavor symmetry $SU(2)\times SU(2)$.  In ABJM, 
we decompose
\[SU(4)_R\simeq SO(6)_R\supset SO(4)\times SO(2)\simeq SU(2)\times SU(2)\times SO(2).\]
$(\mathbf2,\mathbf2)$ refers to representation of the $SO(4)$ factor, while $q_3$ denotes the charge under the remaining $SO(2)$, 
normalized so that $Q\equiv (Q_{34})_-$ has charge $+1$.  The spin notation also differs between the theories: $J$ is the three-dimensional Lorentz spin in ABJM, whereas $(j_1,j_2)$ are the two four-dimensional Lorentz spins in KW.

We shall first list the number of operators and their properties, and then 
specify their forms below. 
In both theories, there are 8 linearly independent gauge-invariant 
operators (4 single-traces and 4 double-traces). All these $8$ operators are 
constructed with the letters $a_i$, $\psi_{ai}$, $b_{\hat{i}}$, $\psi_{b\hat{i}}$ 
on which the $Q$-action is identical between the two theories.
Among these $8$ operators, there are $6$ $Q$-closed operators in both theories.
To check whether some of them are $Q$-exact, $O=Q\Lambda$, the two theories may 
use different letters in $\Lambda$ so the results differ between the two theories.
Let us denote by $Q_{\mathrm{KW}}$ and $Q_{\mathrm{ABJM}}$ the supercharges of the 
KW and ABJM theories respectively. 
$2$ of the $6$ $Q$-closed operators are $Q_{\rm KW}$-exact in the KW theory, leaving $4$ cohomologies, 
while $1$ of them is $Q_{\rm ABJM}$-exact in the ABJM theory and leaves 
$5$ cohomologies. In the KW theory, $3$ of the $4$ cohomologies are gravitons 
and one is $F_{i\hat{i}}$ which is a generalized monotone. In the ABJM theory, 
$4$ of the $5$ cohomologies are gravitons and one is $O_{f,i\hat{i}}$, 
which is fortuitous/monotonous in the ordinary/generalized sense 
respectively. 

We further present the operator contents of the $3$ and $4$ graviton cohomologies 
in both theories. The three common graviton operators in KW and ABJM are
\begin{equation}
\begin{aligned}
    \tr \psi_{a(i}a_{j)}\tr a^j b_{\hat{i}},\qquad
    \tr \psi_{b(\hat{i}}b_{\hat{j})}\tr a_i b^{\hat{j}},\qquad
    \tr (\psi_{aj}a^j-b^{\hat{j}}\psi_{b\hat{j}})\tr a_i b_{\hat{i}},
\end{aligned}\label{eq:grav2,2}
\end{equation}
which are all double-traces. The two $Q_{\rm KW}$-exact operators of the KW theory 
mentioned in the previous paragraph are given by the two linearly independent 
combinations of the following three operators, 
\begin{equation}\label{QKW-exact-charge-sector}
\begin{aligned}
    Q_{\mathrm{KW}}\tr a_i f_1 b_{\hat i},\qquad
    &Q_{\mathrm{KW}}\tr f_2 a_i b_{\hat i},\qquad
    Q_{\mathrm{KW}}\tr \psi_{ai}\psi_{b\hat{i}},\\
    Q_{\mathrm{KW}} \tr &\Big[(a_i f_1 +f_2 a_i)b_{\hat i}-\psi_{b\hat{i}}\psi_{ai}\Big]=0,
\end{aligned}
\end{equation}
where the last equation holds because the expression inside $Q_{\rm KW}$ 
is a $Q_{\rm KW}$-closed graviton in the KW theory. On the other hand, 
$Q_{\mathrm{ABJM}}\tr \psi_{ai}\psi_{b\hat{i}}$ is a $Q_{\mathrm{ABJM}}$-exact 
operator in this sector, analogous to $Q_{\rm KW}\tr\psi_{ai}\psi_{b\hat{i}}$ 
in (\ref{QKW-exact-charge-sector}), 
but another $Q_{\rm KW}$-exact operator in (\ref{QKW-exact-charge-sector}) 
has no analogue in the ABJM theory since it does not have the 
letters $f_1,f_2$. This explains why there 
is only one $Q_{\rm ABJM}$-exact operator in this charge sector.\footnote{One 
may think that there are $3$ $Q_{\mathrm{ABJM}}$-exact operators, i.e.  
$Q_{\mathrm{ABJM}}\tr \psi_{ai}\psi_{b\hat{i}}$, 
$Q_{\mathrm{ABJM}}\tr a_iDb_{\hat i}$, $Q_{\mathrm{ABJM}}\tr b_{\hat i}Da_i$, 
but $2$ linear combinations of them vanish since they are 
$Q_{\mathrm{ABJM}}$-variations of gravitons \cite{Belin:2025hsg}.} 
The extra graviton in the ABJM which does not have 
the KW analogue can be taken, for example, to be  
$Q_{\mathrm{KW}}\tr f_2 a_i b_{\hat{i}}$: 
\begin{align}\label{eq:QKW-exact ABJM-grav}
    Q_{\mathrm{KW}}\tr f_2 a_i b_{\hat{i}}
    =\tr (\psi_{b\hat{j}}b^{\hat j}-a^j\psi_{aj})a_ib_{\hat i}
    -\frac{1}{2}\tr (\psi_{b\hat{j}}b^{\hat j}-a^j\psi_{aj})\tr a_ib_{\hat i},
\end{align}
The right-hand side is a nontrivial graviton cohomology in ABJM, which is 
not $Q_{\rm ABJM}$-exact. 
(The denominator of the second term is $N$ for general rank $N$ theory.)
The second double-trace term is one of the gravitons in equation \eqref{eq:grav2,2}. 
The first single-trace term
\begin{align}
    \tr (\psi_{b\hat{j}}b^{\hat j}-a^j\psi_{aj})a_ib_{\hat i},\label{eq:ABJM grav2,2}
\end{align}
is an independent graviton in the ABJM theory, in addition to \eqref{eq:grav2,2}.
\eqref{eq:ABJM grav2,2} is also a graviton cohomology 
of KW theory, but it is not independent of \eqref{eq:grav2,2}.

At $N=2$, the single-trace terms in \eqref{eq:F-N2-expansion} contain four bifundamental letters; for example,
$\tr(a_i\psi_a^j a_jb_{\hat i})$, while the corresponding double-trace term is
$\tr(a_i\psi_a^j)\tr(a_jb_{\hat i})$.  The first $N=2$ trace relation that can relate single- and multi-trace terms involves six bifundamental letters. Hence no $N=2$ trace relation can relate the single- and double-trace terms in \eqref{eq:F-N2-expansion}, and it is sufficient to compare their single-trace parts. The three common graviton representatives in \eqref{eq:grav2,2} are double-traces. Comparing the single-trace part of $F_{i\hat i}$ with the single-trace parts of the $Q_{\rm KW}$-exact operators in \eqref{QKW-exact-charge-sector} (or equivalently the remaining ABJM graviton \eqref{eq:ABJM grav2,2} and $Q_{\rm ABJM}\tr \psi_{ai}\psi_{b\hat{i}}$) shows that $F_{i\hat i}$ is ordinarily fortuitous in both ABJM and KW. By contrast, replacing the plus sign between the two terms in \eqref{eq:general-ABJM-fortuitous} by a minus sign gives an operator cohomologous to a graviton in both theories. Its explicit rewriting at $N=2$ is given in Appendix~\ref{appendix:general-N-fortuity}. Thus both combinations are generalized monotones, while the sum is ordinarily fortuitous and the difference is an ordinary mesonic monotone.

The same distinction persists at general rank $N$.  Denote the two terms in
\eqref{eq:general-ABJM-fortuitous} by $m_1$ and $m_2$, so that
$F_N=m_1+m_2$, and define $F_N^{(-)}=m_1-m_2$.  Both combinations are
generalized monotones by construction.  In
Appendix~\ref{appendix:general-N-fortuity}, we show that $F_N$ is non-exact
and independent of ordinary mesonic monotones, whereas $F_N^{(-)}$ is a
nontrivial ordinary mesonic monotone.

\section{Fortuitous cohomologies at $N=2$}\label{Section:fortuitous}

In this section, we study the BMN sector of the $SU(2) \times SU(2)$ KW theory. We first compute the exact BMN index and analyze the contribution from generalized monotonous cohomologies. Comparing the full index with the monotone index, we identify the 
contributions from the strongly fortuitous cohomologies. We then construct explicit representatives of the strongly fortuitous cohomologies.

In the $SU(2) \times SU(2)$ KW theory, it is convenient to recast the bifundamental chiral superfields as complex vectors of $SO(4) \simeq SU(2) \times SU(2)$ as follows:
\begin{equation}
    A_{i\mu} \equiv \frac{1}{2} \tr (\bar\sigma_\mu A_i)\ , \quad B_{\hat{i}\mu} \equiv  \frac{1}{2} \tr (\sigma_\mu B_{\hat{i}})\ ,
\end{equation}
where $\sigma_\mu\equiv(i\vec{\sigma},\mathbf 1_2)$, $\bar\sigma_\mu\equiv(-i\vec\sigma,\mathbf 1_2)$ with $\mu=1, \cdots, 4$. In what follows, we use $\mu, \nu, \rho, \sigma, \cdots = 1,\cdots,4$ for these $SO(4)$ vector indices, rather than the Lorentz spinor indices.
Note that $\tr\left(\bar\sigma_\mu\sigma_\nu\right)=2\delta_{\mu\nu}$. 
The chiral field-strength superfields of the two $\mathcal{N}=1$ vector multiplets transform as $(\mathbf 3,\mathbf 1)$ and $(\mathbf 1,\mathbf 3)$ under the gauge group. Writing $W^{(1)}=W^{(1)a}\sigma^a/2$ and $W^{(2)}=W^{(2)a}\sigma^a/2$, and suppressing their common Lorentz spinor index, we combine them into a rank-two antisymmetric tensor of $SO(4)$ as follows:
\begin{equation}
    \mathcal W_{\mu\nu} = -\mathcal W_{\nu\mu} \equiv -\frac{1}{2} \, \tr \left(\sigma_{\mu\nu}W^{(1)} + \bar\sigma_{\mu\nu}W^{(2)}\right)\ ,
\end{equation}
where $\sigma_{\mu\nu}\equiv\frac{1}{4}\left(\sigma_\mu\bar\sigma_\nu-\sigma_\nu\bar\sigma_\mu\right)$ and $\bar\sigma_{\mu\nu}\equiv\frac{1}{4}\left(\bar\sigma_\mu\sigma_\nu-\bar\sigma_\nu\sigma_\mu\right)$. With these conventions, $W^{(1)}$ and $W^{(2)}$ correspond respectively to the self-dual and anti-self-dual components of $\mathcal W_{\mu\nu}$.
Here and throughout, we choose the orientation $\epsilon_{1234}=+1$.

At $N=2$, the IR KW SCFT has an enhanced flavor symmetry $SU(4) \supset SU(2) \times SU(2) \times U(1)_B$ \cite{Klebanov:1998hh,Forcella:2008bb,Maruyoshi:2009uk}. 
$A$ and $B$ can be organized into $\Phi_I \equiv (A_i, B_{\hat i})$, where $I=1,2,3,4$, which transforms in the anti-fundamental representation of $SU(4)$. Note that the enhanced $SU(4)$ mixes mesonic and baryonic operators, since the extra generators in $SU(4)$ are charged under $U(1)_B$. Note also that $\mathcal W$ is a singlet under $SU(4)$. The BPS scalars and fermions associated with $A$ and $B$ can be organized as $\phi^I \equiv (a^i, b^{\hat{i}})$ and $\psi_I \equiv (\psi_{ai}, \psi_{b\hat{i}})$, which transform in the fundamental and anti-fundamental representations of $SU(4)$, respectively. We also denote by $f$ the BPS field strength associated with $\mathcal{W}$, i.e. $f_{\mu\nu} \equiv -\frac{1}{2} \, \tr(\sigma_{\mu\nu}f_1 + \bar\sigma_{\mu\nu}f_2)$.

This provides an $SO(4)\times SU(4)$-covariant description of the BMN letters.

The $Q$-action\footnote{From this point onward, we normalize $Q$ to be 
$-\frac{1}{4}$ times the $Q$ used in the preceding sections.} 
on the BPS letters in the BMN sector is given by
\begin{equation}
    \begin{aligned}
    &Q\phi^I_\mu=0\ ,\\
    &Q\psi_{I\mu} = - \frac{1}{12} \epsilon_{IJKL} \epsilon_{\mu\nu\rho\lambda} \phi^J_\nu \phi^K_\rho \phi^L_\lambda\ ,\\
    &Qf_{\mu\nu} = \frac{1}{2} \phi^I_{[\mu|} \psi_{I|\nu]}\ .
    \end{aligned}
\end{equation}
The BMN cohomologies constructed with these letters admit the following useful structure. 
We define the following six fermionic generators $Q^{IJ} = -Q^{JI}$:
\begin{equation}
\begin{aligned}
    &Q^{IJ}\phi^K_\mu = \epsilon^{IJKL}\psi_{L\mu}\ ,\\
    &Q^{IJ}\psi_{K\mu}= -2 \epsilon_{\mu\nu\rho\sigma}\delta^{[I}_K \phi^{J]}_\nu f_{\rho\sigma}\ ,\\
    &Q^{IJ}f_{\mu\nu} =0\ .
\end{aligned}
\end{equation}
These generators satisfy
\begin{equation}
    \{Q,Q^{IJ}\} = 0\ , \quad \{Q^{IJ},Q^{KL}\}= 0\ ,
\end{equation}
on gauge-invariant operators. So $Q^{IJ}$ can act on a cohomology to 
generate its `descendants.'

In the $N=2$ theory, the generalized monotone generators constructed in Section \ref{Section: monotone} can be organized into irreducible representations of the enhanced $SU(4)$ symmetry. We focus on the BMN sector, in which $j_2=0$. As shown in Section \ref{sec:mesonic-monotones}, the mesonic monotone generators admit a complete classification for general $N$. The situation is more subtle for the baryonic monotone generators, for which the complete set is not known, as discussed in Section \ref{sec:baryonic-monotones}. Nevertheless, we shall see below that, at $N=2$, the baryonic monotones presented in Section \ref{sec:baryonic-monotones} are sufficient to construct the complete set of baryonic monotone generators, thanks to the simplifications in the mesonic sector and the $SU(4)$ enhancement.

At $j_1=0$, under $SU(2) \times SU(2) \times U(1)_B$, the mesons $\tr (a^ib^{\hat j})$ transform as $(\mathbf{2},\mathbf{2})_0$, while the baryons $\frac{1}{2}\epsilon_{\hat p_1 \hat p_2} \epsilon^{p_1 p_2} (a^i)^{\hat p_1}{}_{p_1} (a^j)^{\hat p_2}{}_{p_2}$ and $\frac{1}{2}\epsilon_{p_1 p_2} \epsilon^{\hat p_1 \hat p_2}  (b^{\hat i})^{p_1}{}_{\hat p_1} (b^{\hat j})^{p_2}{}_{\hat p_2}$ transform as $(\mathbf{3},\mathbf{1})_{-2}$ and $(\mathbf{1},\mathbf{3})_{+2}$, respectively. Here, the subscripts denote the $U(1)_B$ charge. These representations combine to 
yield the irreducible representation $\mathbf{10}$ of $SU(4)$,
\begin{equation}
    \mathbf{10} \longrightarrow (\mathbf{3},\mathbf{1})_{-2} \oplus  (\mathbf{2},\mathbf{2})_0 \oplus (\mathbf{1},\mathbf{3})_{+2}\ .
\end{equation}
This is a rank $2$ symmetric representation, which we call 
$u^{IJ}=u^{JI}$. In terms of the $SO(4)\times SU(4)$ covariant letters, 
$u^{IJ}$ are given by 
\begin{equation}
    u^{IJ} \equiv \phi^I_\mu \phi^J_\mu = \begin{pmatrix}
    \frac{1}{2}\epsilon_{\hat p_1 \hat p_2} \epsilon^{p_1 p_2} (a^i)^{\hat p_1}{}_{p_1} (a^j)^{\hat p_2}{}_{p_2} & \frac{1}{2} \tr (a^ib^{\hat j})\\
    \frac{1}{2} \tr (a^jb^{\hat i}) & \frac{1}{2}\epsilon_{p_1 p_2} \epsilon^{\hat p_1 \hat p_2}  (b^{\hat i})^{p_1}{}_{\hat p_1} (b^{\hat j})^{p_2}{}_{\hat p_2}
    \end{pmatrix}\ .
\end{equation}

At $j_1=\frac{1}{2}$, the mesons
$\frac{1}{2}\tr(a^i\psi_{aj}) -\frac{1}{4}\delta^i_j\tr(a^k\psi_{ak}) \in (\mathbf{3},\mathbf{1})_0$, 
$\frac{1}{2}\tr(b^{\hat i}\psi_{b\hat j}) -\frac{1}{4}\delta^{\hat i}_{\hat j} 
\tr(b^{\hat k}\psi_{b\hat k})\in (\mathbf{1},\mathbf{3})_0$,  
$\tr(a^i\psi_{ai}-\psi_{b\hat i}b^{\hat i}) \in (\mathbf{1},\mathbf{1})_0$
and the baryons 
$\frac{1}{2} \epsilon_{\hat p_1\hat p_2}\epsilon^{p_1p_2} (a^i)^{\hat p_1}{}_{p_1} (\psi_{b\hat j})^{\hat p_2}{}_{p_2} \in (\mathbf{2},\mathbf{2})_{-2}~$,
$\frac{1}{2} \epsilon_{p_1p_2}\epsilon^{\hat p_1\hat p_2} (b^{\hat i})^{p_1}{}_{\hat p_1} (\psi_{aj})^{p_2}{}_{\hat p_2} \in (\mathbf{2},\mathbf{2})_{+2}$
combine to yield the adjoint representation 
$\mathbf{15}$ of $SU(4)$,
\begin{equation}
    \mathbf{15}
    \longrightarrow
    (\mathbf{3},\mathbf{1})_0
    \oplus
    (\mathbf{1},\mathbf{3})_0
    \oplus
    (\mathbf{1},\mathbf{1})_0
    \oplus
    (\mathbf{2},\mathbf{2})_{-2}
    \oplus
    (\mathbf{2},\mathbf{2})_{+2}\ .
\end{equation}
They can be written as an $SU(4)$ traceless tensor ${v^I}_J$, given by 
\begin{equation}
\begin{aligned}
    {v^I}_J
    &\equiv
    \phi^I_\mu\psi_{J\mu}
    -\frac{1}{4}\delta^I_J\phi^K_\mu\psi_{K\mu}
    \\
    &=
    \begin{pmatrix}
    \frac{1}{2}\tr(a^i\psi_{aj})
    -\frac{1}{8}\delta^i_j
    \left[
        \tr(a^k\psi_{ak})
        +\tr(b^{\hat k}\psi_{b\hat k})
    \right]
    &
    \frac{1}{2}
    \epsilon_{\hat p_1\hat p_2}\epsilon^{p_1p_2}
    (a^i)^{\hat p_1}{}_{p_1}
    (\psi_{b\hat j})^{\hat p_2}{}_{p_2}
    \\
    \frac{1}{2}
    \epsilon_{p_1p_2}\epsilon^{\hat p_1\hat p_2}
    (b^{\hat i})^{p_1}{}_{\hat p_1}
    (\psi_{aj})^{p_2}{}_{\hat p_2}
    &
    \frac{1}{2}\tr(b^{\hat i}\psi_{b\hat j})
    -\frac{1}{8}\delta^{\hat i}_{\hat j}
    \left[
        \tr(a^k\psi_{ak})
        +\tr(b^{\hat k}\psi_{b\hat k})
    \right]
    \end{pmatrix}\ .
\end{aligned}
\end{equation}

At $j_1=1$, the mesons $\mathcal M^{i \hat j} \equiv \tr[ (a^if_1+f_2a^i)b^{\hat j} -\epsilon^{ik}\epsilon^{\hat j\hat l} \psi_{b\hat l}\psi_{ak} ] \in (\mathbf{2},\mathbf{2})_0$ and the baryons 
$\mathcal B_- \equiv \frac{1}{2} \epsilon_{\hat p_1\hat p_2}\epsilon^{p_1p_2} [ (a^if_1+f_2a^i)^{\hat p_1}{}_{p_1} (a_i)^{\hat p_2}{}_{p_2} - \epsilon^{\hat k\hat l} (\psi_{b\hat l})^{\hat p_1}{}_{p_1} (\psi_{b\hat k})^{\hat p_2}{}_{p_2} ] \in(\mathbf{1},\mathbf{1})_{-2}$, $\mathcal B_+ \equiv \frac{1}{2} \epsilon_{p_1p_2}\epsilon^{\hat p_1\hat p_2} [ (f_1b^{\hat i}+b^{\hat i}f_2)^{p_1}{}_{\hat p_1} (b_{\hat i})^{p_2}{}_{\hat p_2} + \epsilon^{kl} (\psi_{al})^{p_1}{}_{\hat p_1} (\psi_{ak})^{p_2}{}_{\hat p_2} ] \in(\mathbf{1},\mathbf{1})_{+2}$ combine to yield the rank $2$ 
antisymmetric representation $\mathbf{6}$ of $SU(4)$,
\begin{equation}
    \mathbf{6}
    \longrightarrow
    (\mathbf{1},\mathbf{1})_{-2}
    \oplus
    (\mathbf{2},\mathbf{2})_0
    \oplus
    (\mathbf{1},\mathbf{1})_{+2}\ .
\end{equation}
We write them as $w^{IJ}=-w^{JI}$, which are given by  
\begin{equation}
\begin{aligned}
    w^{IJ}
    &\equiv
    \epsilon_{\mu\nu\rho\sigma}
    f_{\mu\nu}\phi^I_\rho\phi^J_\sigma
    -\frac{1}{2}\epsilon^{IJKL}
    \psi_{K\mu}\psi_{L\mu}
    =
    \frac{1}{2}
    \begin{pmatrix}
    -\epsilon^{ij} \mathcal B_-
    &
    \mathcal M^{i\hat j}
    \\
    -\mathcal M^{j \hat i}
    &
    \epsilon^{\hat i\hat j} \mathcal B_+
    \end{pmatrix}\ .
\end{aligned}
\end{equation}

The generalized monotone generators $u^{KL}$, $v^K{}_L$, and $w^{KL}$ transform under
$Q^{IJ}$ as follows:
\begin{equation}\label{extra-Q}
\begin{aligned}
    &Q^{IJ}u^{KL}=-2\epsilon^{IJM(K}{v^{L)}}_M\ ,\\
    &Q^{IJ}{v^{K}}_{L}=2\delta^{[I}_L w^{J]K}+\frac{1}{2}\delta^K_Lw^{IJ}\ ,\\
    &Q^{IJ}w^{KL}=-\epsilon^{IJKL} Q(\epsilon_{\mu\nu\rho\sigma}f_{\mu\nu}f_{\rho\sigma})\ ,
\end{aligned}
\end{equation}
where the last line shows that $Q^{IJ}$ maps $w^{KL}$ to a $Q$-exact operator.

As these constructions show, in the $SO(4) \times SU(4)$-covariant notation, not only mesonic operators but also baryonic operators can be represented as single-trace operators
in the ordinary sense. This is possible because contractions with the $SU(2)$ $\epsilon$ tensors identify 
the fundamental and anti-fundamental representations, reflecting the pseudoreality of the \(SU(2)\) fundamental representation. Here, by a ``trace'' we mean contractions of the $SO(4)$ vector indices.

It is important to note that every 
\(SU(4)\) multiplet of $SO(4)$ gauge-singlets contains operators with zero baryon number. To see this, consider an operator containing \(n_\phi\) letters of \(\phi^I_\mu\) type, 
\(n_\psi\) letters of \(\psi_{I\mu}\) type, and an arbitrary number of \(f_{\mu\nu}\)'s. Under the central element \(-{\bf 1}_4\in SO(4)\), both \(\phi^I_\mu\) and \(\psi_{I\mu}\) are odd, whereas \(f_{\mu\nu}\) is even. Gauge invariance therefore requires
\begin{equation}
n_\phi+n_\psi=0\qquad (\mathrm{mod}\ 2)\ .
\end{equation}
Now consider the $SU(2)_X \subset SU(4)$ subgroup whose Cartan subgroup is $U(1)_B$. Under this subgroup, \(\phi^I_\mu \in \mathbf{4}\) and \(\psi_{I\mu}\in \overline{\mathbf{4}}\) decompose into the spin-\(\frac12\) representations, 
\begin{equation}
\mathbf{4} \longrightarrow \mathbf{2} \oplus \mathbf{2}\ ,\qquad
\overline{\mathbf{4}} \longrightarrow \mathbf{2} \oplus \mathbf{2}\ ,
\end{equation}
where the Cartan weights of each doublet are \(\frac{B}{2} = \pm \frac{1}{2}\). 
Since \(n_\phi+n_\psi\) is even, their tensor product decomposes only into integer-spin representations of $SU(2)_X$. Every integer-spin representation of $SU(2)_X$ contains a state with zero Cartan weight, whose $U(1)_B$ charge is also zero. Thus, there are no purely baryonic \(SU(4)\) multiplets, i.e., every baryonic operator belongs to an irreducible representation of the \(SU(4)\) symmetry that also contains mesons.

The single-trace mesonic operators classified in Section \ref{sec:mesonic-monotones}, including those containing more letters within the trace, can be reduced to polynomials in the mesonic operators contained in $u^{IJ}$, $v^I{}_J$, and $w^{IJ}$ using trace relations. The enhanced $SU(4)$ symmetry requires these trace relations to be promoted to $SU(4)$-covariant forms. The $SU(4)$-covariant completion of the trace relations for the mesonic operators then generates the corresponding trace relations for the baryonic operators in the same $SU(4)$ multiplets. Namely, every single-trace baryonic operator can likewise be reduced to a polynomial in the baryonic operators contained in $u^{IJ}$, $v^I{}_J$, and $w^{IJ}$. Therefore, we conclude that $u^{IJ}$, $v^I{}_J$, and $w^{IJ}$ exhaust the primitive single-trace cohomologies among the generalized monotones in the BMN sector thanks to the $SU(4)$ enhancement. In Section \ref{sec:monotone-index}, we 
shall count these multi-trace generalized monotonous cohomologies.

Finally, we recast the consistent truncation to $SU(2)$ $\mathcal{N}=4$ SYM discussed in Section \ref{subsec:KW-to-SYM} in the $SO(4)\times SU(4)$-covariant notation. We first turn on the background $\langle\phi^4_4\rangle=\hat\phi\neq0$, which is equivalent to setting $\langle\bar B^{\hat 2}\rangle = \hat \phi \, \mathbf{1}_2$. Around this background, we take the light letters to be
\begin{equation}
    L=\{\phi^m_i\ ,\quad \psi_{m i}\ ,\quad f_i \equiv \epsilon_{ijk} f_{jk}\}\ ,
\end{equation}
and the heavy letters to be
\begin{equation}
    H=\left\{\phi^4_i\ ,\; \phi^m_4\ , \varphi_H \equiv \phi^4_4-\hat \phi\ , \quad \psi_{4 i}\ ,\; \psi_{m 4}\ ,\; \chi_H \equiv \psi_{4 4} - \frac{1}{3\hat\phi}\phi^m_i\psi_{m i}\ ,\quad f_{i4}\right\}\ ,
\end{equation}
where $m,n,p=1,2,3$ and $i,j,k=1,2,3$ 
are respectively $SU(3)\subset SU(4)$ and $SO(3)\subset SO(4)$ indices. One can verify that $\left. Qh\right|_{H=0}= 0$ for every $h \in H$. The action of $Q$ on the light letters at $H=0$ is given by
\begin{equation}
\begin{aligned}
&\left.  Q\phi_i^m \right|_{H=0} =0\ ,\\
&\left.  Q\psi_{mi} \right|_{H=0} =-\frac{\hat{\phi}}{4} \epsilon_{mnp}\epsilon_{ijk} \phi_j^n\phi_k^p\ ,\\
&\left.  Qf_{i} \right|_{H=0} =\frac{1}{2} \epsilon_{ijk} \phi^m_{j}\psi_{mk}\ ,
\end{aligned}
\end{equation}
which is equivalent, up to normalization, to the classical $Q$-action of $SU(2)\simeq SO(3)$ $\mathcal{N}=4$ SYM on its BMN letters \cite{Grant:2008sk,Choi:2023znd}. Namely, we have checked that $TQ_\textrm{KW} = Q_{\mathcal{N}=4} T$, where $T$ denotes the above truncation obtained by setting all heavy letters to zero.

Under this truncation, $u^{IJ}$, $v^I{}_J$, and $w^{IJ}$ reduce to
\begin{equation}
    \begin{aligned}
    &u^{IJ} \longrightarrow \left\{ u^{mn} \equiv \phi^m_i\phi^n_i = \phi^m \cdot \phi^n\ , \quad u^{44}=\hat\phi^2 \right\}\ , \\
    &v^I{}_J \longrightarrow   v^m{}_n \equiv \phi^m_i\psi_{ni}-\frac{1}{3}\delta^m_n \phi^p_i\psi_{pi} = \phi^m \cdot \psi_{n}-\frac{1}{3}\delta^m_n \phi^p \cdot \psi_{p}\ , \\
    &w^{IJ} \longrightarrow w^{m4} = - w^{4m} \equiv w^m \equiv \hat{\phi}f_i\phi^m_i-\frac{1}{2}\epsilon^{mnp}\psi_{ni}\psi_{pi} = \hat{\phi}f \cdot \phi^m-\frac{1}{2}\epsilon^{mnp}\psi_{n}\cdot \psi_{p}\ ,
    \end{aligned}
\end{equation}
where all other components vanish. Here, $u^{mn}$, $v^m{}_n$, and $w^m$ are, up to normalization, graviton generators of $SU(2)$ $\mathcal N=4$ SYM \cite{Grant:2008sk,Choi:2023znd}.

The fermionic generators $Q^{IJ}$ decompose under the truncation as
\begin{equation}
    Q^{IJ} \to \left\{ Q^{m4}, \quad \widetilde Q_m \equiv \frac{1}{2} \epsilon_{mnp} Q^{np} \right\}\ ,
\end{equation}
where $m,n,p=1,2,3$. On the light letters $\ell \in L$, their projected actions are
\begin{equation}
    T Q^{m4} \ell = Q^{m}_+ T \ell \ , \qquad T\widetilde Q_m \ell = 0\ ,
\end{equation}
where $Q^m_+$ are fermionic generators that anticommute with 
$Q_{\mathcal{N}=4}$.\footnote{In $\mathcal{N}=4$ SYM, $Q^m_+$ belong to 
the $\mathcal{N}=4$ superconformal algebra $PSU(2,2|4)$. In contrast, 
in the full KW theory, $Q^{IJ}$ are not exact symmetry generators.} 
For a general operator $O$ containing heavy letters, however, $TQ^{m4}O$ can receive an additional contribution from the $\chi_H$-dependence of $O$, and $T\widetilde Q_m O$ need not vanish.

\subsection{BMN index}
In this subsection, we analyze the $N=2$ BMN index, which counts $Q$-cohomologies in the BMN sector as defined in \eqref{eq:BMN-index-definition}. We introduce $SU(4)$ fugacities $\xi_I$, subject to
\begin{equation}
    \prod_{I=1}^{4}\xi_I=1\ .
\end{equation}
In terms of the $SU(2)\times SU(2)\times U(1)_B$ fugacities introduced in \eqref{eq:index-definition} and used in App. \ref{appendix:finite-N-index}, the $SU(4)$ fugacities can be parametrized as
\begin{equation}
    (\xi_1,\xi_2,\xi_3,\xi_4) = \left( \frac{u}{b},\frac{1}{ub},\frac{b}{v},bv \right)\ .
\end{equation}
The single-letter contributions from the BMN letters to the index are
\begin{equation}
    \phi^I:\ t^{3/2}\xi_I\ ,\qquad
    \psi_{I}:\ -t^{9/2}\xi_I^{-1}\ ,\qquad
    f:\ t^6\ ,
\end{equation}
where $t$ is the fugacity for $3(2j_1-r)$, as introduced in \eqref{eq:index-definition}. There is no dependence on the fugacity $y$, since all BMN letters have $j_2=0$.

As explained earlier, every gauge-singlet operator must contain an even total number of $\phi^I_\mu$ and $\psi_{I\mu}$.
It follows that 
the contribution from every gauge singlet operator is an integral power of $t^3$. We thus introduce $x=t^3$ and henceforth regard the index as a function of $x$. Then, from \eqref{eq:BMNindex}, the $N=2$ BMN index is
\begin{align}
    \mathcal I^{\mathrm{BMN}}_{N=2}(x;\boldsymbol{\xi})
    =& \frac{1}{4} \oint\frac{dz_1}{2\pi iz_1} \oint\frac{dz_2}{2\pi iz_2}
    \prod_{k=1}^{2} \frac{(1-z_k^2)(1-z_k^{-2})}{(1-x^2)(1-x^2 z_k^2)(1-x^2 z_k^{-2})} \nonumber\\
    &\times \prod_{I=1}^{4} \prod_{p_1,p_2=\pm 1} 
    \frac{1-x^{3/2}\xi_I^{-1} z_1^{p_1}z_2^{p_2}}{1-x^{1/2}\xi_I z_1^{p_1}z_2^{p_2}}\ .
\end{align}
The contour integrals can be evaluated using the residue theorem.

For convenience, we define
\begin{equation}
    \mathcal D_{[a,b,c]}(x) \equiv \operatorname{PE}\left[-\chi_{[a,b,c]}x\right]\ ,
\end{equation}
where the plethystic exponential is defined by $\operatorname{PE}[f(x,\boldsymbol{\xi})] \equiv \exp \left[\sum_{n=1}^\infty \frac{1}{n}f(x^n,\boldsymbol{\xi}^n)\right]$. Here, $\chi_{[a,b,c]}\equiv \chi_{[a,b,c]}(\boldsymbol{\xi})$ denotes the $SU(4)$ character with Dynkin labels $[a,b,c]$.
For example,
\begin{equation}\label{D-examples}
    \begin{aligned}
    &\mathcal D_{[2,0,0]}(x) = \operatorname{PE}\left[- \chi_{[2,0,0]} x \right] = \prod_{1\leq I \leq J\leq4} \left(1-\xi_I\xi_J x\right) \ , \\
    &\mathcal D_{[0,1,0]}(x) = \operatorname{PE}\left[- \chi_{[0,1,0]} x \right] = \prod_{1\leq I<J\leq4} \left(1-\xi_I\xi_J x\right) \ .
    \end{aligned}
\end{equation}

The index can then be written in the factorized form
\begin{align}\label{eq:BMN-index-N2}
    \mathcal I^{\mathrm{BMN}}_{N=2} (x;\boldsymbol{\xi})
    &=\frac{\mathcal P_{59}(x;\boldsymbol{\xi})\mathcal D_{[0,1,0]}(x^5)}{\mathcal D_{[2,0,0]}(x)\mathcal D_{[0,1,0]}(x^3)(1-x^4)^2 \mathcal D_{[2,0,0]}(x^5)} \nonumber\\
    &=\mathcal P_{59}(x;\boldsymbol{\xi})\,
    \operatorname{PE}\left[
        \chi_{[2,0,0]} x
        +\chi_{[0,1,0]} x^3
        +2x^4
        +\chi_{[2,0,0]} x^5
        -\chi_{[0,1,0]} x^5
    \right]\ ,
\end{align}
where $\mathcal D_{[2,0,0]}(x^5)$ in the denominator is partially canceled by $\mathcal D_{[0,1,0]}(x^5)$ in the numerator (as is clear from (\ref{D-examples})), while all other denominator factors remain uncanceled. Here, $\mathcal P_{59}(x;\boldsymbol{\xi})$ is a degree-$59$ polynomial in $x$ with $SU(4)$-character coefficients,
\begin{equation}
    \mathcal P_{59}(x;\boldsymbol{\xi}) = \sum_{n=0}^{59} p_n(\boldsymbol{\xi})x^n\ ,
    \qquad
    p_n(\boldsymbol{\xi}) = \sum_{a,b,c\geq 0} m^{(n)}_{a,b,c}\, \chi_{[a,b,c]}(\boldsymbol{\xi})\ .
\end{equation}
Its unrefined form is
\begingroup
\allowdisplaybreaks
\begin{align}
\mathcal P_{59}(x;\boldsymbol{1})
={}&1-15x^2-4x^3+148x^4-156x^5-410x^6
+1076x^7-924x^8-546x^9+4214x^{10}
\nonumber\\
&-9022x^{11}+10398x^{12}-3996x^{13}-13604x^{14}
+39304x^{15}-57925x^{16}+56276x^{17}
\nonumber\\
&-28254x^{18}-27704x^{19}+98702x^{20}-170450x^{21}
+230913x^{22}-254634x^{23}
\nonumber\\
&+222657x^{24}-123552x^{25}-46700x^{26}+252228x^{27}
-443766x^{28}+577850x^{29}
\nonumber\\
&-606519x^{30}+519594x^{31}-340376x^{32}+100808x^{33}
+140804x^{34}-332876x^{35}
\nonumber\\
&+450324x^{36}-482164x^{37}+448370x^{38}-376284x^{39}
+280519x^{40}-183048x^{41}
\nonumber\\
&+101498x^{42}-44452x^{43}+19196x^{44}-17874x^{45}
+24177x^{46}-28294x^{47}
\nonumber\\
&+25328x^{48}-16760x^{49}+7879x^{50}-1428x^{51}
-1514x^{52}+1494x^{53}-751x^{54}
\nonumber\\
&+358x^{55}-112x^{56}-26x^{57}+30x^{58}-6x^{59}\ .
\label{eq:P59-unrefined}
\end{align}
\endgroup
The complete set of integer coefficients $m^{(n)}_{a,b,c}$ is provided in App.~\ref{appendix:BMN-index-coefficients}.

\subsection{Monotone index}\label{sec:monotone-index}
We now consider the contribution from the generalized monotonous cohomologies to the BMN index. For simplicity, we shall refer to these generalized monotones simply as monotones, including both mesonic and baryonic cohomologies, throughout this section. As discussed earlier, the monotones are generated by the following single-trace operators:
\begin{equation}\label{monotone generators}
    u^{(IJ)} \equiv \phi^{I}_\mu \phi^J_\mu\ , \quad {v^I}_J \equiv \phi^I_\mu \psi_{J\mu} - \frac{1}{4} \delta^I_J \phi^K_\mu \psi_{K\mu}\ , \quad w^{[IJ]} \equiv \epsilon_{\mu\nu\rho\sigma}f_{\mu\nu}\phi^I_\rho \phi^J_\sigma - \frac{1}{2} \epsilon^{IJKL} \psi_{K\mu} \psi_{L\mu}\ .
\end{equation}
Products of \eqref{monotone generators} form an overcomplete set of monotones. The trace relations make certain linear combinations of multi-trace operators $Q$-exact, and such combinations should therefore not be counted as independent cohomologies.

Before analyzing the trace relations in detail, let us first examine the general structure of the index counting monotones in the BMN sector. The single-trace monotone generators contribute
\begin{equation}
    u^{IJ}:\  \chi_{[2,0,0]}x\ ,\qquad
    {v^{I}}_J: - \chi_{[1,0,1]} x^2\ ,\qquad
    w^{IJ}: \chi_{[0,1,0]} x^3\ .
\end{equation}
If there were no relations among these generators, so that the monotonous cohomology ring were freely generated, its index would be
\begin{equation}
    \operatorname{PE}\left[-\chi_{[1,0,1]} x^2 + \chi_{[2,0,0]}x + \chi_{[0,1,0]} x^3 \right] = \frac{\operatorname{PE}\left[-\chi_{[1,0,1]} x^2\right]}{\mathcal D_{[2,0,0]}(x) \mathcal D_{[0,1,0]}(x^3)}\ .
\end{equation}
However, trace relations among multi-trace monotone operators, together with their higher syzygies (relations among the trace relations, relations among those relations, and so forth), modify this freely generated index. Their contributions affect only the numerator. 
The $N=2$ BMN monotone index therefore takes the general form
\begin{equation}\label{eq:formal-BMN-mono-index-N2}
    \mathcal I^{\mathrm{BMN,mono}}_{N=2} (x;\boldsymbol{\xi}) = \frac{\mathcal P^{\mathrm{mono}}_{N=2}(x;\boldsymbol{\xi})}{\mathcal D_{[2,0,0]}(x) \mathcal D_{[0,1,0]}(x^3)}\ ,
\end{equation}
where $\mathcal P^{\mathrm{mono}}_{N=2}(x;\boldsymbol{\xi})$ is a finite polynomial in $x$, with $\mathcal P^{\mathrm{mono}}_{N=2}(0;\boldsymbol{\xi})=1$, whose coefficients are integral linear combinations of $SU(4)$ characters. It incorporates the contributions of the fermionic generators, the trace relations, and their higher syzygies. More formally, this rational form follows from the Hilbert--Serre theorem applied to the monotonous cohomology algebra. Some factors in the numerator may cancel against factors in the denominator, but no additional denominator factors are required.

Having established this general structure, we now turn to a detailed analysis of the trace relations. We shall follow the strategy used in \cite{Choi:2023znd,Choi:2023vdm}. We first restrict the BPS letters to the locus where they are simultaneously diagonal in the $SU(2) \times SU(2)$ notation. (We will return to a subtle issue associated with this restriction shortly.) Note that the space of $2\times 2$ diagonal matrices is spanned by $\sigma_3$ and $\sigma_4 = \mathbf 1_2$, and that $\sigma_{14}, \sigma_{23} \propto \sigma_1$, $\sigma_{13}, \sigma_{24} \propto \sigma_2$, while $\sigma_{12}, \sigma_{34} \propto \sigma_3$. Therefore, in the $SO(4)$ notation, this restriction translates to
\begin{equation}
    \phi^I_1 = \phi^I_2= 0 \ , \quad \psi_{I1} = \psi_{I2}=0\ , \quad f_{13} = f_{14} = f_{23} = f_{24} = 0\ .
\end{equation}
On this locus, the monotone operators reduce to eigenvalue polynomials with no dependence on the off-diagonal components. Moreover, all nontrivial $Q$-actions vanish identically except
\begin{equation}\label{Qf34}
    Qf_{34}= \frac{1}{4} \left(\phi^I_{3} \psi_{I4} - \phi^I_{4} \psi_{I3} \right)\ .
\end{equation}
Therefore, the trace relations reduce to polynomial relations among the eigenvalues, modulo $Q$-exact terms generated by $Qf_{34}$.

Next, based on the above analysis, we compute the index coefficients counting the linearly independent monotones on the diagonal locus through order $x^{16}$. At each order $x^d$ and in each sector of fixed $SU(4)$ weight and spin, we enumerate products of the monotone generators in \eqref{monotone generators}. We expand these products in a monomial basis of the diagonal BMN letters modulo the $Q$-exact terms generated by \eqref{Qf34}. The resulting coefficient vectors form the columns of a matrix, whose rank gives the number of linearly independent operators in that sector. Finally, we sum the ranks over the spin sectors with the appropriate fermion-parity signs to obtain the index coefficients. Exact computations using this procedure become technically challenging at higher orders, and we have carried them out only through order $x^{13}$.

For orders $x^{14}$ through $x^{16}$, we instead use finite-field linear algebra. We reduce entries of the same coefficient matrix modulo each of the two distinct large primes $p_1=1000003$ and $p_2=1000033$, and compute the corresponding ranks over $\mathbb F_{p_1}$ and $\mathbb F_{p_2}$. The two ranks agree in every sector. We then sum the common ranks over the spin sectors with the appropriate fermion-parity signs to obtain the index coefficients. The agreement between the two primes provides a strong consistency check, although it does not by itself establish the corresponding ranks over $\mathbb Q$. Indeed, columns that are linearly independent over \(\mathbb Q\) can become dependent modulo a bad prime, and this could in principle occur for both primes. For generic random matrices, one heuristically expects the probability of such a simultaneous rank drop to scale as \(O((p_1p_2)^{-1})\sim 10^{-12}\), though this does not provide a rigorous error bound for our highly structured matrices. For general background on probabilistic modular rank computation and rank certification, see, e.g., \cite{Saunders:2004mrc}. Similar modular rank computations have recently been employed to check the $Q$-exactness of certain operators in \cite{Kim:2026rnx}. In this paper, we use modular rank computations only to determine the monotone index coefficients. All other computations, including checks of the $Q$-exactness of various operators, are exact.

The resulting unrefined monotone index computed on the diagonal locus is
\begin{equation}
\label{diag-grav}
\begin{aligned}
&\mathcal I^{\mathrm{BMN,diag\text{-}mono}}_{N=2}(x;\boldsymbol{1}) \\
 &\;\;\qquad=1+10x+40x^2+72x^3+60x^4+70x^5+107x^6+8x^7 +96x^8+174x^9 \\
&\;\;\qquad\quad-142x^{10}+192x^{11} +296x^{12}-460x^{13}+419x^{14}+486x^{15}-1035x^{16}+O(x^{17})\ .
\end{aligned}
\end{equation}
We also obtain the refined index explicitly, although the expression is too lengthy to present here. The refined result can be reconstructed from \eqref{eq:master-complete-inventory}, \eqref{fortuitous-index-factorized}, and App. \ref{appendix:BMN-index-coefficients}. 

In fact, \eqref{diag-grav} does not provide the full BMN monotone index. This is because there exist monotone operators that vanish on the diagonal locus but are nevertheless not $Q$-exact. Such operators are called non-Coulomb operators \cite{Chang:2025mqp,Choi:2025bhi}. We have explicitly checked that no non-Coulombic 
monotones appear through order $x^{9}$. At order $x^{10}$, we find a non-Coulomb operator
\begin{equation}
    O_\mathrm{nC} = \textrm{tr} \, v^5 = {v^{I}}_{J} {v^{J}}_{K} {v^{K}}_{L} {v^{L}}_{M} {v^{M}}_{I}\ ,
\end{equation}
which vanishes on the diagonal locus but is not $Q$-exact, as we show shortly. So we must include the contributions from such non-Coulomb operators to the BMN monotone index. Note that the truncation of $O_\mathrm{nC}$ to the $\mathcal N =4$ cohomology is $Q_{\mathcal N=4}$-exact: $TO_\mathrm{nC} = Q_{\mathcal N=4} \widetilde O$ for some $\widetilde O$.

Despite this limitation, there is a simple heuristic argument based on a semiclassical picture that motivates the restriction to the diagonal locus \cite{Choi:2023znd,Choi:2023vdm}.
If one goes `deep inside' the moduli space of supersymmetric vacua, the scalar matrices can generically be taken to be mutually commuting and hence simultaneously diagonal, with their eigenvalues providing coordinates on the moduli space modulo Weyl permutations. Gauge-invariant operators restricted to this locus therefore become symmetric polynomials in the eigenvalues. Quantizing these moduli, together with other massless fields on the 
moduli space, naturally yields particle-like BPS excitations.
In the D-brane picture, the eigenvalues describe the positions of the constituent branes, whereas the off-diagonal matrix components describe degrees of freedom transverse to the moduli space, such as strings stretched between distinct branes. Since the monotonous cohomologies are expected to describe multi-particle states built from single-particle states in the AdS gravity dual, it is natural to expect that much of their spectrum and their finite-$N$ relations can be captured by quantizing these diagonal, or eigenvalue, degrees of freedom. This argument, however, does not guarantee that all monotonous cohomologies are visible on the diagonal locus. For instance, explicit non-Coulombic
monotones are found in the $\mathcal N=4$ SYM with $SO(N)$ gauge groups \cite{Chang:2025mqp,Choi:2025bhi} and, as we have just seen, in the KW 
theory.\footnote{We were also informed that similar non-Coulombic monotones exist in
$\mathcal{N}=4$ SYM with $SU(N)$ gauge groups \cite{Kim-Lee-Song}. 
We thank the authors of \cite{Kim-Lee-Song} for informing us of their results prior 
to publication.}

To compute the full BMN monotone index, we should therefore count independent monotone operators constructed from the full matrix-valued fields, thereby incorporating contributions from non-Coulomb operators. This is a technically challenging task, and we have carried out the complete counting through order $x^{9}$ using modular rank computations. As already mentioned, the diagonal-locus prescription correctly reproduces the full monotone index through this order. For orders $x^{10}$ through $x^{16}$, instead of performing the complete counting, we have examined the non-Coulomb operators that can be constructed from $O_{\mathrm{nC}}$. All computations involving the non-Coulomb operators presented below are exact, rather than modular rank computations.

First, we find that $u^{IJ} O_{\mathrm{nC}}$ and ${v^I}_J O_{\mathrm{nC}}$ are $Q$-exact:
\begin{equation}
    \begin{aligned}
        u^{IJ} O_{\mathrm{nC}} = {}& Q \, \frac{1}{3}\left[ -8 r^{(I|K|}(v^4)^{J)}{}_K
+20 r^{KL}v^{(I}{}_K(v^3)^{J)}{}_L
-10 r^{K(I}v^{J)}{}_K\operatorname{tr}(v^3) \right. \\
&\left. \qquad +6 \epsilon^{(J|NPQ|}u^{I)K}r_{KLMN}v^L{}_Pv^M{}_Q
+18 \epsilon^{MNPQ}r^{(I}{}_{KLM}v^{J)}{}_Nv^K{}_Pv ^L{}_Q
\right]\ , \\
    v^I{}_J O_{\mathrm{nC}} ={}& Q \left[ 2\epsilon^{IKLM}r_{KPQJ}v^P{}_Lv^Q{}_Rv^R{}_M +2\epsilon^{IKLM}r_{KPQR}v^P{}_Lv^Q{}_Mv^R{}_J +r(v^3)^I{}_J \right]\ ,
    \end{aligned}
\end{equation}
where the various $r$'s denote $Q$-preimages of the monotone trace relations, defined by
\begin{equation}\label{r-preimage}
    \begin{aligned}
        &r^{IJ} =2 f_{\mu\nu}\phi_\mu^I\phi_\nu^J\ , \qquad Qr^{IJ}=u^{K[I}v^{J]}{}_K\ ,  \\
        &\widetilde r^I{}_{JKL} =-\frac23\epsilon_{\mu\nu\rho\sigma} \phi_\mu^I\psi_{J\nu}\psi_{K\rho}\psi_{L\sigma}\ , \qquad r^I{}_{JKL} = \widetilde r^I{}_{JKL} -\frac12\delta^I_{(J} \widetilde r^M{}_{|M|KL)}\ , \\
        &\widetilde R^I{}_{JKL} = \epsilon_{MNP(J}u^{IM}v^N{}_{K}v^P{}_{L)}\ ,\qquad Qr^I{}_{JKL} = \widetilde R^I{}_{JKL} -\frac12\delta^I_{(J} \widetilde R^M{}_{|M|KL)}  \ ,   \\
        &r_{IJKL} \equiv \frac12\,\epsilon_{\mu\nu\rho\sigma}\, \psi_{I\mu}\psi_{J\nu}\psi_{K\rho}\psi_{L\sigma}\ , \qquad  Q r_{IJKL} =\epsilon_{MNP(I}v^M{}_Jv^N{}_Kv^P{}_{L)}\ , \\
        &r\equiv f_{\mu\nu}\left[ \phi_\mu^I\phi_\nu^J\psi_{I\rho}\psi_{J\rho} +\phi_\rho^I\phi_\rho^J\psi_{I\mu}\psi_{J\nu} -4\phi_\rho^I\phi_\mu^J\psi_{I\nu}\psi_{J\rho} \right]\ , \qquad  Q r = \textrm{tr} \, v^3\ .
    \end{aligned}
\end{equation}

Next, we consider the following $Q$-closed operator:
\begin{equation}
    \Theta^{IJ} \equiv \Pi_{[0,1,0]}^{IJ}
\left[
(Q^{KL})^{\wedge5}
w^{MN} w^{RS} O_{\mathrm{nC}}
\right]
\simeq_{Q}
60\,w^{IJ}\,s^3\ ,
\end{equation}
where $\simeq_{Q}$ denotes equality up to a $Q$-exact term, and \(\Pi_{[0,1,0]}^{IJ}\) denotes the projection onto the \([0,1,0]\) component, normalized such that the \([0,1,0]\) projection of the symmetric product of three \(w\)'s is \(w^{IJ}\,s\). 
Here, we define
\begin{equation}
    \begin{aligned}
        &(Q^{IJ})^{\wedge p}
\equiv
\frac1{p!}\sum_{\sigma\in S_p}(-1)^\sigma
Q^{I_{\sigma(1)}J_{\sigma(1)}}\cdots
Q^{I_{\sigma(p)}J_{\sigma(p)}}\ , \\
        &s \equiv \frac{1}{8}\epsilon_{IJKL} w^{IJ} w^{KL}\ ,
    \end{aligned}
\end{equation}
where $S_p$ is the permutation group of $p$ elements.
A direct expansion yields
\begin{equation}
    \left.60\,w^{IJ}s^3\right|_{\phi=0}\neq0\ .
    \label{eq:pure-psi-endpoint}
\end{equation}
In fact, its zero-$\phi$ component contains four distinct
$\psi^{14}$ monomials.  Since every nonzero $Q$-variation of a
BMN letter contains at least one $\phi$, the zero-$\phi$
projection of every $Q$-exact operator vanishes. Therefore, $\Theta^{IJ}$ is not $Q$-exact.

Since $Q^{IJ}w^{KL}$ is $Q$-exact, as shown in the third line of (\ref{extra-Q}), 
one obtains
\begin{equation}
    (Q^{IJ})^{\wedge5}(w^{KL} w^{MN} O_{\mathrm{nC}})
    \simeq_Q
    w^{KL} w^{MN} (Q^{IJ})^{\wedge5} O_{\mathrm{nC}}\ .
\end{equation}
Therefore, the non-$Q$-exactness of $\Theta^{IJ}$ established
above implies that $(Q^{IJ})^{\wedge5} O_{\mathrm{nC}}$ is not $Q$-exact. Moreover, since the $Q^{IJ}$-action maps $Q$-exact operators
to $Q$-exact operators, it follows that $(Q^{IJ})^{\wedge p} O_{\mathrm{nC}}$ is not $Q$-exact for
$p=0,1,2,3,4,5$. On the other hand, $(Q^{IJ})^{\wedge6} O_{\mathrm{nC}}\simeq_Q 0$. 
To see this, note that $O_{\mathrm{nC}}$ contains five $v$'s, 
and that \(Q^{IJ}\) can act nontrivially on each \(v\) factor at most once (modulo \(Q\)-exact terms: see (\ref{extra-Q})). 
Thus, $(Q^{IJ})^{\wedge6} O_{\mathrm{nC}}$ is $Q$-exact.

$Q^{IJ}w^{KL} O_{\mathrm{nC}}$ transforms in the $[0,1,0]\otimes [0,1,0] =[0,0,0] \oplus [0,2,0] \oplus [1,0,1]$ representation of $SU(4)$. Among these, the singlet and all of its further $Q^{IJ}$-descendants are $Q$-exact. 
Indeed, $\epsilon_{IJKL}Q^{IJ}w^{KL} O_{\mathrm{nC}} \simeq_Q \epsilon_{IJKL}w^{KL} Q^{IJ} O_{\mathrm{nC}} = 0$ since $\epsilon_{IJKL}w^{KL}Q^{IJ}{v^M}_N = 0$ from (\ref{extra-Q}).

On the other hand, $w^{KL} O_{\mathrm{nC}}$, the $[0,2,0] \oplus [1,0,1]$ sector of $Q^{IJ}w^{KL} O_{\mathrm{nC}}$, and all subsequent $Q^{IJ}$-descendants that survive modulo the above $Q$-exact singlet tower are not $Q$-exact. Indeed, each surviving irreducible component has a nonzero image in the surviving part of the $(Q^{IJ})^{\wedge4}$-descendant of the $[0,2,0]\oplus[1,0,1]$ sector, transforming in the $[0,0,0] \oplus [0,2,0]$ representation. After multiplication by $w^{RS}$ and projection onto $[0,1,0]$, each of these two components yields a nonzero multiple of $\Theta^{IJ}$. Since $\Theta^{IJ}$ is not $Q$-exact, none of the operators in this tower can be $Q$-exact.

Finally, $w^{IJ}w^{KL}O_{\mathrm{nC}}$ decomposes into the $[0,0,0]\oplus [0,2,0]$ representation of $SU(4)$. Each of these components has a
$(Q^{IJ})^{\wedge5}$-descendant whose projection onto
$[0,1,0]$ is a nonzero multiple of $\Theta^{IJ}$. Therefore, neither component of $w^{IJ}w^{KL}O_{\mathrm{nC}}$ is $Q$-exact.

In summary, through order $x^{16}$, we have identified the following non-Coulombic monotones that are not $Q$-exact\footnote{$(Q^{IJ})^{\wedge 3} w O_\textrm{nC}$ contains two copies of \([1,0,1]\). Viewing $[0,1,0]$ as the $SO(6)$ vector, so that $(Q^{IJ})^{\wedge 3}$ is written as $Q_{abc}=Q_{[abc]}$, the two copies correspond to $Q_{abc}w_c O_\textrm{nC}$ and $Q_{[abc}w_{d]} O_\textrm{nC}$. The one listed here is the latter.}:
\begin{equation}
\begin{array}{c|c|c|c}
    Q^{IJ}\text{-primary}&Q^{IJ}\text{-descendant level}&x^{2j_1-r}&SU(4)\text{ representation}
    \\ \hline
    O_{\mathrm{nC}}&0&x^{10}&[0,0,0]
    \\
    O_{\mathrm{nC}}&1&x^{11}&[0,1,0]
    \\
    O_{\mathrm{nC}}&2&x^{12}&[1,0,1]
    \\
    O_{\mathrm{nC}}&3&x^{13}&[2,0,0]\oplus[0,0,2]
    \\
    O_{\mathrm{nC}}&4&x^{14}&[1,0,1]
    \\
    O_{\mathrm{nC}}&5&x^{15}&[0,1,0]
    \\ \hline
    wO_{\mathrm{nC}}&0&x^{13}&[0,1,0]
    \\
    wO_{\mathrm{nC}}&1&x^{14}&[0,2,0]\oplus[1,0,1]
    \\
    wO_{\mathrm{nC}}&2&x^{15}&[1,1,1]\oplus[2,0,0]\oplus[0,0,2]
    \\
    wO_{\mathrm{nC}}&3&x^{16}&[2,1,0]\oplus[0,1,2]\oplus[1,0,1]
    \\ \hline
    w^2O_{\mathrm{nC}}&0&x^{16}&[0,0,0]\oplus[0,2,0]
\end{array}
\label{eq:master-complete-inventory}
\end{equation}
All other operators generated from $O_{\mathrm{nC}}$ through
order $x^{16}$ are $Q$-exact.
Adding the contributions of the non-Coulomb operators listed above to the unrefined diagonal-locus index \eqref{diag-grav}, one obtains
\begin{equation}\label{eq:BMN-mono-index-N2}
\begin{aligned}
\mathcal I^{\mathrm{BMN,mono}}_{N=2}(x;\boldsymbol{1})
={}&1+10x+40x^2+72x^3+60x^4+70x^5+107x^6+8x^7 +96x^8+174x^9 \\
&-143x^{10}+198x^{11}+281x^{12}
-446x^{13}+439x^{14}+408x^{15}-951x^{16}+ O(x^{17})\ .
\end{aligned}
\end{equation}

We do not know whether there are any additional non-Coulomb operators at order $x^{10}$ or higher that are not generated from $O_{\mathrm{nC}}$. Nevertheless, we shall proceed with \eqref{eq:BMN-mono-index-N2}, and compare it with the full BMN index \eqref{eq:BMN-index-N2}. In the unrefined limit, their difference is
\begin{equation}
\begin{aligned}
&\mathcal I^{\mathrm{BMN}}_{N=2}(x;\boldsymbol{1}) - \mathcal I^{\mathrm{BMN,mono}}_{N=2}(x;\boldsymbol{1}) \\
={}&-35 x^{10}+126 x^{11}-151 x^{12}-122 x^{13}+651 x^{14}-886
   x^{15}+197 x^{16}
+O(x^{17})\ .
\label{eq:N2-BMN-unrefined}
\end{aligned}
\end{equation}
All the nonzero coefficients of (\ref{eq:N2-BMN-unrefined}) indicate the  
existence of strongly fortuitous cohomologies in the corresponding charge sectors, 
subject to possible contamination from yet-unsubtracted non-Coulombic monotone 
contributions that we might have missed. In the next subsection, we shall construct 
the strongly fortuitous cohomologies 
that partly account for (\ref{eq:N2-BMN-unrefined}). In particular, using 
the explicitly constructed operators, we shall argue that they cannot be cohomologous 
to any monotones, neither Coulombic nor non-Coulombic (if they exist), showing that 
they are indeed strongly fortuitous.

\subsection{Strongly fortuitous cohomologies}\label{subsec:strongly-fortuitous}
In this subsection we study new strongly fortuitous cohomologies which will partly account for \eqref{eq:N2-BMN-unrefined}. 
We will construct them explicitly using the ansatz of \cite{Choi:2023vdm}, which is based on relations of the monotone trace relations. Then, applying the method of \cite{deMelloKoch:2024pcs}, adapted to our context through the truncation to the $\mathcal{N}=4$ SYM cohomology discussed in Sec.~\ref{subsec:KW-to-SYM}, we will show that they are not $Q$-exact.

We first examine the general structure of the fortuitous index, defined here as the difference between the full BMN index and the monotone contribution identified in the previous subsection. 
From \eqref{eq:BMN-index-N2} and \eqref{eq:formal-BMN-mono-index-N2}, we find that the fortuitous index takes the form
\begin{equation}\label{eq:formal-BMN-fort-index-N2}
\begin{aligned}
&\mathcal I^{\mathrm{BMN,fort}}_{N=2}(x;\boldsymbol{\xi})
=\mathcal I^{\mathrm{BMN}}_{N=2}(x;\boldsymbol{\xi})
-\mathcal I^{\mathrm{BMN,mono}}_{N=2}(x;\boldsymbol{\xi})
= \frac{\mathcal P^{\mathrm{fort}}_{N=2} (x;\boldsymbol{\xi})}{\mathcal D_{[2,0,0]}(x)\mathcal D_{[0,1,0]}(x^3)(1-x^4)^2 \mathcal D_{[2,0,0]}(x^5)}\ ,
\end{aligned}
\end{equation}
where $\mathcal P^{\mathrm{fort}}_{N=2} (x;\boldsymbol{\xi}) \equiv \mathcal P_{59}(x;\boldsymbol{\xi})\mathcal D_{[0,1,0]}(x^5) - \mathcal P^{\mathrm{mono}}_{N=2} (x;\boldsymbol{\xi}) (1-x^4)^2 \mathcal D_{[2,0,0]}(x^5)$ is a finite polynomial in $x$ with $SU(4)$-character coefficients and satisfies $\mathcal P^{\mathrm{fort}}_{N=2} (0;\boldsymbol{\xi})=0$. The factors $\mathcal D_{[2,0,0]}(x)$ and $\mathcal D_{[0,1,0]}(x^3)$ may be partially or completely canceled by the numerator $\mathcal P^{\mathrm{fort}}_{N=2} (x;\boldsymbol{\xi})$. In contrast, $\mathcal D_{[2,0,0]}(x^5)$ is only partially canceled, whereas neither of the two $(1-x^4)$ factors is canceled by the numerator.

The denominator factors $\mathcal D_{[2,0,0]}(x)$ and $\mathcal D_{[0,1,0]}(x^3)$ in \eqref{eq:formal-BMN-fort-index-N2} can be accounted for by the towers of operators 
generated from any given `core' fortuitous cohomology $O_{\rm fort}$, by dressing it 
with $u^{IJ}$ and $w^{IJ}$:
\begin{equation}
    u^{m} w^{n} O_{\mathrm{fort}}\ , \qquad m,n=0,1,2,\ldots\ .
\end{equation}
$O_{\mathrm{fort}}$ will contribute to the numerator $\mathcal P^{\mathrm{fort}}_{N=2} (x;\boldsymbol{\xi})$. Possible cancellations with the numerator reflect the fact that trace relations may render some linear combinations of these dressed operators $Q$-exact, thereby reducing or even eliminating the corresponding independent towers. See \cite{Choi:2023vdm} for an analogous phenomenon.
The denominator factor $(1-x^4)^{2}$ in \eqref{eq:formal-BMN-fort-index-N2} suggests two towers analogous to those found in \cite{Choi:2023znd}. Schematically, the corresponding dressed operators would take the form
\begin{equation}\label{eq:f2-dressed-fortuitous}
    (f_{\mu\nu}f_{\mu\nu})^m (\epsilon_{\rho\sigma\lambda\tau}f_{\rho\sigma}f_{\lambda\tau})^n O_{\mathrm{fort}} + \cdots \ ,  \qquad m,n=0,1,2,\ldots\ ,
\end{equation}
where $\cdots$ denotes additional terms required to make the operator $Q$-closed. The remaining denominator factor $\mathcal D_{[2,0,0]}(x^5)$ in \eqref{eq:formal-BMN-fort-index-N2} suggests another tower of the form 
\begin{equation}
    z^{n} O_{\mathrm{fort}} + \cdots\ ,  \qquad n=0,1,2,\ldots\ , \quad
    \textrm{where}\quad 
    z^{(IJ)} \equiv \phi^I_\mu
    \left(f_{\mu\rho}f_{\rho\nu}
    -\frac{1}{4}\delta_{\mu\nu}f_{\rho\sigma}f_{\sigma\rho}\right)
    \phi^J_\nu\ .
\end{equation}
Here, $\cdots$ again denotes additional terms required to make the operators $Q$-closed. Since neither $(1-x^4)^2$ nor $\mathcal D_{[2,0,0]}(x^5)$ can be completely canceled by the numerator, the corresponding towers cannot be eliminated entirely. Although trace relations may render some linear combinations of the dressed operators $Q$-exact, infinitely many nontrivial classes must survive in the fortuitous cohomology. Note further that the above $Q$-closed completions are not guaranteed to exist for arbitrary $Q$-closed operators. Indeed, neither of the two \(f^2\)-dressings of \(O_{\mathrm{nC}}\) 
in (\ref{eq:f2-dressed-fortuitous}) admits a \(Q\)-closed completion.

In addition, there may exist a finite tower generated by $Q^{IJ}$,
\begin{equation}
    \left(Q^{IJ}\right)^{\wedge n} O_{\mathrm{fort}}\ ,
    \qquad n=0,1,\ldots,6\ .
\end{equation}
Provided that all of these $Q^{IJ}$-descendants remain nontrivial in the $Q$-cohomology, their contribution to the index would be
\begin{equation}
\begin{aligned}
&\sum_{n=0}^{6}
(-1)^n
\chi_{\wedge^n[0,1,0]}(\boldsymbol{\xi})\,
x^n \, \mathfrak i_{\mathrm{fort}}
=
\operatorname{PE}\!\left[-\chi_{[0,1,0]}x\right]
\mathfrak i_{\mathrm{fort}} \\
=&{}
\left[
1-\chi_{[0,1,0]}x
+\chi_{[1,0,1]}x^2
-\left(\chi_{[2,0,0]}+\chi_{[0,0,2]}\right)x^3
+\chi_{[1,0,1]}x^4
-\chi_{[0,1,0]}x^5+x^6
\right]\mathfrak i_{\mathrm{fort}}
\ ,
\end{aligned}
\end{equation}
where $\mathfrak i_{\mathrm{fort}}$ denotes the contribution of $O_{\mathrm{fort}}$ to the index. 
We shall assume that all the $Q^{IJ}$-descendants of fortuitous cohomologies remain nontrivial, as is indeed the case in $\mathcal N=4$ SYM \cite{Choi:2023vdm}. This need not hold for general $Q$-cohomologies, as we have seen that $(Q^{IJ})^{\wedge6} O_{\mathrm{nC}}$ is $Q$-exact. Later, we prove this for the threshold fortuitous cohomology and some of its dressed operators. For the other fortuitous cohomologies, we assume this property without proof.

In what follows, we denote the Cartan product by $\odot$. For irreducible
$SU(4)$ representations $V_\lambda$ and $V_\mu$, we define
\begin{equation}
    V_\lambda\odot V_\mu
    \equiv
    V_{\lambda+\mu}
    \subset
    V_\lambda\otimes V_\mu\ ,
\end{equation}
where $V_{\lambda+\mu}$ is the irreducible component generated
by the tensor product of the two highest-weight states.  Equivalently,
in terms of the $SU(4)$ Dynkin labels,
\begin{equation}
    [a,b,c]\odot[a',b',c']
    =
    [a+a',b+b',c+c']\ .
\end{equation}
For operators $A\in V_\lambda$ and $B\in V_\mu$, $A\odot B$
denotes the projection of $A\otimes B$ onto
$V_{\lambda+\mu}$.

The fortuitous index can then be written in the factorized form
\begin{equation}\label{fortuitous-index-factorized}
\begin{aligned}
\mathcal I^{\mathrm{BMN,fort}}_{N=2}(x;\boldsymbol{\xi})
={}&
\operatorname{PE}\!\left[-\chi_{[0,1,0]}x\right]
\left(
    \mathcal T_{10}\left(x;\boldsymbol{\xi}\right)
   +\mathcal T_{12}\left(x;\boldsymbol{\xi}\right)
   +\mathcal T_{14}\left(x;\boldsymbol{\xi}\right)
   +\mathcal T_{16}\left(x;\boldsymbol{\xi}\right)
\right)
+O\!\left(x^{17}\right)\ ,
\end{aligned}
\end{equation}
where
\begin{equation}
\begin{aligned}
\mathcal T_{10}\left(x;\boldsymbol{\xi}\right)
={}&
-\sum_{r=0}^{6}
 \chi_{[2r+4,0,0]}x^{10+r}
-\sum_{r=0}^{3}
 \chi_{[2r+4,1,0]}x^{13+r}
-\sum_{r=0}^{1}
 \chi_{[2r+6,0,0]}x^{15+r}
-\chi_{[4,2,0]}x^{16}\ ,
\\
\mathcal T_{12}\left(x;\boldsymbol{\xi}\right)
={}&
+\sum_{r=0}^{4}
 \chi_{[2r+4,0,0]}x^{12+r}
+\sum_{r=0}^{1}
 \chi_{[2r+4,1,0]}x^{15+r} + \chi_{[4,0,0]} x^{16}\ ,
\\
\mathcal T_{14}\left(x;\boldsymbol{\xi}\right)
={}&
-\sum_{r=0}^{2}
 \chi_{[2r,0,0]}x^{14+r},
\\[4pt]
\mathcal T_{16}\left(x;\boldsymbol{\xi}\right)
={}& \chi_{[0,0,0]} x^{16}\ .
\end{aligned}
\end{equation}
Note that we have split the vanishing combination $\left(+\chi_{[4,0,0]}-\chi_{[4,0,0]}\right) x^{16}$ into $+\chi_{[4,0,0]}x^{16}$ and $-\chi_{[4,0,0]}x^{16}$, which are assigned to $\mathcal T_{12}$ and $\mathcal T_{14}$, respectively. This splitting turns out to be natural from the perspective of the dressed fortuitous operators, as we discuss below.

Each of the four $\mathcal T_{d}$ terms can be interpreted as the contribution from a fortuitous core primary operator at order $x^d$, together with its dressed operators. Here, ``primary'' refers to a $Q^{IJ}$-primary, while ``core'' means that it cannot be obtained by dressing another operator. Let the four fortuitous core primaries be denoted by $O_{10}$, $O_{12}$, $O_{14}$, and $O_{16}$, which transform in the $[4,0,0]$, $[4,0,0]$, $[0,0,0]$, and $[0,0,0]$ representations of $SU(4)$, respectively. Note that $O_{10}$ and $O_{14}$ are fermionic, whereas $O_{12}$ and $O_{16}$ are bosonic. The four terms in $\mathcal T_{10}$ can be understood as the contributions from the following four types of bosonic dressings:
\begin{equation}\label{O10-dressings}
    u^{\odot r} \odot O_{10}\ , \qquad u^{\odot r} \odot w \odot O_{10}\ , \qquad u^{\odot r} \odot z \odot O_{10}+\cdots\ , \qquad w^{\odot 2} \odot O_{10}\ ,
\end{equation}
respectively. Similarly, the three terms in $\mathcal T_{12}$ can be understood as the contributions from three bosonic dressings:
\begin{equation}
    u^{\odot r} \odot O_{12}\ , \qquad u^{\odot r} \odot w \odot O_{12}\ , \qquad F O_{12}+\cdots\ ,
\end{equation}
respectively. Here, $F$ denotes either $F_s \equiv f_{\mu\nu}f_{\mu\nu}$ or $F_p \equiv \epsilon_{\mu\nu\rho\sigma}f_{\mu\nu}f_{\rho\sigma}$. $\mathcal T_{14}$ can be understood as the contribution from the dressed tower 
\begin{equation}
    u^{\odot r} \odot O_{14}\ .
\end{equation}
The last term $\mathcal T_{16}$ is simply the contribution from $O_{16}$.

In principle, \eqref{fortuitous-index-factorized} may receive contributions from 
unnoticed/unsubtracted non-Coulombic monotones. We have explicitly checked that no non-Coulombic monotones can match the charges and fermion parities of the fortuitous core primaries $O_{10}$, $O_{12}$, $O_{14}$, and $O_{16}$. 
Although non-Coulombic monotone contributions elsewhere in the dressed towers cannot be completely ruled out, our analysis strongly suggests that the fortuitous core primaries, together with their dressings and $Q^{IJ}$-descendants, fully account for \eqref{fortuitous-index-factorized}. (For the $O_{10}$ tower, more concrete arguments ruling out this caveat 
will be provided below.)

From the above observations, we expect that, for a given fortuitous core primary $O_{\mathrm{fort}}\in[a,b,c]$ at order $x^d$, there exists a family of non-$Q$-exact towers obtained by dressing $O_{\mathrm{fort}}$ with Cartan products of $u^{IJ}$, $w^{IJ}$, and $z^{IJ}$, as well as with $F_{s}$ and $F_p$, together with their $Q^{IJ}$-descendants:
\begin{equation}
    (Q^{IJ})^{\wedge p} O^{(r,s,t,u_s,u_p)}_{\mathrm{fort}} \equiv (Q^{IJ})^{\wedge p} \left(u^{\odot r} \odot w^{\odot s} \odot \left(z^{\odot t} \odot F_s^{u_s} F_p^{u_p} O_{\mathrm{fort}} + \cdots \right) \right)\ , 
\end{equation}
where
\begin{equation}
    r,s=0,1,2,\ldots,  \quad  (t,u_s,u_p) \in \mathfrak{D}(O_{\mathrm{fort}}) \subset \mathbb{Z}^3_{\geq 0}\ , \quad  p=0,1,\ldots 6\ .
\end{equation}
Here, $\mathfrak{D}(O_{\mathrm{fort}})$ denotes the set of allowed $z$- and $F_{s,p}$-dressing numbers for \(O_{\mathrm{fort}}\), and $\cdots$ denotes additional terms required to make the operators $Q$-closed when the dressing involves \(z^{IJ}\) or $F_{s,p}$. The dressed operator transforms in the $\wedge^p [0,1,0] \otimes [a+2(r+t),b+s,c]$ representation of $SU(4)$ at order $x^{d+r+3s+5t+4u_s+4u_p+p}$.
The contribution of these towers to the index is
\begin{equation}
\mathcal I_{O_{\mathrm{fort}}}(x;\boldsymbol{\xi})
={}
(-1)^{F\left(O_{\mathrm{fort}}\right)}
\operatorname{PE}\!\left[-\chi_{[0,1,0]}x\right]
\sum_{r,s=0}^{\infty} \sum_{t,u_s,u_p} 
\chi_{[a+2(r+t),b+s,c]}\,
x^{d+r+3s+5t+4u_s+4u_p}\ ,
\end{equation}
where $(t,u_s,u_p)$ are summed over $\mathfrak{D}(O_{\mathrm{fort}})$, and $F\left(O_{\mathrm{fort}}\right)$ denotes the fermion number of $O_{\mathrm{fort}}$. We further expect that other dressings of $O_{\mathrm{fort}}$ are $Q$-exact and therefore do not contribute to the index. The full fortuitous index can then be expressed as a sum of contributions from towers generated by fortuitous core primaries:
\begin{equation}
    \mathcal I^{\mathrm{BMN,fort}}_{N=2}(x;\boldsymbol{\xi}) = \sum_{O_{\mathrm{fort}}} \mathcal I_{O_{\mathrm{fort}}}(x;\boldsymbol{\xi})\ .
\end{equation}

We now provide an explicit realization of the above conjectured structure for the threshold fortuitous operator \(O_{10}\). $O_{10}$ is given by
\begin{equation}
\begin{aligned}
O_{10}^{IJKL}
={}&
4 r^{MN} u^{P(I}v^J{}_M v^K{}_N v^{L)}{}_P
-2r^{MN} u^{(IJ}v^K{}_P v^{L)}{}_M v^P{}_N \\
&+2r^{M(I} u^{J|P}v^Q{}_M v^{|K}{}_P v^{L)}{}_Q 
+2r^{M(I} u^{J|P}v^{|K}{}_M v^{L)}{}_Q v^Q{}_P \\
&+\frac92\epsilon^{MNP(I} r^J{}_{MQR} u^{KL)}v^Q{}_Nv^R{}_P
-9\epsilon^{MNP(I} r^J{}_{MQR} u^{K|Q}v^R{}_Nv^{|L)}{}_P\ ,
\end{aligned}
\end{equation}
where $r^{IJ}$ ,$r^I{}_{JKL}$, and their $Q$-images are defined in \eqref{r-preimage}.
The $Q$-closedness of $O_{10}$ follows from relations of the monotone trace relations $Qr^{IJ}$ and $Qr^I{}_{JKL}$. This is the KW analog of the ansatz for fortuitous cohomology based on relations of the graviton trace relations in $\mathcal{N}=4$ SYM, which was analyzed in detail in \cite{Choi:2023vdm}. 
We should also verify that $O_{10}$ is not $Q$-exact. To do so, we apply the method of \cite{deMelloKoch:2024pcs}, adapted to our context through the truncation to $\mathcal N =4$ cohomology discussed in Sec.~\ref{subsec:KW-to-SYM}. Since the truncation maps $Q_\textrm{KW}$-exact operators to $Q_{\mathcal{N}=4}$-exact operators, it suffices to show that the truncation of $O_{10}$ is not $Q_{\mathcal{N}=4}$-exact. 
We consider the truncation of its $I=J=K=L=4$ component:
\begin{equation}\label{eq:O10toO0}
    T O_{10}^{4444} = 3\hat{\phi}^3 \epsilon^{p_1p_2p_3}v^m\!_{p_1}v^n\!_{p_2}
    (\psi_m\cdot\psi_n\times\psi_{p_3}) = 3\hat{\phi}^3 O_0\ ,
\end{equation}
where $(A \times B)_i = \epsilon_{ijk} A_j B_k$. Here, $O_0$ is the threshold fortuitous cohomology of $SU(2)$ $\mathcal{N}=4$ SYM, and is not $Q_{\mathcal{N}=4}$-exact \cite{Choi:2023znd}. Therefore, $O_{10}$ is not $Q_\textrm{KW}$-exact.

We emphasize that $O_{10}$ represents not merely a nontrivial cohomology, but indeed a strongly fortuitous cohomology. This can be seen clearly from the fact that it is not $Q$-cohomologous to any generalized monotone built from $u^{IJ}$, $v^I{}_J$, and $w^{IJ}$ 
(neither Coulombic nor non-Coulombic). If it were, its truncation would also be cohomologous to a graviton cohomology built from $u^{mn}$, $v^m{}_n$, and $w^m$ of $SU(2)$ $\mathcal N=4$ SYM \cite{Grant:2008sk,Choi:2023znd}. However, it is known 
that $O_0$ is not cohomologous to any graviton cohomology \cite{Choi:2022caq,Choi:2023znd}.
Therefore, $O_{10}$ is a strongly fortuitous cohomology.

Next, based on \eqref{O10-dressings}, we consider various dressings of the threshold fortuitous core primary $O_{10}$, analogous to those of the corresponding fortuitous core primary $O_0$ in $SU(2)$ $\mathcal N=4$ SYM \cite{Choi:2023znd}. In $SU(2)$ $\mathcal N =4$ SYM, for each $t \geq 0$, there exists a fortuitous primary $O_t$ of the form $O_t = \hat\phi^{2t}(f \cdot f)^t O_0 + \cdots$, unique  up to normalization. See \cite{Choi:2023znd} for their explicit $Q$-closed completions. One might therefore expect that the $Q$-closed completions
\begin{equation}
    \widehat Z_t \equiv z^{\odot t} \odot O_{10} + \cdots \in [4+2t,0,0]
\end{equation}
also exist for every $t \geq 0$, with their truncations satisfying
\begin{equation}
    T\left(
        \widehat Z_t^{\underbrace{44\cdots4}_{4+2t}}
    \right)
    \simeq_{Q}
    \frac{3}{8^t}\hat\phi^3O_t\ ,
    \label{eq:Zt-truncation}
\end{equation}
which would imply that $\widehat Z_t$ is not $Q$-exact.\footnote{One might wonder whether $(F_s)^{u_s} (F_p)^{u_p} O_{10}+ \cdots$ also admit $Q$-closed completions. We 
checked that they in fact admit $Q$-exact (not just $Q$-closed) completions, 
so they do not represent nontrivial cohomologies.} We have explicitly constructed such $\widehat Z_t$ for $t=1,2$. For $t=1$, its explicit form is given in App.~\ref{appendix:explicit-Zhat1}. For $t=2$, we also obtained an explicit form, but it is too lengthy to present in the paper. Instead, in App.~\ref{appendix:explicit-Zhat1}, we briefly explain the recursive construction used to obtain it. For $t\geq3$, one may in principle continue the same recursive procedure, but the construction becomes increasingly complicated, and we leave it for future work. 
We then define
\begin{equation}
    O_{10}^{(r,s,t)}
    \equiv
    u^{\odot r}\odot w^{\odot s}\odot\widehat Z_t
    \in[4+2(r+t),s,0]\ ,
    \qquad
    r,s,t=0,1,2,\ldots \ ,
\end{equation}
which yields
\begin{align}
    T\left[
        (u^{44})^r(w^{34})^s
    \widehat Z_t^{44\cdots4}
    \right]
    &\simeq_{Q}
    \hat\phi^{2r}(w^3)^s
    \left(
        \frac{3}{8^t}\hat\phi^3O_t
    \right)
    =
    \frac{3}{8^t}
    \hat\phi^{\,2r+3}
    (w^3)^sO_t\ .
    \label{eq:Zt-dressed-truncation}
\end{align}
Since the dressed operator $(w^m)^s O_t$ is not $Q$-exact for any $s,t\geq 0$ \cite{Choi:2023znd}\footnote{Although an explicit proof of this general statement is not given there, it can be shown by an argument analogous to the one used below.}, it follows that $O_{10}^{(r,s,t)}$ is not $Q$-exact for any $r,s\geq 0$ and any $t$ for which $\widehat Z_t$ exists. Furthermore, by the same argument as before, these operators are strongly fortuitous, since their truncations yield fortuitous cohomologies of $SU(2)$ $\mathcal{N}=4$ SYM \cite{Choi:2023znd}.

Finally, we consider the $Q^{IJ}$-descendants of the operators discussed above. In $\mathcal{N}=4$ SYM, $Q^m_+$ is a physical supercharge whose action generates 
descendants within a short superconformal multiplet \cite{Cordova:2016emh,Choi:2023znd}. So the non-$Q$-exactness of a superconformal primary implies the non-$Q$-exactness of all of its nonzero $Q^m_+$-descendants. In the KW theory, by contrast, $Q^{IJ}$ is not a physical supercharge but rather a fermionic generator that anticommutes with $Q$. Thus, while the action of $Q^{IJ}$ preserves both $Q$-closedness and $Q$-exactness, it does not necessarily preserve non-$Q$-exactness. We must therefore explicitly show that the $Q^{IJ}$-descendants 
are also not $Q$-exact.

One might try to prove the non-$Q$-exactness by truncating the $Q^{IJ}$-descendants 
to $\mathcal{N}=4$ SYM. This certainly works for $Q^{m4}=Q^m_+$. 
However, $\widetilde Q_{m}$ does not correspond to any supercharge of $\mathcal{N}=4$ SYM. It turns out that, after truncation, the $\widetilde Q_{m}$-descendants become $Q$-cohomologous to linear combinations of the $\mathcal{N}=4$ primary and its descendants. Since demonstrating this explicitly is cumbersome, we instead adopt a different approach. We will directly show, after truncation, that the $Q^{IJ}$-descendants are not $Q$-exact.

One way to show that an operator $O$ is not $Q$-exact is to consider a $Q^\dagger$-closed operator $P$ and check whether it has a nonzero inner product with $O$. If the inner product is nonzero, $O$ cannot be $Q$-exact, since $\langle P | Q \widetilde O \rangle = \langle Q^\dagger P | \widetilde O \rangle = 0$. However, explicitly constructing such a $Q^\dagger$-closed operator $P$ and computing its inner product with $O$ can be challenging in practice. Instead, in a simple charge sector, we shall construct a dual cocycle $\lambda$, namely a linear functional that annihilates all $Q$-exact operators in that sector. Such a functional is easier to construct, and in some cases $\lambda$ can be realized as $\langle P|$ with a $Q^\dagger$-closed operator $P$.

We now apply this strategy to the $\Omega$-descendant of $O_{10}^{(r,s,t)}$, where 
\begin{equation}
    \Omega\equiv (Q^{IJ})^{\wedge6}=Q^{12}Q^{13}Q^{23}Q^{14} Q^{24}Q^{34}\ .
\end{equation}
Let $E_3$ be the $SU(4)$ lowering generator defined by
\begin{equation}
    E_3 \phi^4 = \phi^3 \equiv Z\ , \qquad E_3 \psi_3 = -\psi_4\ .
\end{equation}
Since $\Omega \in \wedge^6 [0,1,0] = [0,0,0]$, we have $[E_3, \Omega] = 0$.
We define the extremal component as
\begin{equation}
    \left(O_{10}^{(r,s,t)}\right)_{\mathrm{ext}}
    =
    (u^{44})^r(w^{34})^s
    \widehat Z_t^{44\ldots4}\ .
\end{equation}
It follows that
\begin{align}
    E_3\Omega
    \left(O_{10}^{(r,s,t)}\right)_{\mathrm{ext}}
    ={}&
    \Omega(w^{34})^s
    \left[
        2r\,u^{34}(u^{44})^{r-1}
        \widehat Z_t^{44\ldots4}+(4+2t)(u^{44})^r
        \widehat Z_t^{344\ldots4}
    \right]\ ,
    \label{eq:general-t-mixed-top}
\end{align}
where the first term is absent for $r=0$.
Set $S=s+2$ and let $\pi_S^{(t)}$ denote the projection onto the $f^{\,2t+S-1}Z^{\,S-1}\psi^9$ sector after truncation. The gauge-invariant monomials in this sector can be expressed as
\begin{equation}
    \mathcal V_j^{(S,t)}
    =
    (f \cdot f)^{t+j}
    (Z \cdot Z)^j
    \left(f \cdot Z\right)^{S-1-2j}
    \omega_9\ ,
    \qquad
    0\leq j\leq
    \left\lfloor\frac{S-1}{2}\right\rfloor\ ,
    \label{eq:general-t-target-basis}
\end{equation}
where $\omega_9 \equiv \prod_{m,i=1}^3 \psi_{mi}$.
The possible $Q$-preimages of this sector are spanned by
\begin{equation}
    \mathcal U_k^{(S,t)}
    =
    (f \cdot f)^{t+k+1}
    (Z \cdot Z)^k
    \left(f \cdot Z\right)^{S-3-2k}
    (f\times Z)\cdot\eta_3\ ,
    \qquad
    0\leq k\leq
    \left\lfloor\frac{S-3}{2}\right\rfloor\ ,
    \label{eq:general-t-preimage-basis}
\end{equation}
where $\eta_3$ is a vector of the $SO(3)$ gauge symmetry 
at $\psi^8$ order, defined by $\psi_{3i}\eta_{3,i}=\omega_9$ (no sum over $i$). 
For $S=2$, this preimage space is empty.  
A direct computation yields
\begin{equation}
    \pi_S^{(t)}
    \left(Q\mathcal U_k^{(S,t)}\right)
    =
    (k+t+2)\mathcal V_{k+1}^{(S,t)}
    -(k+t+1)\mathcal V_k^{(S,t)}\ .
    \label{eq:general-t-boundary-map}
\end{equation}
We define
\begin{equation}
    \lambda_S^{(t)}
    \equiv
    \Lambda_S^{(t)}\circ\pi_S^{(t)}\ ,
    \qquad
    \Lambda_S^{(t)}
    \left(\mathcal V_j^{(S,t)}\right)
    =
    \frac{1}{j+t+1}\ .
    \label{eq:general-t-functional}
\end{equation}
It then follows that
\begin{align}
    \lambda_S^{(t)}
    \left(Q\mathcal U_k^{(S,t)}\right)
    =0\ .
\end{align}
Thus, $\lambda_S^{(t)}$ annihilates every $Q$-exact operator after projection onto the $f^{\,2t+S-1}Z^{\,S-1}\psi^9$ sector.
We now define the numerical endpoint coefficient $\mathcal C_{r,s,t}$ by
\begin{equation}
    \lambda_{s+2}^{(t)}
    \left[
        T\left(
            E_3\Omega
            \left(O_{10}^{(r,s,t)}\right)_{\mathrm{ext}}
        \right)
    \right]
    =
    \mathcal C_{r,s,t}
    \hat\phi^{\,2r+s+2t}\ .
    \label{eq:general-t-endpoint-pairing}
\end{equation}
The coefficient $\mathcal C_{r,s,t}$ receives contributions from
both the leading term $z^{\odot t}\odot O_{10}$ and all completion
terms in $\widehat Z_t$.  
For $t=0,1,2$, these coefficients are given by
\begin{equation}\label{eq:general-t-endpoint-coefficients}
    \begin{aligned}
    &\mathcal C_{r,s,0} = 160(r+2)(r+3)(s+4)^2(s+5)\neq 0\ , \\
    &\mathcal C_{r,s,1} = 10(r+3)(r+4)(s+4)(s+6)(s+7)\neq 0\ , \\
    &\mathcal C_{r,s,2} = \frac{5}{6}(r+4)(r+5)(s+4)(s+8)(s+9)\neq 0\ .
    \end{aligned}
\end{equation}
Therefore, $\Omega O_{10}^{(r,s,t)}$ is not $Q$-exact for any $r,s \geq 0$ and $t=0,1,2$. Finally, by the perfect exterior pairing, for every nonzero $Q^{IJ}$-descendant at level $p$, there exists a complementary $(Q^{IJ})^{\wedge(6-p)}$-action that maps it to a nonzero multiple of $\Omega O^{(r,s,t)}_{10}$. Since the $Q^{IJ}$-action maps $Q$-exact operators to $Q$-exact operators, it follows that $(Q^{IJ})^{\wedge p} O^{(r,s,t)}_{10}$ is not $Q$-exact for any $r,s\geq 0$, $t=0,1,2$, and $p=0,1,\ldots,6$. More generally, for any $t$ for which $\widehat Z_t$ exists, if $\mathcal C_{r,s,t}\neq 0$, then $(Q^{IJ})^{\wedge p} O^{(r,s,t)}_{10}$ is not $Q$-exact.

Moreover, we can show that these operators are indeed strongly fortuitous. Let $R=\operatorname{diag}(-1,-1,-1,1)$ be a representative of the nontrivial element of $O(4)/SO(4)$. Define an involutive automorphism $\sigma$ on the BMN operator algebra by
\begin{equation}
    \sigma(\phi^I_\mu) = R_\mu{}^\nu \phi^I_\nu\ , \qquad \sigma(\psi_{I\mu}) = R_\mu{}^\nu \psi_{I\nu}\ , \qquad \sigma(f_{\mu\nu})
    =-R_\mu{}^\rho R_\nu{}^\lambda f_{\rho\lambda}\ .
\end{equation}
Note that $\sigma$ maps gauge-invariant operators to gauge-invariant operators. One can see that
\begin{equation}
    \sigma^2 = \mathrm{id}\ , \qquad \sigma Q = -Q \sigma\ , \qquad [\sigma, Q^{IJ}] = 0\ ,
\end{equation}
so $\sigma$ induces an involution on the $Q$-cohomology. For a gauge-invariant tensor contraction $O$, its $\sigma$-parity is
\begin{equation}
    \sigma O
    =(-1)^{n_f+n_{\epsilon}} O\ ,
\label{eq:kw-sigma-contraction}
\end{equation}
where $n_f$ counts the number of $f_{\mu\nu}$ and $n_{\epsilon}$
counts the number of $SO(4)$ gauge $\epsilon$ tensors. Then, we obtain
\begin{equation}
    \sigma u^{IJ}=u^{IJ}\ ,
    \qquad
    \sigma v^I{}_J=v^I{}_J\ ,
    \qquad
    \sigma w^{IJ}=w^{IJ}\ .
\label{eq:kw-sigma-even-generators}
\end{equation}
Therefore, every generalized monotone built from $u^{IJ}$, $v^I{}_J$, and $w^{IJ}$ is $\sigma$-even.
By contrast, both $r^{IJ}$ and $r^I{}_{JKL}$ in
\eqref{r-preimage} are $\sigma$-odd. Thus, $\sigma O_{10}=-O_{10}$. In addition, since $\sigma z^{IJ} = z^{IJ}$, one can always choose $\widehat Z_t$ to be $\sigma$-odd by replacing it with $(\widehat Z_t-\sigma\widehat Z_t)/2$. Consequently, we conclude that $(Q^{IJ})^{\wedge p} O^{(r,s,t)}_{10}$ is $\sigma$-odd for any $p=0,1,\ldots,6$, $r,s\geq 0$, and $t$ for which $\widehat Z_t$ exists, and is therefore strongly fortuitous. This is analogous to the situation in $SU(2)$ $\mathcal N=4$ SYM, where all graviton cohomologies contain an even number of letters, while all known fortuitous cohomologies contain an odd number of letters \cite{Choi:2023znd}.

In summary, we have shown that the operators 
\begin{equation}\label{BH-tower}
    (Q^{IJ})^{\wedge p} O^{(r,s,t)}_{10} \equiv (Q^{IJ})^{\wedge p} \left(u^{\odot r} \odot w^{\odot s} \odot \left(z^{\odot t} \odot O_{10} + \cdots \right) \right)
\end{equation}
are strongly fortuitous for any $r,s\geq0$, $t=0,1,2$, and $p=0,1,\ldots,6$. After truncation to $SU(2)$ $\mathcal N =4$ SYM, $(Q^{IJ})^{\wedge p} O^{(r,s,t)}_{10}$ is $Q$-cohomologous to a linear combination of the fortuitous cohomologies $\hat \phi^{r'} (w^m)^{s'} O_{t'}$ and their $Q^n_+$-descendants \cite{Choi:2023znd}, where $r'\in\mathbb Z$ and $s',t'\geq0$.

Before concluding this section, we comment on a possible large $N$ holographic interpretation of the infinitely many strongly fortuitous cohomologies \eqref{BH-tower} we have found. Turning on a nonzero Higgs VEV $\langle\bar B^{\hat 2}\rangle = v \, \mathbf{1}_N$ induces a nonzero baryonic condensate $\langle\det \bar B^{\hat 2}\rangle$ and spontaneously breaks the baryon number symmetry. On the dual gravity side, this Higgs vacuum is described by the warped resolved-conifold solution of \cite{Klebanov:2007us}, with the resolution scale set by the Higgs VEV. The corresponding fully backreacted geometry interpolates between an asymptotic AdS$_5\times T^{1,1}$ region in the UV and an AdS$_5\times S^5$ throat in the IR. 
The Higgs branch reduction \eqref{eq:O10toO0} of the KW fortuitous primary $O_{10}$ to the $\mathcal N=4$ fortuitous primary $O_0$ then suggests a semiclassical picture in which the $\mathcal N=4$ BPS black hole core resides in the AdS$_5\times S^5$ throat \cite{Gutowski:2004ez,Gutowski:2004yv,Chong:2005hr,Kunduri:2006ek} and is connected to the asymptotic AdS$_5\times T^{1,1}$ region through the Higgs-flow geometry. Since the baryonic operators are dual to D3-branes wrapping internal three-cycles \cite{Gubser:1998fp,Berenstein:2002ke}, it is natural to regard this interpolating region as a baryonic condensate hair surrounding the black hole core. The Higgs scale set by $v$ controls the separation between the two regions. For sufficiently large $v$, 
the black hole core and the baryonic condensate hair can be well-separated, 
suggesting a weakly interacting hairy black hole description 
\cite{Basu:2010uz,Bhattacharyya:2010yg}. As $v$ is decreased, 
the core and the hair can no longer be regarded as separate constituents. In the limit $v\to0$, they should merge into genuine BPS black holes in AdS$_5\times T^{1,1}$.

One may compare the above black hole dressed by a baryonic condensate with 
those found in \cite{deMelloKoch:2024pcs,Choi:2025lck}. There, the black hole core is surrounded by a large D3-brane domain wall, namely a dual giant graviton \cite{McGreevy:2000cw,Grisaru:2000zn,Hashimoto:2000zp}. 
The RR 5-form flux jumps by one unit upon crossing the D3-brane wall. In addition, its local gravitational backreaction is negligible in classical supergravity at leading order in large $N$.
In contrast, in our case, since the baryonic condensate is made of D3-branes wrapping 
an internal three-cycle rather than being a wall in AdS, the RR five-form flux does not change 
between UV and IR. Furthermore, the baryonic condensate backreacts substantially 
on the geometry already at leading order in large $N$ \cite{Klebanov:2007us}.

The various dressings \eqref{BH-tower} of $O_{10}$ also admit a natural physical interpretation. Their Higgs-branch reduction \eqref{eq:Zt-dressed-truncation} suggests that the \(u\)-dressings are naturally associated with additional excitations of the baryonic D3-brane condensate. The \(w\)-dressings may instead be viewed as particle-like BPS excitations surrounding the black hole core. Note that in $SU(2)$ $\mathcal{N}=4$ SYM, $O_0$ can be dressed by $w$ but not by $u$ \cite{Choi:2023znd}. By contrast, the \(z\)-dressings modify the underlying \(\mathcal N=4\) black hole core itself, replacing $O_0$ by $O_t$ \cite{Choi:2023znd}.

The discussion so far concerns the classical $Q$-cohomology. Recently, \cite{Kim:2026rnx} studied the one-loop corrections to fortuitous cohomologies in $SU(2)$ $\mathcal{N}=4$ SYM. In particular, they found that while $O_0$ and $w^m O_0$ are not lifted at one loop, $O_1, O_2$ and $O_3$ are lifted to non-BPS operators. It would be interesting to investigate whether the corresponding operators in the KW theory exhibit an analogous lifting pattern at one loop.

\clearpage

\section{Conclusion and discussion}

In this paper, we formulated a classical cohomology problem for the BPS operators 
in the Klebanov-Witten (KW) theory. We suggested 
a refined fortuity program that classifies the apparently $N$-dependent baryon 
cohomologies into the generalized monotonous class, 
and the rest as the strongly fortuitous class.
We expect the latter class to better characterize 
the holographic BPS black hole states. The ideas in this paper should be 
helpful for studying typical holographic $\mathcal{N}=1$ SCFTs (like those 
dual to AdS$_5\times Y^{p,q}$ or AdS$_5\times L^{p,q,r}$), 
which are usually defined as RG fixed points and contain baryonic operators. 

We also constructed infinitely many strongly fortuitous cohomologies 
for the KW theory at $N=2$. We first detected evidence for their existence 
by computing the index which subtracts the contributions from generalized monotones, 
and then constructed their representatives using the ansatz of \cite{Choi:2023vdm} 
and the methods of \cite{deMelloKoch:2024pcs}. 
We also discussed the possible interpretation of the simplest fortuitous cohomology 
(called $O_{10}$ in Section \ref{subsec:strongly-fortuitous}) as a hairy black hole state, suggesting novel 
BPS hairy black holes surrounded by baryonic condensates.

In light of the more refined notion of monotones established in this paper, 
there might be more refined criteria 
(based on regular patterns as $N$ varies) that narrow the definition of fortuity in other theories and better characterize the erratic/chaotic states for 
black holes. See  \cite{Chen:2024oqv} for a possible notion of chaos in the BPS sector 
related to fortuity.

It is also interesting to consider the `chaos invasion' picture of \cite{Chen:2024oqv}
for the ABJM fortuitous states of Section \ref{section:abjm} 
which become di-baryons at particular 
$N=N_\star$, by following the $N$-dependent states \cite{Budzik:2023vtr} 
as multi-trace operators. 
At $N>N_\star$, the multi-trace operator is non-BPS because it is
fortuitous in the sense of \cite{Chang:2024zqi}. As $N$ decreases and reaches $N_\star$, 
the general scenario of \cite{Chen:2024oqv} is that the BPS sector fortuitously 
enlarges by what used to be non-BPS states, inheriting some of the chaotic properties 
of the typical non-BPS sector. However, for these ABJM states, this is 
precisely where the di-baryon and meson are related by a trace relation and produce 
a null state. The only remaining BPS state at $N=N_\star$ already existed at larger $N$, 
with its solitonic mass scaling as $N\sim G_N^{-1/2}$. 
Therefore, the BPS sector does not acquire more states by these operators 
invading from the non-BPS sector.

Even in the fortuitous class, it may be desirable to distinguish 
more and less erratic states. For instance, various black hole states explored in 
$\mathcal{N}=4$ SYM \cite{Choi:2023znd,deMelloKoch:2024pcs} and in the KW theory 
(this paper) are interpreted as hairy black hole states of the `grey galaxy' type
\cite{Kim:2023sig} or the `dual giant dressed' type \cite{Choi:2024xnv} (whose BPS 
limits are discussed in \cite{Choi:2025lck,Jones:2025gno}), consisting of the core 
black hole part and the hair part. In Section \ref{subsec:strongly-fortuitous}, we suggested yet another 
class of hairy black holes dressed by the baryonic condensates. 
In all these cases, one expects that 
a substantial part of the heavy state would be less erratic/chaotic, 
consisting of gravitons or D-branes. It will be interesting to 
find refined algebraic measures which probe this part of a heavy state.

\vskip 0.5cm

\hspace*{-0.8cm} {\bf\large Acknowledgements}
\vskip 0.2cm

\hspace*{-0.75cm} 
We thank Chi-Ming Chang, Abhijit Gadde, Seunggyu Kim, Eunwoo Lee, Jehyun Lee, 
Siyul Lee, Shiraz Minwalla, Jaewon Song and Alberto Zaffaroni for helpful discussions. 
This work is supported in part by a KIAS Individual Grant PG106601 at the Korea Institute for Advanced Study (JC), WPI Initiative, MEXT, Japan at Kavli IPMU, the University of Tokyo (SC), Basic Science Research Program through the National Research Foundation of Korea (NRF) funded by the Ministry of Education (RS-2025-02663044) (SC) and the NRF grant RS-2026-25477643 (SK). This research used computing resources at Kavli IPMU.

\appendix

\section{Baryons and the moduli space}\label{appendix:KW-review}

\subsection{Scalar baryons}\label{appendix:scalar-baryons}

Baryonic operators are dual to D3-branes wrapping topological three-cycles of $T^{1,1}$.
The minimal-volume wrapped branes correspond to \cite{Gubser:1998fp,Berenstein:2002ke}
\begin{equation}
\begin{aligned}
    (\det A)_{(i_1\cdots i_N)} \equiv
    \frac{1}{N!}\epsilon_{p_1\cdots p_N}\epsilon^{\hat{p}_1\cdots \hat{p}_N}
    (A_{i_1})^{p_1}\!_{\hat{p}_1}\cdots (A_{i_N})^{p_N}\!_{\hat{p}_N}\;,\\
    (\det B)_{(\hat{i}_1\cdots\hat{i}_N)}\equiv
    \frac{1}{N!}\epsilon_{\hat{p}_1\cdots \hat{p}_N}\epsilon^{p_1\cdots p_N}
    (B_{\hat{i}_1})^{\hat{p}_1}\!_{p_1}\cdots (B_{\hat{i}_N})^{\hat{p}_N}\!_{p_N}\;.
\end{aligned}
\end{equation}
The two operators are respectively in the $(N/2,0)$ and $(0,N/2)$ representations 
of the flavor symmetry. Both have R-charge $N/2$ and dimension $\Delta=3N/4$. Their baryon numbers are $+N$ and $-N$, respectively.

The flavor symmetry $SU(2)\times SU(2)$ and the R-symmetry arise from isometries of
\[
T^{1,1}=(SU(2)\times SU(2))/U(1),
\]
whereas $U(1)_B$ is not an isometry.
Instead, it is the baryonic global symmetry whose charge measures D3-brane wrapping on the nontrivial three-cycle of \(T^{1,1}\), whose topology is \(S^2\times S^3\).\footnote{
The Sasaki--Einstein space $T^{1,1}$ is a $U(1)$ fibration over $S^2\times S^2$ with metric \cite{Candelas:1989js}
\begin{align*}
    ds^2_{T^{1,1}}=\frac{1}{9}(d\psi+\cos\theta_1d\phi_1+\cos\theta_2d\phi_2)^2
    +\frac{1}{6}\sum_{i=1}^{2}\big[d\theta_i^2+\sin^2\theta_i d\phi_i^2\big],
\end{align*}
where $\psi\sim\psi+4\pi$. Supersymmetric three-cycles are obtained by fixing either $(\theta_2,\phi_2)$ or $(\theta_1,\phi_1)$. The two representatives are homologous, and $b_3(T^{1,1})=1$.}

A minimally wrapped D3-brane behaves as a charged particle moving on the transverse $S^2$ \cite{Gubser:1998fp}. Because the RR five-form carries $N$ units of flux through $T^{1,1}$, this particle sees $N$ units of magnetic flux on the sphere. Its ground states therefore form the spin-$N/2$ representation of the corresponding $SU(2)$, matching the flavor representation of the determinant operators.

Fluctuations of the wrapped brane were analyzed with the probe D3-brane DBI action in \cite{Berenstein:2002ke}. More generally, quantizing the moduli space of holomorphic surfaces \cite{Mikhailov:2000ya} gives the scalar part of the full BPS brane Hilbert space \cite{Beasley:2002xv}. The associated scalar baryons can be organized using the bifundamental composites
\begin{align}
    A_{I;J}\equiv A_{i_1}B_{j_1}\cdots
    A_{i_m}B_{j_m}A_{i_{m+1}}\;,
\end{align}
Here $I=(i_1,\ldots,i_{m+1})$ and $J=(j_1,\ldots,j_m)$. The indices within $I$ and within $J$ are separately symmetrized by the chiral-ring relation. 
\begin{align}\label{eq:chiralrel}
    A_1B_kA_2=A_2B_kA_1\;,\;\;
    B_1A_iB_2=B_2A_iB_1\;.
\end{align}
A general scalar BPS baryon of charge $N$ is therefore
\begin{align}
    \epsilon_{p_1\cdots p_N}\epsilon^{\hat{p}_1\cdots \hat{p}_N}
    (A_{I_1;J_1})^{p_1}\!_{\hat{p}_1}\cdots (A_{I_N;J_N})^{p_N}\!_{\hat{p}_N}\;.
\end{align}
Charge $-N$ baryons are constructed in the same way from
\[
B_{I;J}=B_{j_1}A_{i_1}\cdots A_{i_m}B_{j_{m+1}}.
\]
Unlike the minimal determinants, these operators need not lie in a single irreducible representation of $SU(2)\times SU(2)$. Their flavor representation is the part of
\[
(I_1,J_1)\otimes\cdots\otimes(I_N,J_N)
\]
that is symmetric under exchange of any two pairs $(I_k,J_k)$ and $(I_l,J_l)$.

Some of these baryons factor into a zero-baryon-number operator times $(\det A)_{i_1\ldots i_N}$. In the gravity dual, such a product describes a BPS Kaluza--Klein excitation placed on top of the BPS wrapped D3-brane. Consider, for example,
\begin{align}\label{eq:baryonexample}
    \epsilon_{p_1\cdots p_N}\epsilon^{\hat{p}_1\cdots \hat{p}_N}
    (A_{i_1}B_kA_{i_2})^{p_1}\!_{\hat{p}_1}
    (A_{i_3})^{p_2}\!_{\hat{p}_2}\cdots
    (A_{i_{N+1}})^{p_N}\!_{\hat{p}_N}\;
\end{align}
This operator factorizes if either $(i_1,i_3,\ldots,i_{N+1})$ or $(i_2,i_3,\ldots,i_{N+1})$ is totally symmetric. In the second case, the identity
\begin{align}
    \epsilon^{\hat{p}_1\cdots \hat{p}_N}
    (A_{(i_2})^{q}\!_{\hat{p}_1}
    (A_{i_3})^{p_2}\!_{\hat{p}_2}\cdots
    (A_{i_{N+1})})^{p_N}\!_{\hat{p}_{N}}
    =\epsilon^{qp_2\cdots p_N}(\det A)_{(i_2i_3\cdots i_{N+1})}\;,
\end{align}
reduces \eqref{eq:baryonexample} to
\begin{align}
    (N-1)!\tr (A_{i_1}B_k) (\det A)_{(i_2\cdots i_{N+1})}\;.
\end{align}
Symmetrizing $i_1$ with the determinant indices gives the $(\frac{N+1}{2},\frac12)$ component, while contracting it with one determinant index using $\epsilon^{i_1i_2}$ gives the $(\frac{N-1}{2},\frac12)$ component.

The contracted component might appear to vanish in the chiral ring, since the epsilon contraction antisymmetrizes flavor indices. A direct diagonal-matrix test shows that it does not. Any scalar operator that is $\bar Q_{\dot\alpha}$-exact must vanish when $A_i$ and $B_k$ are diagonal, because the chiral-ring relations \eqref{eq:chiralrel} are then satisfied. Set
\[
A_1=\mathrm{diag}(a_1,\ldots,a_{N-1},0),\qquad
A_2=\mathrm{diag}(\tilde a_1,\ldots,\tilde a_N),\qquad
B_k=\mathrm{diag}(b_1,\ldots,b_N).
\]
The highest-weight component of the spin-$(N-1)/2$ representation evaluates to
\begin{align*}
    \epsilon^{i_1i_2}\tr(A_{i_1}B_k)(\det A)
    _{(i_2 11\cdots 1)}
    =(a_1b_1+\cdots+a_{N-1}b_{N-1})
    \frac{1}{N}a_1\cdots a_{N-1}\tilde{a}_N\;.
\end{align*}
This expression is generically nonzero. Thus, factorization can occur even when some $A$-indices are antisymmetrized; it always occurs when all of them are symmetrized.

To obtain a nonfactorized operator from \eqref{eq:baryonexample}, one may instead contract $i_1$ and $i_2$ with, for example, $i_3$ and $i_4$ through $\epsilon^{i_1i_3}\epsilon^{i_2i_4}$. This component transforms as $(\frac{N-3}{2},\frac12)$.
It is again nontrivial in the chiral ring.
Indeed, take
\[
A_1=\mathrm{diag}(a_1,\ldots,a_{N-2},0,0),\qquad
A_2=\mathrm{diag}(\tilde a_1,\ldots,\tilde a_N),\qquad
B_k=\mathrm{diag}(b_1,\ldots,b_N).
\]
For $(i_5,\ldots,i_{N+1})=(1,\ldots,1)$, the operator becomes
\begin{align*}
    \epsilon^{i_1i_3}\epsilon^{i_2i_4}
    &\epsilon_{p_1\cdots p_N}\epsilon^{\hat{p}_1\cdots \hat{p}_N}
    (A_{i_1}B_kA_{i_2})^{p_1}\!_{\hat{p}_1}
    (A_{i_3})^{p_2}\!_{\hat{p}_2}
    (A_{i_4})^{p_3}\!_{\hat{p}_3}\cdots
    (A_{i_{N+1}})^{p_N}\!_{\hat{p}_N}\;,
\end{align*}
It is proportional to
\[
a_1\cdots a_{N-2}\tilde a_{N-1}\tilde a_N
(a_1b_1+\cdots+a_{N-2}b_{N-2}),
\]
which is nonzero in general, so this component is not $\bar Q_{\dot\alpha}$-exact. Such nonfactorized scalar baryons exist only for $N\geq3$. The same conclusion can be recovered from the generators of the scalar chiral ring, obtained from the plethystic logarithm of the Hilbert series \cite{Forcella:2007wk}.

\subsection{Moduli space of vacua}\label{appendix:moduli-space}

The baryonic symmetry $U(1)_B$ descends from one of the two abelian gauge factors that decouple in the IR. It is useful to separate the vacuum data into mesonic and baryonic directions. Imposing the $D$-term condition of the decoupled relative $U(1)$, or equivalently quotienting by $U(1)_B$, gives the mesonic branch:
\begin{align}
    \mathcal{M}_{mes}=\mathcal{M}//U(1)_B
\end{align}
Only baryon-neutral operators are well-defined on this quotient and can parametrize $\mathcal M_{\mathrm{mes}}$. The full moduli space $\mathcal M$ retains one additional baryonic direction.

We begin with the mesonic branch. It is the $N$-fold symmetric product of the conifold and has complex dimension $3N$. For a single brane, the singular conifold can be described by
\begin{align}\label{eq:conifold}
    |A_1|^2+|A_2|^2-|B_1|^2-|B_2|^2=0\;,\;\;\;
    (A_1,A_2,B_1,B_2)\in \mathbb{C}^4
\end{align}
together with the $U(1)_B$ quotient. Equivalently, introduce the baryon-neutral combinations
\begin{align}
    z_1=A_1B_1\;,\;\;\;
    z_2=A_2B_2\;,\;\;\;
    z_3=A_1B_2\;,\;\;\;
    z_4=A_2B_1\;,
\end{align}
which obey
\begin{align}
    z_1z_2-z_3z_4=0\;.
\end{align}
These $z_i$ are invariant under the complexified baryonic action
\[
A_i\longrightarrow\mu A_i,\qquad B_k\longrightarrow\mu^{-1}B_k,
\qquad \mu\in\mathbb C^\times.
\]
Thus, the same conifold can be represented as the quotient
\begin{align}
    C=\{(A_1,A_2,B_1,B_2)
    \sim (\mu A_1,\mu A_2,\mu^{-1} B_1,\mu^{-1} B_2),\;\;\; \mu\in\mathbb{C}^{\times}\} ,
\end{align}
To obtain the singular conifold, one must exclude the loci
\[
A_1=A_2=0,\quad (B_1,B_2)\neq(0,0),
\qquad\text{and}\qquad
B_1=B_2=0,\quad (A_1,A_2)\neq(0,0).
\]
After the quotient, each excluded locus is a $\mathbb{CP}^1$. Retaining one of them instead produces a resolution in which the conifold singularity is replaced by $S^2\simeq\mathbb{CP}^1$; equivalently, the right-hand side of \eqref{eq:conifold} is shifted to a nonzero constant.

The base $T^{1,1}$ is obtained by restricting to
\[
|A_1|^2+|A_2|^2=|B_1|^2+|B_2|^2=1.
\]
The complexified $U(1)_B$ action can impose the first equality. Fixing the radial direction, equivalently quotienting by the complexified R-symmetry, imposes the second. The remaining $U(1)_B$ quotient then gives $T^{1,1}=(SU(2)\times SU(2))/U(1)$.

The full moduli space has complex dimension $3N+1$ and is less straightforward to describe. Classically it is the solution of the F-term and D-term equations, modulo gauge transformations:
\begin{align}
    A_1B_kA_2=A_2B_kA_1\;,\;&\;
    B_1A_iB_2=B_2A_iB_1\;,\\
    \bar{A}^iA_i-B_k\bar{B}^k=
    A_i&\bar{A}^i-\bar{B}^kB_k
    =\mathcal{U}\;\mathds{1}_{N}\;,\label{eq:D-term}
\end{align}
Here
\[
\mathcal U=\frac1N\tr(A_i\bar A^i-\bar B^kB_k),
\]
and our Hermitian-conjugation convention is $(A_i)^\dagger=\bar A^i$ and $(B_k)^\dagger=\bar B^k$. One may impose the D-term equations equivalently by quotienting the F-flat solutions by $SL(N,\mathbb C)\times SL(N,\mathbb C)$.

To describe a generic solution, start with the singular-value decomposition of $B_2$:
\begin{align}
    B_2=UDV^{\dag}\;,
\end{align}
Here $U$ and $V$ are unitary, and $D$ is diagonal with nonnegative entries. This follows by diagonalizing the positive-semidefinite Hermitian matrix $B_2^\dagger B_2=VD^2V^{\dag}$. Assuming for simplicity that no eigenvalue of $B_2^{\dag}B_2$ vanishes, $U=B_2VD^{-1}$ is also unitary, so we get the singular value decomposition above.

Because the gauge factors are special unitary rather than unitary, gauge transformations bring $B_2$ to
\begin{align}
    (B_2)_{diag} = U^{\dag}B_2 V
    =e^{i\alpha}\mathrm{diag}(b_1,b_2,\cdots,b_N)\;,
\end{align}
with $U,V\in SU(N)$. The $b_i$ are nonnegative and the common phase $e^{i\alpha}$ cannot be removed by these transformations. We again assume $b_i>0$. A further complexified transformation can then make $B_2$ proportional to the identity:
\begin{align}
    B \mathds{1}_N
    =U^{\dag}B_2 V W\;,
\end{align}
where
\begin{align}
    W=\mathrm{diag}\Big(\frac{b}{b_1},
    \frac{b}{b_2},\cdots,\frac{b}{b_N}\Big)
    \in SL(N,\mathbb{C})\;,
\end{align}
with $b=(b_1\cdots b_N)^{1/N}>0$ and $B=be^{i\alpha}$.

In the gauge $B_2=B\mathds1_N$, the F-term equations imply that $A_1,A_2$, and $B_1$ commute. The complexified gauge transformations preserving $B_2$ form the diagonal $SL(N,\mathbb C)$. On $A_1,A_2,B_1$, this is effectively indistinguishable from $GL(N,\mathbb C)$ because the common complex rescaling cancels. The three commuting matrices can therefore be put simultaneously into Jordan form with a common block decomposition, though their eigenvalues may differ. Solutions with nontrivial Jordan blocks can satisfy the D-term equations, but each such block carries only one eigenvalue. They occupy a lower-dimensional subset than the diagonalizable solutions. The generic solution is consequently described by the $3N$ eigenvalues of $A_1,A_2,B_1$ together with the complex parameter $B$, for a total of $3N+1$ complex parameters.

Undoing $W$ returns four diagonal matrices $A_i,B_k$. Their $4N$ eigenvalues obey
\begin{align}\label{eq:modulieq}
    |(A_1)_{ii}|^2+|(A_2)_{ii}|^2
    -|(B_1)_{ii}|^2-|(B_2)_{ii}|^2
    =\mathcal{U}
\end{align}
for $i=1,\ldots,N$, together with a residual $U(1)^{N-1}$ quotient. Summing these equations over $i$ gives an identity by the definition of $\mathcal U$. There are therefore only $N-1$ independent real conditions and $N-1$ independent phase quotients. Starting from $4N$ complex eigenvalues leaves a moduli space of complex dimension $3N+1$.

When \(\mathcal U=0\), each set of four eigenvalues describes a D3-brane on the singular conifold. The residual Weyl group \(S_N\) permutes the \(N\) sets simultaneously. After quotienting by the complexified \(U(1)_B\) action, one obtains the mesonic branch \(\mathcal M_{\mathrm{mes}}\), namely the \(N\)-fold symmetric product of the conifold.

When $\mathcal U\neq0$, \eqref{eq:modulieq} instead describes the resolved conifold, in which the singular point is replaced by a smooth $S^2$.

In summary, $3N$ complex moduli specify the positions of the $N$ D3-branes, while one further modulus controls the complexified resolution of the conifold. The real parameter $\mathcal U$ plays the role of a Fayet--Iliopoulos parameter and describes the baryonic or resolution direction. 
The corresponding moment-map operator is the bottom component of the \(U(1)_B\) current multiplet.

\section{The superconformal index}\label{appendix:finite-N-index}

The index of the $Y^{p,q}$ quiver theories was computed in \cite{Gadde:2010en}. For $Y^{1,0}=T^{1,1}$ and gauge group $SU(N)\times SU(N)$, the result is
\begin{align}\label{eq:SCI}
    \mathcal{I}(t,y,u,v,b)&=\tr (-1)^F t^{3(2j_1-r)}y^{2j_2}
    u^{2m_1}v^{2m_2}b^{B}
    \nonumber\\
    &=\prod_{k=1}^{2}\Bigg[\frac{\kappa^{N-1}}{N!}
    \oint\prod_{i=1}^{N-1}
    \frac{dz_i^{(k)}}{2\pi i z_i^{(k)}}
    \frac{1}{\prod_{i\neq j}\Gamma(z^{(k)}_{i}/z^{(k)}_{j};p,q)}
    \Bigg]
    \nonumber\\
    &\quad\times
    \prod_{i,j=1}^{N}
    \Gamma(t^{3/2}u^{\pm}b^{-1}z^{(2)}_{i}/z^{(1)}_{j};p,q)
    \Gamma(t^{3/2}v^{\pm}b z^{(1)}_{i}/z^{(2)}_{j};p,q).
\end{align}
The notation $\Gamma(u^\pm)$ means $\Gamma(u)\Gamma(u^{-1})$. As in the main text, $j_1,j_2$ are the half-integrally normalized Cartan generators of the Lorentz group $\mathrm{Spin}(4)=SU(2)\times SU(2)$, acting on undotted and dotted indices, respectively. The charge $r$ is the IR superconformal R-charge. The charges $m_1,m_2$ generate the Cartan subgroup of the flavor $SU(2)\times SU(2)$, and $B$ is baryon number. For each gauge factor, $z_i^{(k)}$ are gauge fugacities constrained by $\prod_{i=1}^N z_i^{(k)}=1$.

Only local operators satisfying
\begin{align}
    \{Q_-,S^-\}=D-2j_1+\frac{3}{2}r=0
\end{align}
contribute, where
\[
\{Q_\alpha,S^\beta\}
=D\delta_\alpha^\beta+2M^\beta{}_\alpha+\frac32r\,\delta_\alpha^\beta.
\]
The weights $2j_1-r$ and $j_2$ commute with $Q_-$. A third combination of $D,j_1,j_2,r$ also commutes with $Q_-$, but it vanishes on contributing states by the condition above; an additional fugacity would therefore be redundant.

We use the abbreviations
\begin{align}
    p\equiv t^3y\;,\;\;\;q\equiv t^3y^{-1},\;\;\;
    \kappa\equiv (p;p)(q;q),\;\;\;
    (a;b)=\prod_{k=0}^{\infty}(1-a b^k),
\end{align}
and define the elliptic gamma function by
\begin{align}
    \Gamma(z;p,q)=\prod_{k,m=0}^{\infty}\frac{1-p^{k+1}q^{m+1}z^{-1}}{1-p^kq^mz}.
\end{align}
To see how the UV letters assemble into \eqref{eq:SCI}, consider one numerator factor:
\begin{align}
    \Gamma (t^{3/2}u^{-1}b^{-1}z^{(2)}_p/z^{(1)}_q;t^3y,t^3y^{-1})&=
    \prod_{k,m=0}^{\infty}\frac{1-t^{3k+3m+6}y^{k-m}t^{-3/2}ubz^{(1)}_q/z^{(2)}_p}
    {1-t^{3k+3m}y^{k-m}t^{3/2}u^{-1}b^{-1}z^{(2)}_p/z^{(1)}_q}\\
    &={\rm PE} \Bigg[\frac{t^{3/2}u^{-1}b^{-1}z_p^{(2)}/z_q^{(1)}-t^{9/2}ubz_q^{(1)}/z_p^{(2)}}{(1-t^3y)(1-t^3y^{-1})}\Bigg].
\end{align}
It is generated by $(\bar A^1)^{\hat p}{}_{q}$, $(\psi_{A1+})^{q}{}_{\hat p}$, and arbitrary $D_{+\dot\pm}$ dressings, with $\psi_{Ai\alpha}=[Q_\alpha,A_i]$, $
\psi_{B\hat i\alpha}=[Q_\alpha,B_{\hat i}]$, and $
D_{\alpha\dot\beta}=(\sigma^\mu)_{\alpha\dot\beta}D_\mu$.
Their charges are listed in Table \ref{tb:uv letter charges}.
\begin{table}[h]
\begin{center}
\begin{tabular}{ |p{1cm}||p{0.8cm}|p{0.8cm}|p{0.8cm}|p{0.8cm}||p{0.8cm}|p{0.8cm}|p{0.8cm}|p{0.8cm}|p{0.8cm}||p{1.6cm}|  }
 \hline
 letters& $E_{UV}$ & $j_1$ & $j_2$ & $r_{UV}$ & $E_{IR}$ & $r_{IR}$ & $m_1$ & $m_2$ & $B$ & \\
 \hline \hline && && && && &&\\[-1.5ex]
 $\bar{A}^1$   & $1$    & $0$ & $0$ & $-2/3$ & $3/4$ & $-1/2$ & $-1/2$ & $0$ & $-1$ & $t^{3/2}u^{-1}b^{-1}$\\[1ex]
 \hline && && && && &&\\[-1.5ex]
 $\psi_{A1+}$&   $3/2$  & $1/2$  & $0$ & $-1/3$ & $5/4$ & $-1/2$ & $1/2$ & $0$ & $1$ & $-t^{9/2}ub$\\[1ex]
 \hline && && && && &&\\[-1.5ex]
 $D_{+\dot{\pm}}$&   $1$  & $1/2$  & $\pm 1/2$ & $0$ & 1 & $0$ & $0$ & $0$ & $0$ & $t^3 y^{\pm}$\\[1ex]
 \hline
\end{tabular}
\end{center}
\vspace{-0.5cm}
\caption{Charges of the UV BPS letters contributing to one elliptic gamma function.}\label{tb:uv letter charges}
\end{table}
The other three numerator factors arise from $(\bar A_2,\psi_{A_2+})$, $(\bar B_1,\psi_{B_1+})$, and $(\bar B_2,\psi_{B_2+})$, again with derivative dressings. Although the chiralini are BPS letters in the UV description, they are not BPS at the IR fixed point: $E_{\mathrm{IR}}-2j_1+\frac32r_{\mathrm{IR}}=-\frac12\neq0$.
For each vector multiplet, the relevant letters are $\bar\lambda_{\dot\alpha}^{(k)}$, $F_{++}^{(k)}$, and their $D_{+\dot\pm}$ dressings, with $k=1,2$. Our spinor convention is
\[
F_{\alpha\beta}^{(k)}=(\sigma^{\mu\nu})_{\alpha\beta}F_{\mu\nu}^{(k)},\qquad
(\sigma^{\mu\nu})_\alpha{}^\beta
=\frac14\left(\sigma_{\alpha\dot\alpha}^{\mu}\bar\sigma^{\nu\dot\alpha\beta}
-\sigma_{\alpha\dot\alpha}^{\nu}\bar\sigma^{\mu\dot\alpha\beta}\right),
\]
and $(\sigma^{\mu\nu})_{\alpha\beta}=\epsilon_{\beta\gamma}(\sigma^{\mu\nu})_\alpha{}^\gamma$. One must also remove the UV gaugino equation of motion; for the first gauge factor, $D_{+\dot{\alpha}}\bar{\lambda}^{(1)\dot{\alpha}}\sim g_{YM}(\psi_{A_i+}\bar{A}^i-\bar{B}^i\psi_{B_i+})-(trace)$.
The BPS condition $E_{\mathrm{UV}}-2j_1+\frac32r_{\mathrm{UV}}=0$ is satisfied
for both $\bar\lambda_{\dot\alpha}^{(k)}$ and $F_{++}^{(k)}$ (for example, $r[\bar\lambda_{\dot\alpha}^{(k)}]=-1$ and $E[\bar\lambda_{\dot\alpha}^{(k)}]=3/2$). Moreover, $E_{\mathrm{UV}}=E_{\mathrm{IR}}$ and $r_{\mathrm{UV}}=r_{\mathrm{IR}}$ in the vector multiplet, since the gauge field keeps its canonical dimension one and R-charge zero throughout the flow. The vector-multiplet contribution is therefore
\begin{align}
    \prod_{i\neq j}^{N}\frac{1}{\Gamma(z^{(k)}_{i}/z^{(k)}_{j};t^3y,t^3y^{-1})}
    &=\prod_{i\neq j}^{N}\prod_{k,m=0}^{\infty} \frac{1-t^{3k+3m}y^{k-m} z^{(k)}_{j}/z^{(k)}_{i}}
    {1-t^{3k+3m+6}y^{k-m}z^{(k)}_{i}/z^{(k)}_{j}}
    \nonumber\\
    &=\mathrm{PE}\Bigg[
    \frac{t^6-1}{(1-t^3y)(1-t^3y^{-1})}\sum_{i\neq j}z^{(k)}_{i}/z^{(k)}_{j}
    \Bigg]
    \nonumber\\
    &=\mathrm{PE}\Bigg[
    \bigg(
    \frac{2t^6-t^3y-t^3y^{-1}}{(1-t^3y)(1-t^3y^{-1})}-1
    \bigg)
    \sum_{i\neq j}z^{(k)}_{i}/z^{(k)}_{j}
    \Bigg].
\end{align}
The $-1$ in the last plethystic exponential is the contribution of the Haar measure. Finally, the Cartan components of the vector multiplet produce $\kappa^{N-1}$, since
\begin{align}
    \mathrm{PE}\Bigg[
    \frac{2t^6-t^3y-t^3y^{-1}}{(1-t^3y)(1-t^3y^{-1})}
    \Bigg]
    &=\prod_{k,m=0}^{\infty}
    \frac{(1-t^{3k+3m+3}y^{k+1-m})(1-t^{3k+3m+3}y^{k-m-1})}{(1-t^{3k+3m+6}y^{k-m})^2}
    \nonumber\\
    &=\prod_{k=0}^{\infty}(1-t^{3k+3}y^{k+1})
    \prod_{m=0}^{\infty}(1-t^{3m+3}y^{-m-1})=\kappa.
\end{align}

In the BMN sector, the operators are made of letters $\bar{A}_i$, $\bar{B}_{\hat{i}}$, $\psi_{Ai+}$, $\psi_{B\hat{i}+}$ and $F_{++}^{(k)}$. 
\begin{align}
    \bar{A}^{1,2}\xrightarrow{}t^{3/2}u^{\mp}b^{-1}\;,\;
    &\bar{B}^{1,2}\xrightarrow{}t^{3/2}v^{\mp}b\;,\;
    \nonumber\\
    \psi_{A1,2+}\xrightarrow{}-t^{9/2}u^{\pm}b\;,\;
    \psi_{B1,2+}\xrightarrow{}&-t^{9/2}v^{\pm}b^{-1}\;,\;
    F_{++}^{(k)}\xrightarrow{}t^{6}\;,\;
\end{align}
The BMN index is therefore, 
\begin{align}\label{eq:BMNindex}
    \mathcal{I}_{\mathrm{BMN}}(t,u,v,b)&=\tr_{\mathrm{BMN}}
    (-1)^F t^{3(2j_1-r)}
    u^{2m_1}v^{2m_2}b^{B}
    \nonumber\\
    &=\prod_{k=1}^{2}\Bigg[\frac{1}{(1-t^6)^{N-1}N!}
    \oint_{\mathbb{T}}\prod_{i=1}^{N-1}
    \frac{dz_i^{(k)}}{2\pi i z_i^{(k)}}
    \prod_{i\neq j}^N
    \frac{1-z^{(k)}_{i}/z^{(k)}_{j}}
    {1-t^6z^{(k)}_{i}/z^{(k)}_{j}}
    \Bigg]
    \nonumber\\
    &\quad\times
    \prod_{i,j=1}^{N}
    \Bigg[
    \frac{(1-t^{9/2}u^{\pm}bz^{(1)}_i/z^{(2)}_j)
    (1-t^{9/2}v^{\pm}b^{-1}z^{(2)}_i/z^{(1)}_j)}
    {(1-t^{3/2}u^{\pm}b^{-1}z^{(2)}_i/z^{(1)}_j)
    (1-t^{3/2}v^{\pm}bz^{(1)}_i/z^{(2)}_j)}
    \Bigg].
\end{align}
where $z^{(k)}_N=1/(z^{(k)}_1\cdots z^{(k)}_{N-1})$, and $(1-xu^{\pm})$ means $(1-xu)(1-xu^{-1})$ and similarly for $v$.

\subsection{Infinite $N$ index}\label{infinite N index}
Let us compute the superconformal index at $N=\infty$ using the KK spectrum in Table \ref{gravity-multiplets}, \cite{Ceresole:1999zs,Ceresole:1999ht}.
\begin{align}
    \mathcal{I}_{\mathrm{KK}}(t,y,u,v,b)=
    \tr_{\mathrm{KK}} (-1)^F t^{3(2j_1-r)}y^{2j_2}u^{2m_1}v^{2m_2}b^{B}
\end{align}
In the Vector$_{\rm{I}}$ family, there are three types of cohomologies \eqref{eq: vector1 family}, whose contributions to the single-particle index are:
\begin{align}
    \mathrm{Vector}_{\rm{I}}(1):&\;\;\;
    \tr (ab)^k 
    \xrightarrow{}
    \sum_{k=1}^{\infty}
    t^{3k}\chi_{\frac{k}{2}}(u)\chi_{\frac{k}{2}}(v)
    \;,
    \\
    \mathrm{Vector}_{\rm{I}}(2):&\;\;\;
    \tr \psi_{b}b
    (ab)^k
    \xrightarrow{}
    -\sum_{k=0}^{\infty}
    t^{6+3k}\chi_{\frac{k}{2}}(u)\chi_{\frac{k+2}{2}}(v)\;,
    \\
    \mathrm{Vector}_{\rm{I}}(3):&\;\;\;
    \tr a\psi_{a}
    (ab)^k
    \xrightarrow{}
    -\sum_{k=0}^{\infty}
    t^{6+3k}\chi_{\frac{k+2}{2}}(u)\chi_{\frac{k}{2}}(v)\;,
\end{align}
(Dressing them with derivatives $D_{+\dot{\pm}}$ yields an overall factor $\frac{1}{(1-t^3y)(1-t^3y^{-1})}$ common to all these operators.)
Here, $\chi_{\frac{k}{2}}(u)\equiv \frac{u^{k+1}-u^{-k-1}}{u-u^{-1}}
=u^{k}+u^{k-2}+\cdots+u^{-k+2}+u^{-k}$ is the character of $SU(2)$.
The forms of cohomology representatives are merely schematic, and the two sets of global symmetry indices are symmetrized separately.
In the Vector$_{\rm{III}}$ family, the following three types of cohomologies contribute to the single-particle index.
\begin{align}
    \mathrm{Vector}_{\rm{III}}(1):&\;\;\;
    \tr \lambda_{2\dot{\alpha}}
    \lambda_2^{\dot{\alpha}}
    (ab)^k 
    \xrightarrow{}
    \sum_{k=0}^{\infty}
    t^{6+3k}\chi_{\frac{k}{2}}(u)\chi_{\frac{k}{2}}(v)\;,
    \\
    \mathrm{Vector}_{\rm{III}}(2):&\;\;\;
    \tr 
    \lambda_{2\dot{\alpha}}
    \lambda_2^{\dot{\alpha}}
    \psi_b b(ab)^k
    \xrightarrow{}
    -\sum_{k=0}^{\infty}
    t^{12+3k}\chi_{\frac{k}{2}}(u)\chi_{\frac{k+2}{2}}(v)\;,
    \\
    \mathrm{Vector}_{\rm{III}}(3):&\;\;\;
    \tr 
    \lambda_{2\dot{\alpha}}
    \lambda_2^{\dot{\alpha}}
    a\psi_a
    (ab)^k
    \xrightarrow{}
    -\sum_{k=0}^{\infty}
    t^{12+3k}\chi_{\frac{k+2}{2}}(u)\chi_{\frac{k}{2}}(v)\;,
\end{align}
In the Gravitino$_{\rm{I}}$ family, there are three types of cohomologies.
\begin{align}
    \mathrm{Gravitino}_{\rm{I}}(1):&\;\;\;
    \tr \lambda_{2\dot{\alpha}}
    (ab)^k 
    \xrightarrow{}
    -(y+y^{-1})
    \sum_{k=1}^{\infty}
    t^{3+3k}\chi_{\frac{k}{2}}(u)\chi_{\frac{k}{2}}(v)
    \;,
    \\
    \mathrm{Gravitino}_{\rm{I}}(2):&\;\;\;
    \tr 
    \lambda_{2\dot{\alpha}}
    \psi_b b(ab)^k
    \xrightarrow{}
    (y+y^{-1})
    \sum_{k=0}^{\infty}
    t^{9+3k}\chi_{\frac{k}{2}}(u)\chi_{\frac{k+2}{2}}(v)\;,
    \\
    \mathrm{Gravitino}_{\rm{I}}(3):&\;\;\;
    \tr 
    \lambda_{2\dot{\alpha}}
    a\psi_a(ab)^k
    \xrightarrow{}
    (y+y^{-1})
    \sum_{k=0}^{\infty}
    t^{9+3k}\chi_{\frac{k+2}{2}}(u)\chi_{\frac{k}{2}}(v)\;,
\end{align}
In the Gravitino$_{\rm{III}}$ family, there is one type of cohomology.
\begin{align}
    \mathrm{Gravitino}_{\rm{III}}:&\;\;\;
    \tr f_2
    (ab)^k
    \xrightarrow{}
    \sum_{k=1}^{\infty}
    t^{6+3k}\chi_{\frac{k}{2}}(u)\chi_{\frac{k}{2}}(v)
    \;,
\end{align}
In the Gravitino$_{\rm{IV}}$ family, there is one type of cohomology.
\begin{align}
    \mathrm{Gravitino}_{\rm{IV}}:&\;\;\;
    \tr f_2
    \lambda_{2\dot{\alpha}}
    \lambda_2^{\dot{\alpha}}
    (ab)^k
    \xrightarrow{}
    \sum_{k=0}^{\infty}
    t^{12+3k}\chi_{\frac{k}{2}}(u)\chi_{\frac{k}{2}}(v)
    \;,
\end{align}
In the Graviton family, there is one type of cohomology.
\begin{align}
    \mathrm{Graviton}:&\;\;\;
    \tr f_2
    \lambda_{2\dot{\alpha}}
    (ab)^k
    \xrightarrow{}
    -(y+y^{-1})
    \sum_{k=0}^{\infty}
    t^{9+ 3k}\chi_{\frac{k}{2}}(u)\chi_{\frac{k}{2}}(v)
    \;,
\end{align}

Now, let us compute the single-particle graviton index.
First, the terms independent of $u$ and $v$ come from the $k=0$ terms of Vector$_{\rm{III}}(1)$, Gravitino$_{\rm{IV}}$, and Graviton.
\begin{align}
    t^6+t^{12}-t^9(y+y^{-1})
    =t^6(1-t^3y)(1-t^3y^{-1})\;,
\end{align}
Second, from particles with $s_1=s_2$, i.e. Vector$_{\rm{I}}(1)$, Vector$_{\rm{III}}(1)$, Gravitino$_{\rm{I}}(1)$, Gravitino$_{\rm{III}}$, Gravitino$_{\rm{IV}}$, Graviton,
\begin{align}
    &(1+t^6-t^3(y+y^{-1})+t^6+t^{12}-t^9(y+y^{-1}))
    \sum_{k=1}^{\infty}t^{3k}
    \chi_{\frac{k}{2}}(u)\chi_{\frac{k}{2}}(v)
    \nonumber\\
    &=(1-t^3y)(1-t^3y^{-1})(1+t^6)
    \sum_{k=1}^{\infty}t^{3k}
    \chi_{\frac{k}{2}}(u)\chi_{\frac{k}{2}}(v)
    \;
\end{align}
Finally, from particles with $s_1\neq s_2$, i.e. Vector$_{\rm{I}}(2),(3)$, Vector$_{\rm{III}}(2),(3)$, Gravitino$_{\rm{I}}(2),(3)$,
\begin{align}
    &(-t^6-t^{12}+t^9(y+y^{-1}))
    \sum_{k=0}^{\infty}t^{3k}
    \Big(\chi_{\frac{k}{2}}(u)\chi_{\frac{k+2}{2}}(v)
    +\chi_{\frac{k+2}{2}}(u)\chi_{\frac{k}{2}}(v)\Big)
    \nonumber\\
    &=-t^6(1-t^3y)(1-t^3y^{-1})
    \sum_{k=0}^{\infty}t^{3k}
    \Big(\chi_{\frac{k}{2}}(u)\chi_{\frac{k+2}{2}}(v)
    +\chi_{\frac{k+2}{2}}(u)\chi_{\frac{k}{2}}(v)\Big)
\end{align}
Therefore, the derivative-dressing factor $1/(1-t^3y)(1-t^3y^{-1})$ cancels completely.
Note that due to this cancellation, there is no $y$ dependence left.
The single-particle index of the KK particles then simplifies to
\begin{align}
    \mathcal{I}^{\mathrm{KK}}_{\text{single particle}}(t,y,u,v,b)
    =\frac{t^3 uv}{1-t^3uv}
    +\frac{t^3 \frac{1}{uv}}{1-t^3\frac{1}{uv}}
    +\frac{t^3 \frac{u}{v}}{1-t^3\frac{u}{v}}
    +\frac{t^3 \frac{v}{u}}{1-t^3\frac{v}{u}}\;.
\end{align}
This simplification is nontrivial and occurs only after all the contributions are summed together.
The superconformal index at $N=\infty$ is
\begin{align}
    \mathcal{I}_{\mathrm{KK}}(t,y,u,v,b)&=\mathrm{PE}[\mathcal{I}^{\mathrm{KK}}_{\text{single particle}}] \nonumber\\
    &=\prod_{n=1}^{\infty}\frac{1}{(1-t^{3n}u^nv^n)(1-t^{3n}u^{-n}v^{-n})(1-t^{3n}u^{n}v^{-n})(1-t^{3n}u^{-n}v^{n})}
\end{align}
This index precisely reproduces the computation by \cite{Nakayama:2006ur} up to the decoupled $U(1)$ sector, and matches with that of \cite{Gadde:2010en}.
Note that we did not include the contributions from the baryonic current and one linear combination of the gaugino bilinears, because they cancel each other in the index.

\section{Generalized monotones at arbitrary $N$}
\label{appendix:general-N-fortuity}

This appendix gives the details behind the general-rank statement in
Section~\ref{section:abjm}.  We first record the explicit rewriting of the $N=2$ difference
of the two baryon--antibaryon products.  Although it has a single-trace part,
it is cohomologous to a graviton:
\begin{equation}
\begin{aligned}
    F^{(-)}_{i\hat{i}}&\equiv\epsilon\epsilon a_i a_j \epsilon\epsilon \psi_a^j b_{\hat i}
    -\epsilon\epsilon \psi_{b}^{\hat j}a_i \epsilon\epsilon b_{\hat j}b_{\hat i}\\
    &=-\tr a_{(i}\psi_{aj)}\tr a^jb_{\hat{i}}
    +\tr b_{(\hat{i}}\psi_{b\hat{j})}\tr b^{\hat{j}}a_i
    -\frac{3}{2}\tr a_ib_{\hat i}\tr C_1
    +\tr b_{\hat i}a_i C_1
    -\tr a_i b_{\hat i}C_2\\
    &=-\tr a_{(i}\psi_{aj)}\tr a^jb_{\hat{i}}
    +\tr b_{(\hat{i}}\psi_{b\hat{j})}\tr b^{\hat{j}}a_i
    -\frac{1}{2}\tr a_ib_{\hat i}\tr C_1
    +Q_{\mathrm{KW}}\tr (a_if_1+f_2a_i)b_{\hat i},
\end{aligned}
\end{equation}
where $C_1=\psi_{aj}a^j-b^{\hat j}\psi_{b\hat{j}}$ and $C_2=\psi_{b\hat{j}}b^{\hat{j}}-a^j\psi_{aj}$. The second line is a graviton in ABJM, while the third line is a graviton in KW theory, up to $Q_{\mathrm{KW}}$-exactness. When converting the second line to the third line, the coefficient $-1/2$ in the third line becomes $(2/N-3/2)$ for general $N$.
Also, $\tr b_{\hat i}a_i C_1+\tr a_ib_{\hat i}C_2=-Q\tr \psi_{ai}\psi_{b\hat{i}}$, so there is only one single-trace ABJM graviton.

At general rank $N$, we now show that
$F_{i_1\cdots i_{N-1}\hat i_1\cdots\hat i_{N-1}}$ is non-exact and
independent of ordinary mesonic monotones in both KW and ABJM.  It is enough to
consider the highest-weight component,
\begin{equation}
    i_1=\cdots=i_{N-1}=1,
    \qquad
    \hat i_1=\cdots=\hat i_{N-1}=1.
\end{equation}
With $a=a_1$ and $b=b_1$, define
\begin{equation}\label{eq:m1-m2-definition}
\begin{aligned}
    m_1&=\mathbf A_{1\cdots1i}\mathbf B^i_{\psi,1\cdots1},
    &m_2&=\mathbf A^{\hat i}_{\psi,1\cdots1}
          \mathbf B_{\hat i1\cdots1},\\
    F_N&=m_1+m_2,
    &F_N^{(-)}&=m_1-m_2.
\end{aligned}
\end{equation}
Each $m_i$ is a product of two baryonic monotonous cohomologies with opposite
baryon numbers and is therefore $Q$-closed.

Before proceeding, it is useful to distinguish relation-dependent $Q$-closure
from fortuity.  The lift of an ordinary fortuitous cohomology to the ordinary
mesonic covering space has a nonzero $Q$-variation which vanishes only after a
finite-rank trace relation is imposed.  The converse is not true in general: a
representative of a graviton cohomology may be shifted by a finite-rank
relation, making its $Q$-closure appear to use that relation.  Relation-dependent
$Q$-closure alone therefore does not prove independence from gravitons.  For
the operators above, we shall separately establish non-exactness and show below
that no rank-$N$ trace relation exists among operators with their field
content.  This will allow their behavior at larger rank to distinguish the
ordinary fortuitous and graviton combinations.

We first show that $m_1$ and $m_2$ are non-exact.  Evaluate the matter fields on
\begin{equation}\label{eq:central-evaluation}
    a_i=x_i\mathbf 1_N,\qquad b_{\hat i}=y_{\hat i}\mathbf 1_N,
    \qquad \psi_a^i=\eta_a^i\mathbf 1_N,\qquad
    \psi_b^{\hat i}=\eta_b^{\hat i}\mathbf 1_N,
\end{equation}
where the $\eta$'s are Grassmann variables, and set the remaining letters to
zero.  Spin, flavor charges, and fermion number restrict the relevant bosonic
preimages in KW to operators containing either $f_{1,2}$ or two matter
fermions, together with scalar dressings.  The
variations of the matter fermions vanish because the scalar
products in $Q\psi_a$ and $Q\psi_b$ are antisymmetric in flavor.  Moreover,
$Qf_{1,2}$ is proportional to the identity before its trace part is subtracted,
and hence also vanishes on \eqref{eq:central-evaluation}.  Thus every possible
$Q_{\rm KW}$-exact operator at these quantum numbers evaluates to zero.

The derivative letters require a separate check in the zero-monopole ABJM
sector considered here.  At the quantum
numbers of \eqref{eq:m1-m2-definition}, a bosonic preimage that can contribute
to the one-fermion, derivative-free sector contains either two matter fermions
or one BPS derivative, together with scalar dressings.  The first type again
vanishes after applying $Q$ on \eqref{eq:central-evaluation}.  For the second
type, the ABJM transformations take the form, up to an overall normalization
and flavor-index conventions \cite{Belin:2025hsg},
\begin{equation}
\begin{aligned}
Q(Da_i)&\ \sim\
\big(a_{[1}\psi_{a2]}+\psi_{b[\hat1}b_{\hat2]}\big)a_i
-a_i\big(\psi_{a[2}a_{1]}+b_{[\hat2}\psi_{b\hat1]}\big),\\
Q(Db_{\hat i})&\ \sim\
\big(\psi_{a[2}a_{1]}+b_{[\hat2}\psi_{b\hat1]}\big)b_{\hat i}
-b_{\hat i}\big(a_{[1}\psi_{a2]}+\psi_{b[\hat1}b_{\hat2]}\big).
\end{aligned}
\end{equation}
On \eqref{eq:central-evaluation}, the matrix coefficients obey
$a_{[1}\psi_{a2]}=\psi_{a[2}a_{1]}$ and
$\psi_{b[\hat1}b_{\hat2]}=b_{[\hat2}\psi_{b\hat1]}$.  All of them also
commute with $a_i$ and $b_{\hat i}$, so the two terms in each line cancel.
Hence the relevant $Q_{\rm ABJM}$-exact operators also evaluate to zero.

On the other hand, the two operators in \eqref{eq:m1-m2-definition} evaluate,
up to nonzero normalizations, to
\begin{equation}
    m_1\ \longmapsto\
    x_1^{N-1}y_1^{N-1}x_i\eta_a^i,\qquad
    m_2\ \longmapsto\
    x_1^{N-1}y_1^{N-1}\eta_b^{\hat i}y_{\hat i}.
\end{equation}
These are nonzero for generic coefficients.  By setting one set of fermionic
coefficients to zero, the same evaluation shows separately that $F_N$ and
$F_N^{(-)}$ are not $Q$-exact in either theory.

We next compare these cohomologies with ordinary mesonic monotones.  Converting
the gauge epsilon tensors into antisymmetrized Kronecker deltas gives
\begin{equation}\label{eq:m1-m2-mesonic}
\begin{aligned}
    m_1&=N!^2a^{\hat p_1}{}_{p_1}\cdots
    a^{\hat p_{N-1}}{}_{p_{N-1}}(a_i)^{\hat p_N}{}_{p_N}
    (\psi_a^i)^{[p_1}{}_{[\hat p_1}b^{p_2}{}_{\hat p_2}\cdots
    b^{p_N]}{}_{\hat p_N]},\\
    m_2&=N!^2a^{\hat p_1}{}_{p_1}\cdots
    a^{\hat p_{N-1}}{}_{p_{N-1}}(\psi_b^{\hat i})^{\hat p_N}{}_{p_N}
    (b_{\hat i})^{[p_1}{}_{[\hat p_1}b^{p_2}{}_{\hat p_2}\cdots
    b^{p_N]}{}_{\hat p_N]}.
\end{aligned}
\end{equation}
Expanding the antisymmetrizations rewrites $m_1$ and $m_2$ as mesonic
multi-traces.  Their single-trace terms are
\begin{equation}\label{eq:m1-m2-single-trace}
\begin{aligned}
    \left.m_1\right|_{\rm st}
    &=(N-1)!^2(-1)^{N-1}\tr\psi_a^i\Big[
    (ab)^{N-1}a_i+(ab)^{N-2}a_i ba+\cdots
    +ab a_i(ba)^{N-2}+a_i(ba)^{N-1}\Big],\\
    \left.m_2\right|_{\rm st}
    &=(N-1)!^2(-1)^{N-1}\tr\psi_b^{\hat i}\Big[
    (ba)^{N-1}b_{\hat i}+(ba)^{N-2}b_{\hat i}ab+\cdots
    +ba b_{\hat i}(ab)^{N-2}+b_{\hat i}(ab)^{N-1}\Big].
\end{aligned}
\end{equation}
Group each adjacent pair of oppositely charged bifundamental letters into one adjoint matrix.
Each term in \eqref{eq:m1-m2-single-trace} contains $N$ such matrix factors, or
$2N$ elementary bifundamental letters.  The first rank-$N$ trace relations
involve $N+1$ adjoint matrix factors, so no trace relation at this degree can
relate the single-trace terms in \eqref{eq:m1-m2-single-trace} to the
higher-trace terms.  The identity \eqref{eq:m1-m2-mesonic} between the
baryonic products and their mesonic expressions is not a vanishing relation
among mesonic multi-traces and does not alter this conclusion.
The single-trace terms in \eqref{eq:m1-m2-single-trace} are not separately
$Q$-closed.  Their variations contain $N+1$ adjoint matrix factors and are cancelled
by the variations of the higher-trace terms only after the first rank-$N$
trace relation is imposed.  Thus there is no trace relation among the terms of
$m_i$ themselves, even though their $Q$-closure as mesonic expressions uses one.

To follow the same mesonic expressions away from rank $N$, keep the integer
$N$ in \eqref{eq:m1-m2-mesonic} fixed and evaluate its right-hand sides on
$N'\times N'$ matrices.  In the language of Section 3, these fixed multi-trace
polynomials are elements of the ordinary mesonic covering space, and this
evaluation is their projection to rank $N'$.  We denote the resulting
polynomials by $M_1$ and $M_2$.  At $N'=N$ they agree with $m_1$ and $m_2$,
while they vanish for $N'<N$.  This continuation should not be confused with
replacing $N$ by $N'$ in the original factorized baryonic operators.  In the
latter family the number of fields grows with $N'$, whereas $M_1$ and $M_2$
retain the energies $E_{\rm KW}=\frac32N+\frac12$ and
$E_{\rm ABJM}=N+\frac12$.

We next show directly that the difference $M_1-M_2$ is $Q$-closed without
using trace relations.  Introduce the completely antisymmetrized scalar
polynomial underlying $M_1$ and $M_2$,
\begin{equation}
P_N(a,b)=N!^2a^{[\hat p_1}{}_{[p_1}\cdots
a^{\hat p_N]}{}_{p_N]}b^{p_1}{}_{\hat p_1}\cdots
b^{p_N}{}_{\hat p_N}.
\end{equation}
It is gauge invariant at every matrix size.  Therefore, for arbitrary adjoint
matrices $X$ and $\widehat X$,
\begin{equation}\label{eq:PN-gauge-invariance}
\left[\tr(\widehat Xa-aX)\partial_a
+\tr(Xb-b\widehat X)\partial_b\right]P_N(a,b)=0,
\end{equation}
where $\partial_a$ and $\partial_b$ act componentwise.  This is simply the
infinitesimal gauge variation of $P_N$.  A direct variation of $M_1$ and $M_2$
can be written as
\begin{equation}\label{eq:QM1-QM2-derivatives}
\begin{aligned}
QM_1={}&\frac{1}{N}\tr(ba_2b_2-b_2a_2b)\partial_bP_N
+\frac{1}{N^2}
 \big[\tr(b_2ab-bab_2)\partial_b\big]
 \big[\tr(a_2\partial_a)\big]P_N,\\
QM_2={}&\frac{1}{N}\tr(a_2b_2a-ab_2a_2)\partial_aP_N
-\frac{1}{N^2}
 \big[\tr(a_2ba-aba_2)\partial_a\big]
 \big[\tr(b_2\partial_b)\big]P_N.
\end{aligned}
\end{equation}

For the two-derivative term in $QM_1$, choose
$\widehat X=ab_2$ and $X=b_2a$ in
\eqref{eq:PN-gauge-invariance}, and then act with
$\tr(a_2\partial_a)$.  For the corresponding term in $QM_2$, choose
$\widehat X=a_2b$ and $X=ba_2$, and then act with
$\tr(b_2\partial_b)$.  The product rule gives
\begin{equation}\label{eq:QM1-QM2-reduction-identities}
\begin{aligned}
0={}&\Big\{\tr(b_2a_2b-ba_2b_2)\partial_b
+\big[\tr(b_2ab-bab_2)\partial_b\big]
 \big[\tr(a_2\partial_a)\big]\Big\}P_N,\\
0={}&\Big\{\tr(a_2b_2a-ab_2a_2)\partial_a
+\big[\tr(a_2ba-aba_2)\partial_a\big]
 \big[\tr(b_2\partial_b)\big]\Big\}P_N.
\end{aligned}
\end{equation}
Substituting these identities into \eqref{eq:QM1-QM2-derivatives} reduces the
two variations to
\begin{equation}\label{eq:QM1-QM2-reduced}
\begin{aligned}
QM_1&=\left(\frac{1}{N}+\frac{1}{N^2}\right)
\tr(ba_2b_2-b_2a_2b)\partial_bP_N,\\
QM_2&=\left(\frac{1}{N}+\frac{1}{N^2}\right)
\tr(a_2b_2a-ab_2a_2)\partial_aP_N.
\end{aligned}
\end{equation}
Finally, \eqref{eq:PN-gauge-invariance} with
$\widehat X=a_2b_2$ and $X=b_2a_2$ states that the two expressions on the
right-hand side of \eqref{eq:QM1-QM2-reduced} are equal.  Therefore,
\begin{equation}\label{eq:Mplus-Mminus-Q}
    Q(M_1-M_2)=0,\qquad Q(M_1+M_2)=2QM_1.
\end{equation}
This proof uses only gauge invariance, not any finite-rank trace relation.  It
follows from the non-exactness of $F_N^{(-)}$ shown above that $M_1-M_2$ is a
nontrivial graviton cohomology.

By contrast, $M_1+M_2$ is closed at $N'=N$ but not at larger rank.  This can be
checked already at $N'=N+1$.  Label the additional direction by $0$ and set
\begin{equation}
\begin{aligned}
    a&=E_{00}+\sum_{r=2}^N E_{rr},&\qquad a_2&=E_{11},\\
    b&=E_{10}+\sum_{r=2}^N E_{rr},& b_2&=E_{01},
\end{aligned}
\end{equation}
with all other scalar components zero.  Here $E_{ij}$ has a single nonzero
entry in row $i$ and column $j$.  On this configuration, $QM_1$ is a nonzero
multiple of $N!^2(1/N+1/N^2)$.  Together with
\eqref{eq:Mplus-Mminus-Q}, this proves that $Q(M_1+M_2)\neq0$ at $N'=N+1$.
Embedding the same configuration into larger matrices gives the same result
for every $N'>N$.

Finally, suppose that $F_N$ were cohomologous at rank $N$ to an ordinary
mesonic monotone.  Their difference would be $Q$-exact.  Every term that can
contribute to this equality is below the first rank-$N$ trace relation; they
contain only the $N$ mesonic matrix factors inside traces.  Thus, there is no
finite-rank relation with which to change the lift of this equality.  The same
equality would therefore hold before imposing the rank-$N$ relations and would
make the fixed polynomial $M_1+M_2$ $Q$-closed at arbitrary rank.  This
contradicts its nonzero $Q$-variation at $N'=N+1$.
Hence, $F_N$ is independent of ordinary mesonic monotones in both KW and ABJM.
It is an ordinary fortuitous cohomology, while its factorization in
\eqref{eq:m1-m2-definition} displays the generalized-monotone pattern.

\clearpage

\section{Refined $N=2$ BMN index}
\label{appendix:BMN-index-coefficients}
This appendix records the complete refined numerator polynomial $\mathcal P_{59}(x;\boldsymbol{\xi})$ appearing in \eqref{eq:BMN-index-N2}. We use the notation $\chi_{[a,b,c]}(\boldsymbol{\xi})$ for the
irreducible $SU(4)$ character with Dynkin labels $[a,b,c]$ and write
\begin{equation}
  \mathcal P_{59}(x;\boldsymbol{\xi})
  =\sum_{n=0}^{59}p_n(\boldsymbol{\xi})x^n\ ,
  \qquad
  p_n(\boldsymbol{\xi})
  =\sum_{a,b,c\geq0}m^{(n)}_{a,b,c}\,\chi_{[a,b,c]}(\boldsymbol{\xi})\ .
\end{equation}
Below, we suppress the argument $\boldsymbol{\xi}$ of $p_n$ and $\chi_{[a,b,c]}$.
The coefficient of each displayed character is $m^{(n)}_{a,b,c}$. Every character not displayed at a given order has zero coefficient.
\begingroup
\allowdisplaybreaks[4]

\endgroup

As a consistency check, setting all flavor fugacities to unity and using
\begin{equation}
  \chi_{[a,b,c]}(\boldsymbol{1})=\dim[a,b,c]
  =\frac{(a+1)(b+1)(c+1)(a+b+2)(b+c+2)(a+b+c+3)}{12}
\end{equation}
reproduces the unrefined polynomial \eqref{eq:P59-unrefined}.

\section{Recursive construction of $\widehat{Z}_{t}$}
\label{appendix:explicit-Zhat1}

In this appendix, we describe a recursive ansatz for constructing
$Q$-closed completions $\widehat Z_t \equiv z^{\odot t} \odot O_{10} + \cdots \in [4+2t,0,0]$. Suppose that there exist $A_j\in[4+2j,0,0]$, with $A_0 = O_{10}$, that satisfy
\begin{equation}
 QA_0=0\ ,
 \qquad
 QA_j=-(Qz)\odot A_{j-1}
 \quad (j\geq1)\ .
 \label{eq:universal-Aj-descent}
\end{equation}
Then, for every $t$ for which $A_0,\ldots,A_t$ have been constructed,
\begin{equation}
 \widehat Z_t
 =\sum_{j=0}^{t}\frac{t!}{(t-j)!}\,
  z^{\odot(t-j)}\odot A_j
\end{equation}
is $Q$-closed. In addition, imposing the truncation condition
\begin{equation}\label{eq:Zt-truncation2}
    T\left(
        \widehat Z_t^{\underbrace{44\cdots4}_{4+2t}}
    \right)
    \simeq_{Q}
    \frac{3}{8^t}\hat\phi^3O_t
\end{equation}
ensures that $\widehat Z_t$ is not $Q$-exact, where $O_t = \hat\phi^{2t}(f \cdot f)^t O_0 + \cdots$ is a fortuitous primary of $SU(2)$ $\mathcal N=4$ SYM \cite{Choi:2023znd}.

We have explicitly solved \eqref{eq:universal-Aj-descent} through
$A_2$, thereby constructing $\widehat Z_1$ and $\widehat Z_2$.
The resulting explicit expression for $\widehat Z_1$ is presented
below. Although the full expression for $\widehat Z_2$ is too lengthy
to display, we have used it to compute $\mathcal C_{r,s,2}$, as
described in \eqref{eq:general-t-endpoint-coefficients}.
For $t\geq3$, one must continue solving
\eqref{eq:universal-Aj-descent} recursively. The resulting systems
become increasingly cumbersome, and their solvability may be subject
to further cohomological obstructions. We leave this problem for
future work.

Now, we present the explicit form of $\widehat{Z}_1^{I_1I_2I_3I_4I_5I_6} \equiv z^{(I_1I_2}O_{10}^{I_3I_4I_5I_6)}+\cdots$, where $\cdots$ denotes the additional terms required for $Q$-closure. Here, $A,B,C,\ldots$ label $SU(4)$ indices, while $\mu,\nu,\rho,\ldots$ label $SO(4)$ gauge indices.
\begingroup
\allowdisplaybreaks[4]

\endgroup
One may verify directly that this expression is \(Q\)-closed and that it satisfies \eqref{eq:Zt-truncation2} for \(t=1\).

\bibliography{References}

\end{document}